\documentclass[12pt,a4paper,superscriptaddress,preprint,groupedaddress]{revtex4-1}
\usepackage{graphicx} 
\usepackage{hyperref}
\usepackage{caption}
\usepackage{color}
\usepackage{stackengine}
\usepackage[position=t,singlelinecheck=off]{subcaption}
\usepackage{array}
\usepackage{multirow}
\usepackage{enumerate}
\usepackage[nameinlink]{cleveref} 
\usepackage{amsmath}
\usepackage{amssymb}
\usepackage{amsthm}
\usepackage{amsfonts}
\usepackage{amssymb}
\usepackage{ascmac}
\usepackage{url}
\usepackage{bm}
\usepackage[utf8]{inputenc}
\usepackage[T1]{fontenc}
\usepackage[scaled]{helvet}%
\usepackage{lmodern}
\def\A{{\bm A}}
\def\B{{\bm B}}

\def\x{{\bm x}}
\def\Tunc{Tun\symbol{231}}
\def\c{{\bm c}}

\def\ER{Erd\H{o}s-R\'{e}nyi\ }
\crefname{equation}{Eq.\!}{Eqs.\!}
\crefname{figure}{Fig.\!}{Figs.\!}
\usepackage{blindtext}
\usepackage{xcolor}
\hypersetup{
  colorlinks   = true, 
  urlcolor     = black, 
  linkcolor    = blue, 
  citecolor   = blue 
}

\usepackage{microtype} 

\definecolor{mygreen}{RGB}{76, 175, 80}
\definecolor{myblue}{RGB}{33,150,243}
\definecolor{myred}{RGB}{254,0,0}
\def\ER{Erd\H{o}s-R\'{e}nyi\ }
\makeatletter
\newcommand*{\rom}[1]{\expandafter\@slowromancap\romannumeral #1@}
\makeatother

\begin{document}
\date{\today}
\title{Finding multiple core-periphery pairs in networks}
\author{Sadamori Kojaku}
\author{Naoki Masuda}
\email{naoki.masuda@bristol.ac.uk}
\affiliation{Department of Engineering Mathematics,
Merchant Venturers Building, University of Bristol,
Woodland Road, Clifton, Bristol BS8 1UB, United Kingdom}
\begin{abstract}
With a core-periphery structure of networks, core nodes are densely interconnected, peripheral nodes are connected to core nodes 
to different extents, and peripheral nodes are sparsely interconnected.
Core-periphery structure composed of a single core and periphery has been identified for various networks.
However, analogous to the observation that many empirical networks are composed of densely interconnected groups of nodes, i.e., communities, a network may be better regarded as a collection of multiple cores and peripheries.
We propose a scalable algorithm to detect multiple nonoverlapping groups of core-periphery structure in a network.
We illustrate our algorithm using synthesised and empirical networks. 
For example, we find distinct core-periphery pairs with different political leanings in a network of political blogs and 
separation between international and domestic subnetworks of airports in some single countries in a worldwide airport network.   
\end{abstract}
\maketitle
\section{Introduction}
Many complex systems can be expressed as networks in which a node represents an object (e.g., person, web page, protein) and an edge represents the relationship between two objects (e.g., friendship, hyperlink, physical interaction).
A network can be characterised by microscale, mesoscale and macroscale structural patterns such as the degree (i.e., the number of edges that a node has), clustering coefficient, and diameter \cite{Newman2010,Barabasi2016}.
Among various structural properties of networks, community structure is a representative mesoscale structure of networks \cite{Fortunato2010}.
A community is a group of nodes that are densely interconnected and sparsely connected to nodes in different communities.
Nodes in the same community often share a role \cite{Girvan2002,Newman2004,Adamic2005,Guimera2005,Sales-Pardo2007,Fortunato2010,Karrer2011} (for an exceptional case, see Ref.~\cite{Guimera2005a}), and therefore identifying communities aids classification of nodes and visualisation of networks \cite{Fortunato2010}.

Core-periphery structure is another mesoscale structure of networks, with which we view a network as being composed of two groups of nodes called the core and periphery.
Although the definition varies, a core is often defined as a group of densely interconnected nodes, and a periphery as a group of nodes that are densely connected (i.e., adjacent) to the core nodes but not to other peripheral nodes \cite{Borgatti2000,Holme2005,Boyd2010,Csermely2013,Lee2014,Rombach2014,Yang2014,Ma2015,Tunc2015,Zhang2015,Cucuringu2016,Gamble2016,Verma2016}.
A core and community are both groups of densely interconnected nodes but have a difference; a core connects densely to its periphery, whereas a community is not densely connected to other nodes outside it.
Core-periphery structure has been found in various networks including brain networks \cite{Bassett2013}, metabolic networks \cite{DaSilva2008}, protein interaction networks \cite{Bruckner2015}, social networks \cite{Borgatti2000,Rombach2014,Yang2014,Gamble2016}, international trade networks \cite{Boyd2010,Rossa2013,Ma2015}, financial networks \cite{Craig2014,Lee2014,Fricke2015} and transportation networks \cite{Holme2005,Lee2014,Rombach2014}.
For example, in a coauthorship network among researchers, leading researchers often publish papers with other leading researchers, forming a core, 
while other researchers tend to publish papers with particular leading researchers such as those in the same research group, forming a periphery \cite{Rombach2014}. 

Borgatti and Everett introduced the first quantitative formulation of core-periphery structure \cite{Borgatti2000}.
In the discrete version of core-periphery structure, which we will focus on in this paper, they introduced an idealised core-periphery structure in which core nodes are adjacent to all other nodes, and peripheral nodes are adjacent to all core nodes but not to any peripheral nodes.
Although it is also realistic to assume that the core-periphery connectivity is sparser than the core-core connectivity \cite{Borgatti2000}, we will focus only on the idealised core-periphery structure in the present study.
Borgatti and Everett sought for the assignment of all nodes in a given network to a core or periphery such that the network is as close as possible to an idealised core-periphery structure.
Following their study, many core-periphery detection algorithms have been developed \cite{Borgatti2000,Holme2005,DaSilva2008,Boyd2010,Rossa2013,Lee2014,Rombach2014,Yang2014,Ma2015,Zhang2015,Cucuringu2016,Gamble2016}.
These algorithms aim to identify a single core-periphery pair embedded in a network. 
However, a network may be better regarded as a collection of multiple core-periphery pairs \cite{Borgatti2000,Csermely2013,Rombach2014,Tunc2015,Zhang2015,Cucuringu2016}, which is the focus of the present study.
For example, co-authorship networks would be composed of multiple groups of researchers.
Researchers would often collaborate with the leading researchers in the same group but not with other researchers in the same group, which may lead to core-periphery structure within the group \cite{Rombach2014}.
Previous studies in this direction have not provided a tailored scalable algorithm to this end. 
A study focused on a related but different type of multiple core-periphery structure \cite{Yan2016}.
Other algorithms aim to detect multiple cores but do not assume that peripheral nodes are sparsely connected to each other \cite{Yang2014,Sardana2016,Xiang2016}.
A network can also have multiple disjoint cores in the form of $k$-cores \cite{Alvarez-Hamelin2005}, $k$-trusses \cite{Cohen2008} or dense subgraphs \cite{Zhao2011,Chen2012}. However, the corresponding algorithms do not tell how densely peripheral nodes are connected to each other or to which core a peripheral node belongs.
An algorithm to find various mesoscale structure of networks including multiple core-periphery pairs \cite{Tunc2015} is computationally costly and only feasible for small networks (Appendix~\ref{sec:birkan}).

We present a scalable algorithm to detect multiple nonoverlapping core-periphery pairs in networks, each of 
which is as close as possible to an idealised core-periphery structure. 
Our algorithm automatically determines the number and the size of the core-periphery pairs. 
Various algorithms to detect core-periphery structure in networks are classified as density-based and transport-based algorithms \cite{DaSilva2008,Lee2014,Cucuringu2016}. 
Densely-based algorithms posit that a core is a densely connected group of nodes, whereas transport-based algorithms posit that a core is a group of nodes that can be reached from other nodes along short paths. In the present study, we focus on density-based algorithms.

\section{Methods}
\label{sec:method}
\subsection{Algorithm}
We extend the idealised core-periphery structure introduced by Borgatti and Everett \cite{Borgatti2000} to the case of multiple pairs of a core and a periphery. 
In the Borgatti--Everett (BE) algorithm, one considers a graph (i.e., network) composed of $N$ nodes and $M$ edges.
Let $\A=(A_{ij})$ be the adjacency matrix, i.e., $A _{ij}=1$ if nodes $i$ and $j$ are adjacent by an edge, and $A_{ij}=0$ otherwise.
We assume an undirected and unweighted network without self-loops, i.e., $A_{ij}=A_{ji}$ and $A_{ii}=0$ for all $i$ and $j$. 
Let $\x=(x_1,x_2,\ldots,x_N)$ be a vector of length $N$, where $x_{i}=0$ if node $i$ is a peripheral node, and $x_i=1$ if node $i$ is a core node.
We define the idealised core-periphery structure as the network where each core node is adjacent to all core and peripheral nodes, and each peripheral node is adjacent to all core nodes but no peripheral node.
The corresponding adjacency matrix, $\B(\x) = (B_{ij}(\x))$, is given by
\begin{align}
    \label{eq:bg}
    B _{ij} (\x) =  
    \left\{
    \begin{array}{ll}
        1 & \mbox{($x_i=1$ or $x_j=1$, and $i \neq j$)}, \\
        0 & \mbox{(otherwise)}.
    \end{array}
    \right. 
\end{align}
The discrete version of the BE algorithm, which we consider in the present study, seeks $\x$ that maximises similarity between $\A$ and $\B(\x)$. We will explain the similarity measure in Section \ref{sec:stest}.

We extend the idealised core-periphery structure to the case of multiple core-periphery pairs.
Let $C$ be the number of core-periphery pairs and $\c=(c_1,c_2,\ldots,c_N)$ be a vector of length $N$, where $c_{i} \in \{1,2,\ldots,C\}$ is the index of the core-periphery pair to which node $i$ belongs.
We exclude overlaps between core-periphery pairs, and between the core and periphery in each core-periphery pair.
The corresponding adjacency matrix, $\B(\c,\x)$, is given by 
\begin{align}
    \label{eq:ideal2}
    B_{ij} (\c,\x) =  
    \left\{
    \begin{array}{ll}
        \delta_{c_i,c_j} & \mbox{($x_i=1$ or $x_j=1$, and $i\neq j$)},\\
        0 & \mbox{(otherwise)}, 
    \end{array}
    \right.
\end{align}
where $\delta$ is Kronecker delta. 

We seek $(\c,\x)$ that makes $\B(\c,\x)$ the closest to $\A$ by maximising 
\begin{align}
    Q^{\text{cp}}(\c,\x) &= \sum_{i=1}^N \sum_{j=1}^{i-1}A_{ij} B_{ij}(\c,\x) - \sum_{i=1}^N \sum_{j=1}^{i-1}p B_{ij} (\c,\x) \nonumber \\
      &= \sum_{i=1}^N \sum_{j=1}^{i-1} (A_{ij} - p) (x_i + x_j-x_i x_j) \delta_{c_i,c_j}, \label{eq:proposed}
\end{align}
where $p=M/[N(N-1)/2]$ is the density of edges in the network.
The term $\sum_{i=1}^N \sum_{j=1}^{i-1}A_{ij}\\B_{ij}(\c,\x)$ represents the number of edges that are present in both the given network and the idealised core-periphery structure.
The null-model term $\sum_{i=1}^N \sum_{j=1}^{i-1} p B_{ij} (\c,\x)$ is the expected number of edges that are present in both the idealised core-periphery structure and an \ER random graph \cite{erdHos1959random}, in which each pair of nodes is adjacent with probability $p$.
The $Q^{\text{cp}}$ ranges between $-M$ and $M$. 
A large $Q^{\text{cp}}$ value indicates that the given network and the idealised core-periphery structure share more edges than by chance.
The \ER random graph model is widely used in the analysis of core-periphery structure \cite{Boyd2006,Boyd2010,Fricke2015,Bassett2013,Craig2014,IntVeld2014}. 
Similar to modularity for community detection, our formulation permits the use of different null models such as the configuration model.
See Section~\ref{sec:discussion} for further discussion.%

\subsection{Maximisation of $Q^{\text{cp}}$}
\label{sec:algorithm}
We use a label switching heuristic \cite{Raghavan2007,Blondel2008} to maximise $Q^{\text{cp}}$.
First, we assign each node to a different core by setting $(c_i,x_i) = (i,1)$ ($1 \le i \le N$). Then we scan all nodes in a random order.
For each scanned node $i$, we calculate the increment in $Q^{\text{cp}}$ when we tentatively update $(c_i, x_i)$ to the core of the core-periphery pair that a neighbour of node $i$, denoted by $j$, belongs to, i.e., $(c_j, 1)$. We also calculate the increment in $Q^{\text{cp}}$ when we tentatively update $(c_i, x_i)$ to $(c_j, 0)$. Note that we experiment on these two cases regardless of whether $x_j = 0$ or $x_j=1$. 
We carry out this procedure for all neighbours of $i$ to measure the increment in $Q^{\text{cp}}$ in each case. 
Finally, we update $(c_i, x_i)$ to the label that has yielded the largest tentative increment in $Q^{\text{cp}}$ (i.e., $(c_j, 0)$ or $(c_j, 1)$ for a neighbour $j$).
If any relabelling does not increase $Q^{\text{cp}}$, we do not update $(c_i,x_i)$.
When we have scanned all nodes, we stop the entire procedure if no node has changed its label in the present round. 
Otherwise, we draw a new random order of nodes and scan all nodes again according to the new random order.

The increment in $Q^{\text{cp}}$ by changing node $i$'s label from $(c,x)$ to $(c ', x ')$ is given by  
\begin{align}
    \label{eq:dq}
    &\left[ \tilde{d}_{i,(c' ,1)}  + x' \tilde{d}_{i,(c' ,0)} - p \left( \tilde{N}_{(c' ,1)}  + x' \tilde{N}_{(c' ,0)} - \delta_{c,c'}\right) \right] \nonumber \\
    & -[ \tilde{d}_{i,(c ,1)} + x \tilde{d}_{i,(c ,0)} - p (  \tilde{N}_{(c ,1)}  + x \tilde{N}_{(c ,0)}-x ) ],
\end{align}
where $\tilde{d}_{i,(c,x)}$ is the number of neighbours of node $i$ that have label $(c,x)$, and 
$\tilde{N}_{(c,x)}$ is the number of nodes with label $(c,x)$.
For each scanned node $i$, we calculate Eq.~(\ref{eq:dq}) at most $2d_i$ times, where $d_i$ is the degree of node $i$.   
Therefore, the time complexity for scanning all nodes in one round is ${\cal O}\left(\sum_{i=1}^N d_i\right)={\cal O}(M)$, and that of the entire algorithm is ${\cal O}(rM)$, where $r$ is the number of rounds.
We run this algorithm 20 times starting from the same initial condition stated above and adopt the node labelling that produces the largest value of $Q^{\text{cp}}$.

\subsection{Significance of the core-periphery structure}
\label{sec:stest}
A detected core-periphery structure may be statistically insignificant \cite{Borgatti2000,Boyd2006}.
Therefore, we adapt a statistical test in the case of a single core-periphery pair \cite{Boyd2006} to the case of multiple core-periphery pairs.

In the statistical test for a single core-periphery pair \cite{Boyd2006},
we measure the significance of a core-periphery pair by a quality function based on the Pearson correlation coefficient \cite{Borgatti2000}, which is defined by
\begin{align}
    \label{eq:bgquality}
    Q_{\text{BE}} ^{\text{cp}} = 
    \frac{
        \sum_{i=1}^N\sum_{j=1}^{i-1}(A_{ij}-p)(B_{ij}(\x)-p_B)
    }{
        \sqrt{\sum_{i=1}^N\sum_{j=1}^{i-1}(A_{ij}-p)^2}\sqrt{ \sum_{i=1}^N\sum_{j=1}^{i-1}(B_{ij}(\x)-p_B)^2}
    },
\end{align}
where $p_B=\sum_{i=1}^{N}\sum_{j=1}^{i-1}B_{ij}(\x)/[N(N-1)/2]$.
A core-periphery pair detected for the given network is deemed to be significant if $Q^{\text{cp}} _{\text{BE}}$ is statistically larger than $Q_{\text{BE}} ^{\text{cp}}$ values calculated for the \ER random graph model, in which the number of edges is the same as that of the original network.
One generates many networks using the \ER random graph and maximises $Q_{\text{BE}} ^{\text{cp}}$ for each network. 
The Kernighan-Lin (KL) algorithm \cite{Kernighan1970} is used for maximising $Q_{\text{BE}} ^{\text{cp}}$.
The core-periphery pair detected for the original network is significant at a significance level of $\alpha \in (0,1]$  if the $Q_{\text{BE}} ^{\text{cp}}$ value for the original network is larger than a fraction $1-\alpha$ of the $Q_{\text{BE}} ^{\text{cp}}$ values for the randomised networks.

In the case of multiple core-periphery pairs, we apply essentially the same statistical test to each core-periphery pair detected in the original network.
For each detected core-periphery pair, we first calculate $Q_{\text{BE}} ^{\text{cp}}$.
Second, we generate 3,000 networks using the \ER random graph, which have the same number of nodes and edges as those of the core-periphery pair.
In counting the number of edges, we only consider the edges connecting nodes within the core-periphery pair. 
Third, we detect a single core-periphery pair in each randomised network by maximising $Q_{\text{BE}} ^{\text{cp}}$ using the KL algorithm.
Fourth, we compare the obtained $Q_{\text{BE}} ^{\text{cp}}$ values between the original and randomised networks.
If a core-periphery pair is judged to be insignificant, we call the corresponding nodes the residual nodes, i.e., those not belonging to any significant core-periphery pair. 

If we test $C$ core-periphery pairs at a significance level of $\alpha$, then the probability of making at least one false positive (i.e., an insignificant core-periphery pair is judged to be significant) is $1-(1-\alpha)^C$, which increases as $C$ increases.
To remedy this multiple comparison problem, we adopt the {\v{S}}id{\'{a}}k correction, with which we test each core-periphery pair at a significance level of $\alpha_1 = 1-(1-\alpha)^{1/C}$, which is equivalent to $1-(1-\alpha_1)^C=\alpha$ \cite{Sidak1967}. 
We set $\alpha=0.01$.

We have decided to use $Q_{\text{BE}} ^{\text{cp}}$ maximised by the KL algorithm as the test statistic to compare the original and randomised networks. 
However, we can also use different algorithms to maximise $Q_{\text{BE}} ^{\text{cp}}$. 
We can also use a different test statistic such as $Q^{\text{cp}}$ restricted to the case of the one core-periphery pair (i.e., $c_i = 1, 1\le i\le N$).

\section{Variation of Information}
For the synthetic networks with planted core-periphery structure, we measure the difference between the true core-periphery structure $(\c,\x)$ and the inferred core-periphery structure  $(\hat \c,\hat \x)$ by the variation of information (VI) \cite{Meila2007}. The VI is given by
\begin{align}
    \label{eq:vi}
    \text{VI}=
     -\sum_{(c,x)}\sum_{(\hat c, \hat x)} 
    P(c,x;\hat c, \hat x)
    \log 
    \frac{
        \left[P(c,x;\hat c, \hat x)\right]^2
    }{
        \left[\sum_{(\hat c',\hat x')}P(c,x;\hat c',\hat x')\right] \times
        \left[\sum_{(c',x')}P(c',x';\hat c,\hat x)\right]
    } 
    , 
\end{align}
where $P(c,x;\hat c,\hat x)$ is the fraction of nodes that have the true label $(c,x)$ and inferred label $(\hat c,\hat x)$.
The VI value is equal to zero if and only if the inferred core-periphery structure is the same as the true one.
We measure the performance of an algorithm by averaging VI values over the 100 generated networks.

\section{Results}
\label{sec:experiment}
We compare the proposed algorithm with the BE algorithm, which detects a single core-periphery pair by maximising $Q_{\text{BE}} ^{\text{cp}}$ using the KL algorithm \cite{Kernighan1970,Boyd2006}. We refer to the latter algorithm as the BE--KL algorithm.
We also compare our algorithm with two other algorithms; two-step and divisive algorithms.  
The two-step algorithm partitions the nodes into core and peripheral nodes using the BE--KL algorithm and also partitions the same nodes into nonoverlapping communities by maximising modularity using the Louvain algorithm \cite{Blondel2008}. 
Then we regard the core and peripheral nodes in each detected community as a core-periphery pair.
The divisive algorithm \cite{Girvan2002,Newman2004,Fortunato2010,Flores2013} partitions the nodes into communities using the Louvain algorithm \cite{Blondel2008} and then partitions the nodes in each community into core and peripheral nodes using the BE--KL algorithm. 
The two-step and divisive algorithms provided similar results.
Therefore, we report the results obtained from the two-step algorithm in this section and 
those obtained from the divisive algorithm in Appendix~\ref{sec:divisive}. 
We apply the statistical test (Section \ref{sec:stest}) to the core-periphery pairs detected by the BE--KL, two-step and our algorithms.   
We do not compare these algorithms with the algorithm introduced by \Tunc\ and Verma \cite{Tunc2015} because of a low speed and insufficient performance 
of their algorithm on model networks with planted core-periphery structure (Appendix~\ref{sec:birkan}). 

\subsection{Synthetic networks}
\label{sec:synthe}
\begin{figure} 
    \centering
    \begin{tabular}{cc}
    	\begin{minipage}{0.33\hsize}	
    		\centering
    		\includegraphics[width=1\hsize]{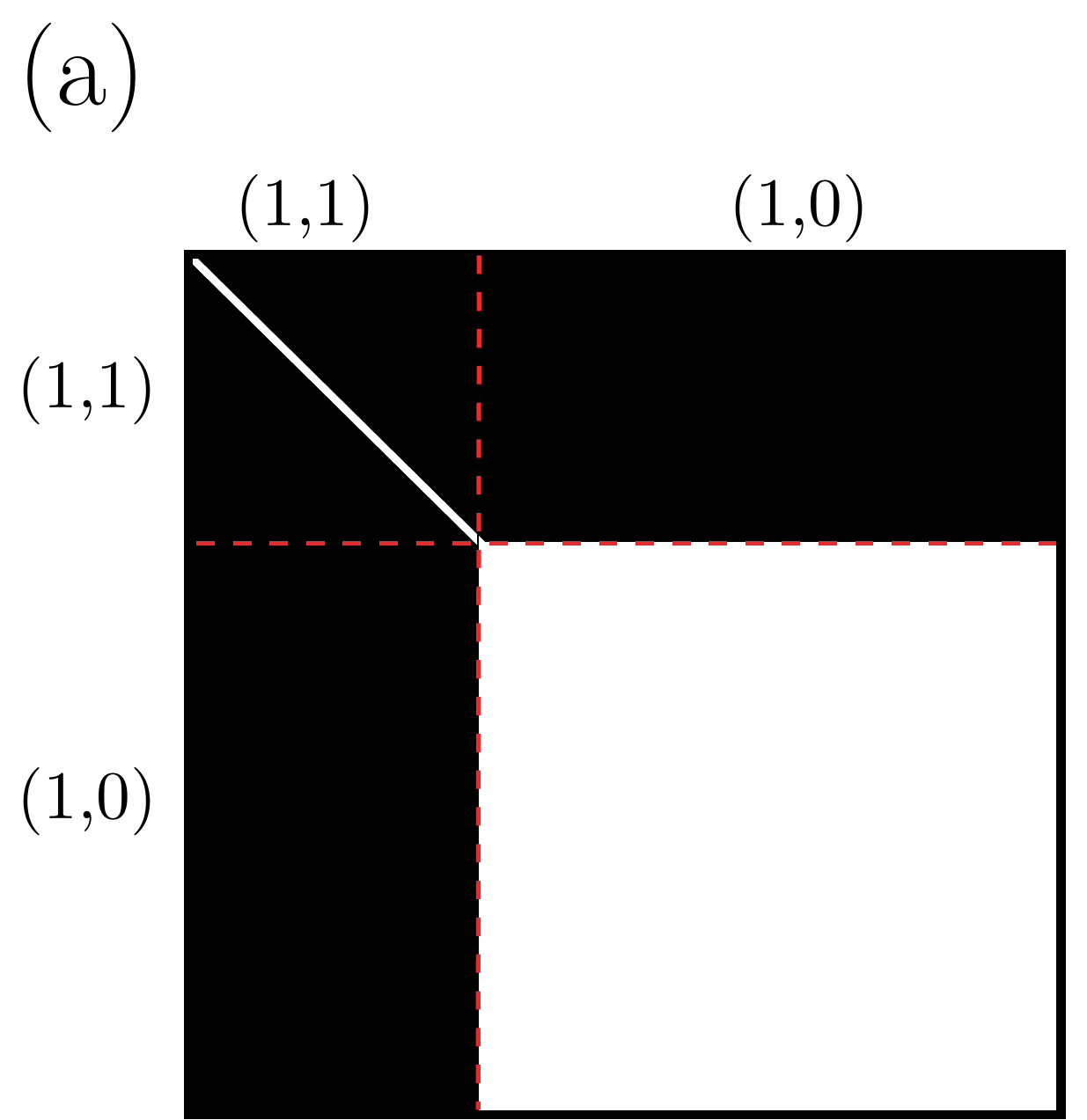}
    	\end{minipage}
    	&
    	\begin{minipage}{0.33\hsize}	
    		\centering
    		\includegraphics[width=1\hsize]{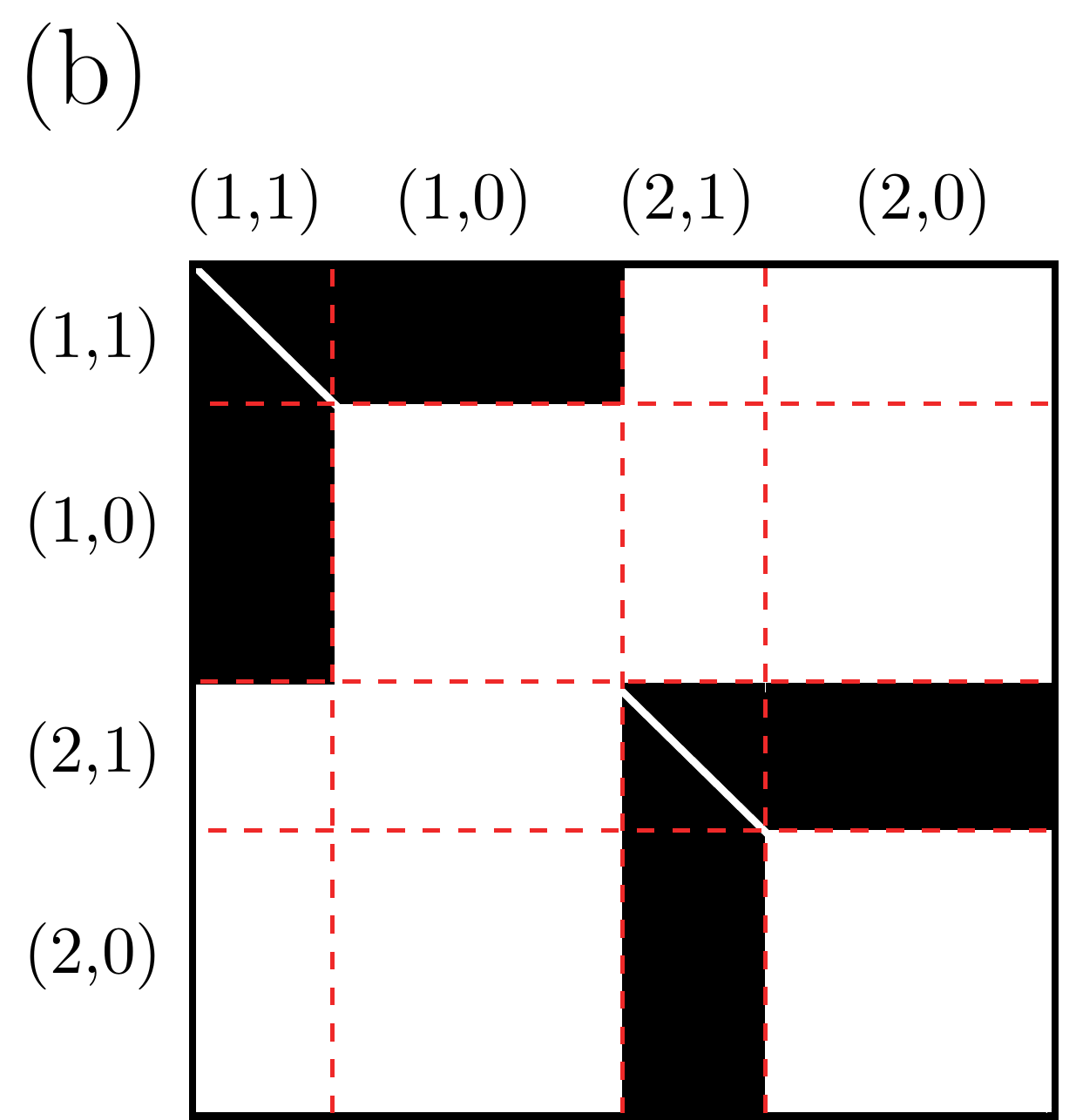}
    	\end{minipage}
    	\\	
    	\\	
    	\begin{minipage}{0.33\hsize}	
    		\centering
    		\includegraphics[width=1\hsize]{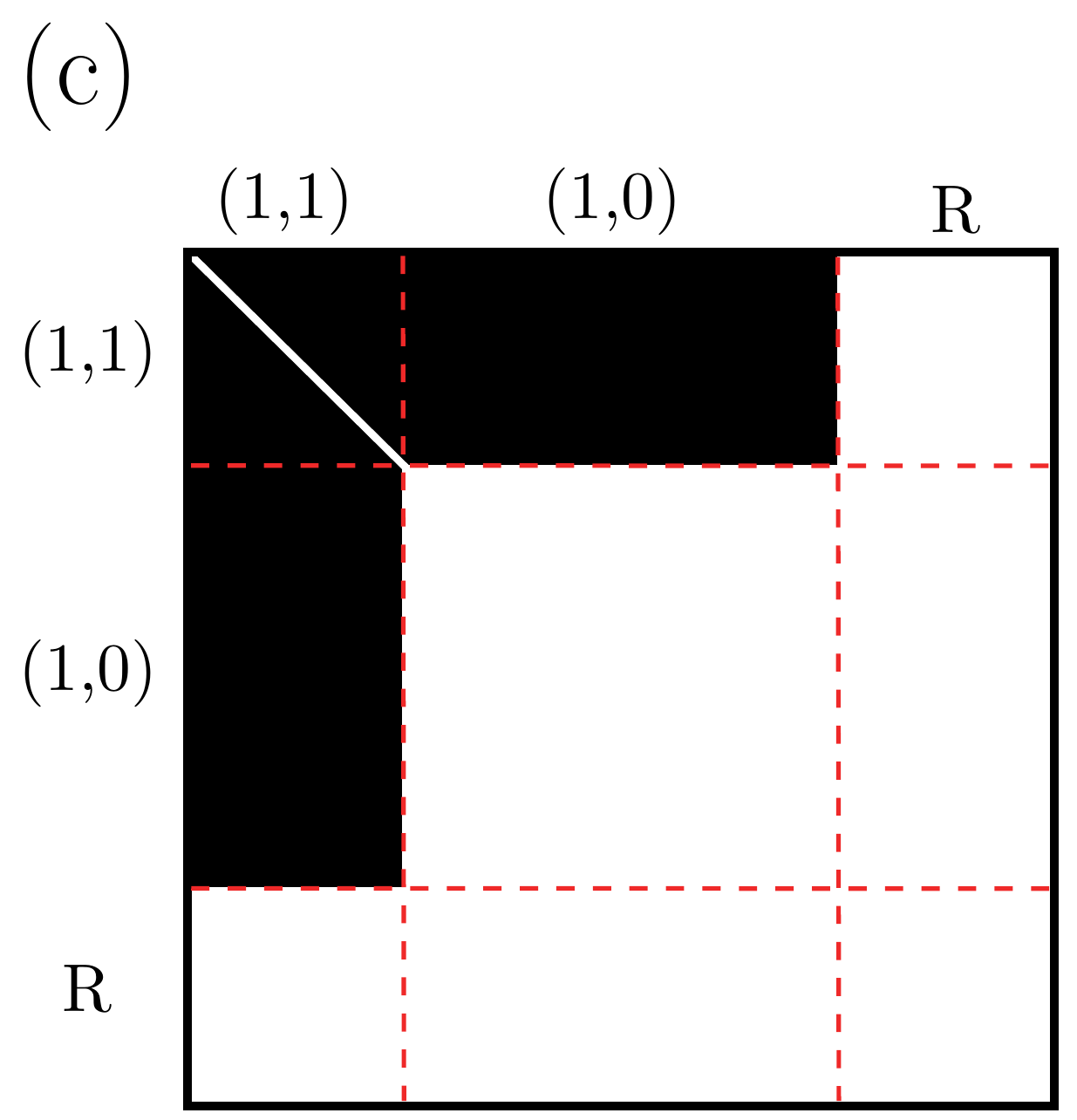}
    	\end{minipage}
    	&
    	\begin{minipage}{0.33\hsize}	
    		\centering
    		\includegraphics[width=1\hsize]{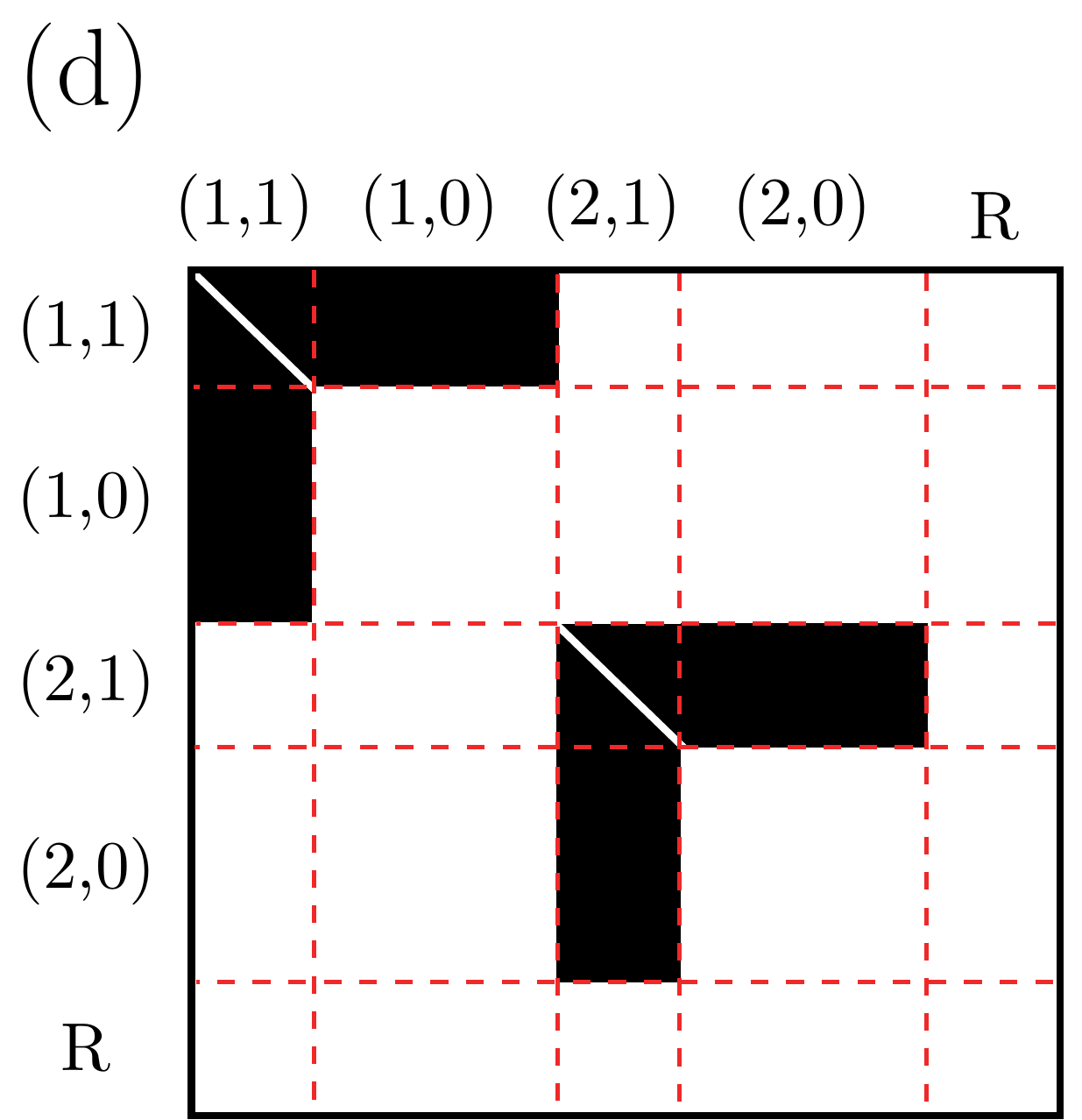}
    	\end{minipage}
    \end{tabular}
    \caption{
	    Schematic illustrations of the adjacency matrices of the networks generated by stochastic block models.
	    The filled blocks correspond to the entries that are equal to 1 with probability $\theta_1$ and zero otherwise. 
	    The empty blocks correspond to the entries that are equal to 1 with probability $\theta_2$ ($\theta_2 < \theta_1$) and zero otherwise. 
	    The diagonal entries are always set to zero and shown as empty entries in the figure for the sake of simplicity. 
	    The dashed lines indicate the borders separating different blocks.
	    The labels $(\c, \x)$ are also indicated at the top and left of the adjacency matrices. Label R represents a block of residual nodes. 
	    The networks are composed of ({a}) a single core-periphery pair, ({b}) two core-periphery pairs, ({c}) a single core-periphery pair and residual nodes, and 
	    ({d}) two core-periphery pairs and residual nodes.
	    In all cases, we set $N=400$.
    }
    \label{fig:sb}
\end{figure}

We compare the performance of the three algorithms on four different types of synthetic networks with a planted core-periphery structure schematically shown in Fig.~\ref{fig:sb}. 
We generate the synthetic networks using stochastic block models \cite{Karrer2011,Rombach2014,Zhang2015,Cucuringu2016,Peixoto2017}.  
We draw label $(c_i,x_i)$ for the $i$th node $(1\leq i \leq N)$ from a distribution $\pi_{(c,x)}$ ($1 \leq c \leq C$ and $x\in\{0,1\}$), where
$\pi_{(c,x)} > 0$ is the probability that $(c_i,x_i)=(c,x)$ and satisfies $\sum_{c=1} ^C \sum_{x=0,1} \pi_{(c,x)}=1$.
Then we place edges between each pair of nodes with label $(c,x)$ and $(c',x')$ with probability $\Theta_{(c,x),(c',x')}$. 
For each type of the stochastic block model, we generate 100 networks and detect core-periphery pairs by the three algorithms.

\begin{figure}
    \centering
    \includegraphics[width=\hsize]{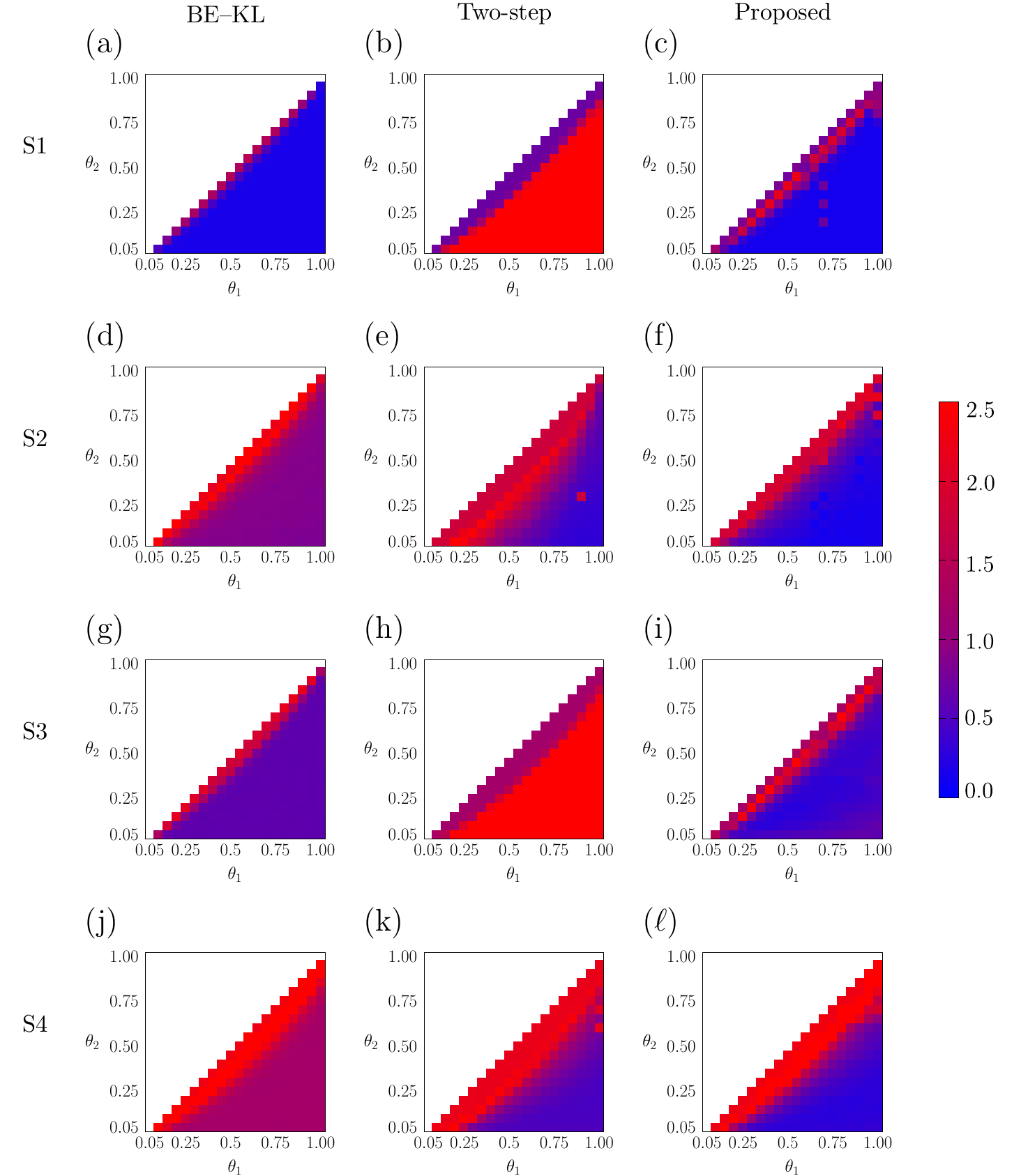}
    \caption{
	    The VI values between the true and inferred core-periphery structure for the three algorithms.
	    Rows S1, S2, S3 and S4 correspond to the networks with planted core-periphery structure shown in Figs.~\ref{fig:sb}(a), \ref{fig:sb}(b), \ref{fig:sb}(c) and \ref{fig:sb}(d), respectively.
	    The colour of each cell indicates the VI value. 
	    The white cells are those for which we did not calculate the VI values, i.e., we only computed them for $\theta_1  > \theta_2$.     
    }
\label{fig:exp}
\end{figure}

As a first example, we consider a network composed of a single core-periphery pair (Fig.~\ref{fig:sb}(a)).
We set $\pi_{(1,1)}=1/4$, $\pi_{(1,0)}=3/4$, $\Theta_{(1,1),(1,1)}=\Theta_{(1,1),(1,0)}=\theta_1$ and $\Theta_{(1,0),(1,0)}=\theta_2$,  where $\theta_1, \theta_2 \in \{0.05, 0.1, 0.15,\ldots,1\}$ and $\theta_1 > \theta_2$.
A generated network has strong core-periphery structure when $\theta_1 \gg \theta_2$. 
If $\theta_1$ is close to $\theta_2$, then the generated network is close to the \ER random graph (i.e., noisy core-periphery structure).
For this network model, the VI value, which quantifies the discrepancy between the true and inferred core-periphery structure, is compared between the three algorithms in Figs.~\ref{fig:exp}(a)--(c).
The VI values for the BE--KL algorithm are approximately equal to zero in the entire parameter region of $\theta_1$ and $\theta_2$. 
The VI values for the two-step algorithm are large even in the case of strong core-periphery structure (i.e., $\theta_1 \gg \theta_2$) because the two-step algorithm divides the single core-periphery pair into communities.
In contrast, the VI values for the proposed algorithm are close to zero for most $\theta_1$ and $\theta_2$ values, as is the case for the BE--KL algorithm.
Therefore, the performance of the proposed algorithm on this network model is comparable to that of the BE--KL algorithm.

As a second example, we examine networks composed of two core-periphery pairs (Fig.~\ref{fig:sb}(b)). 
We set $\pi_{(c,1)}=1/8$, $\pi_{(c,0)}=3/8$, $\Theta_{(c,1),(c,1)}=\Theta_{(c,1),(c,0)}=\theta_1$, and $\Theta_{(c,0),(c,0)}=\Theta_{(1,x),(2,x')}=\theta_2$ for $c\in \{1,2\}$ and $x, x'\in \{0,1\}$.
The VI values for this network are shown in Figs.~\ref{fig:exp}(d)--\ref{fig:exp}(f).
The VI values for the BE--KL algorithm are large in the entire parameter region of $\theta_1$ and $\theta_2$ because the BE--KL algorithm cannot find multiple core-periphery pairs.
The VI values for the two-step algorithm are close to zero in the case of strong core-periphery structure (i.e., $\theta_1$ is considerably larger than $\theta_2$). 
The VI values for the proposed algorithm are smaller than those for the two-step algorithm for most $\theta_1$ and $\theta_2$ values, including the case of noisy core-periphery structure (i.e., when $\theta_1$ is close to $\theta_2$).

In empirical networks, there may be nodes that are better regarded not to belong to any core or periphery. 
Therefore, as a third example, we consider a network composed of a single core-periphery pair and residual nodes (Fig.~\ref{fig:sb}(c)). 
We regard the block of the residual nodes as a single group of nodes, like a core or periphery, when calculating the VI value. 
Let $\text{R}$ be the index for the block of the residual nodes.
We set $\pi_{(1,1)}=\pi_{\text{R}} = 1/5$, $\pi_{(1,0)}=3/5$, $\Theta_{(1,1),(1,1)}=\Theta_{(1,1),(1,0)}=\theta_1$ and $\Theta_{(1,0),(1,0)}=\Theta_{(1,x),\text{R}}=\theta_2$ for $x \in \{0,1\}$.
The VI values for this model are shown in Figs.~\ref{fig:exp}(g)--\ref{fig:exp}(i).
The VI values for the BE--KL algorithm are large even in the case of strong core-periphery structure.
The VI values for the two-step algorithm are large in the entire parameter region of $\theta_1$ and $\theta_2$. 
The VI values for the proposed algorithm are smaller than those for the BE--KL and two-step algorithms for most $\theta_1$ and $\theta_2$ values, including the case of noisy core-periphery structure.

As a fourth example, we consider networks composed of two core-periphery pairs and residual nodes (Fig.~\ref{fig:sb}(d)).
We set $\pi_{(c,1)}=\pi_{\text{R}} = 1/9$, $\pi_{(c,0)}=1/3$, $\Theta_{(c,1),(c,1)}=\Theta_{(c,1),(c,0)}=\theta_1$ for $c \in \{1,2\}$ and 
$\Theta_{(c,0),(c,0)} = \Theta_{(c,x), (c',x')} = \theta_2$, where $c=1, 2$, $c\neq c'$ and $x,x' \in\{0,1\}$.
The VI values for this network model are shown in Figs.~\ref{fig:exp}(j)--($\ell$).
The VI values for the BE--KL algorithm are large in the entire $\theta_1$-$\theta_2$ parameter space.
The VI values for the two-step algorithm are larger than those for the proposed algorithm for most $\theta_1$ and $\theta_2$ values.

\subsection{Empirical networks}
We apply the three algorithms to three empirical networks.
For directed and weighted networks, we disregard the direction and the weight of edges.

\subsubsection{Karate club network}
Consider the karate club network \cite{Zachary1977}, which has $N=34$ nodes and $M=78$ edges (edge density $p=0{.}139$).
A node represents a member of a karate club at a university.
Two members are adjacent if they have socially interacted outside club activities during the observation period. 
During the study, a conflict occurred between the instructor (node 1) and the president (node 34), which fissured the club. 
Based on self-reports, each member was labelled on the instructor's side (15 members), on the president's side (16 members) or neutral (3 members) \cite{Zachary1977}.

\begin{figure}    
    \centering
    \begin{tabular}{cc}
        \begin{minipage}{0.5\hsize}
    	\centering
            \includegraphics[width=\hsize]{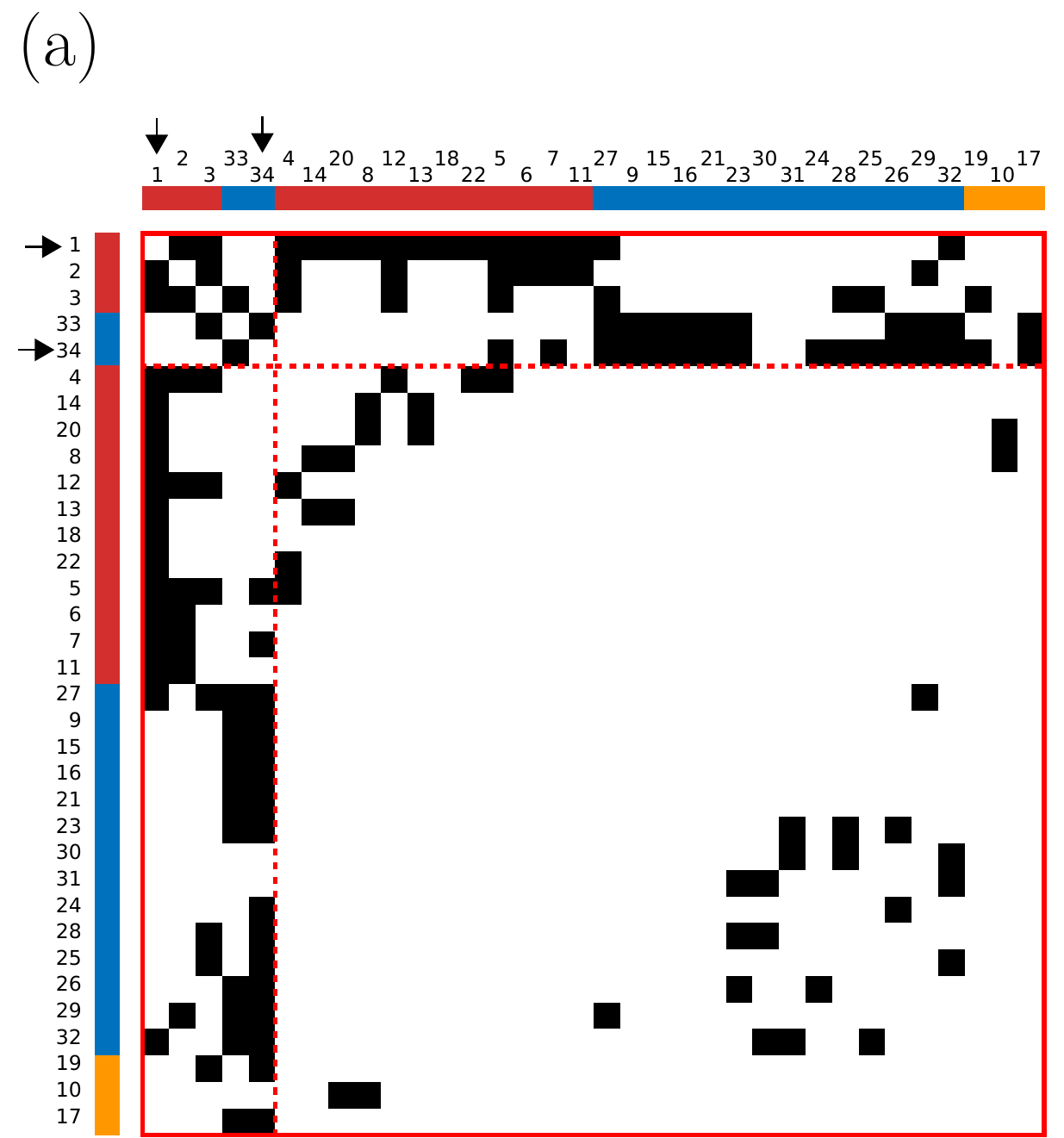}
        \end{minipage}
        & 
        \begin{minipage}{0.5\hsize}
    	\centering
            \includegraphics[width=\hsize]{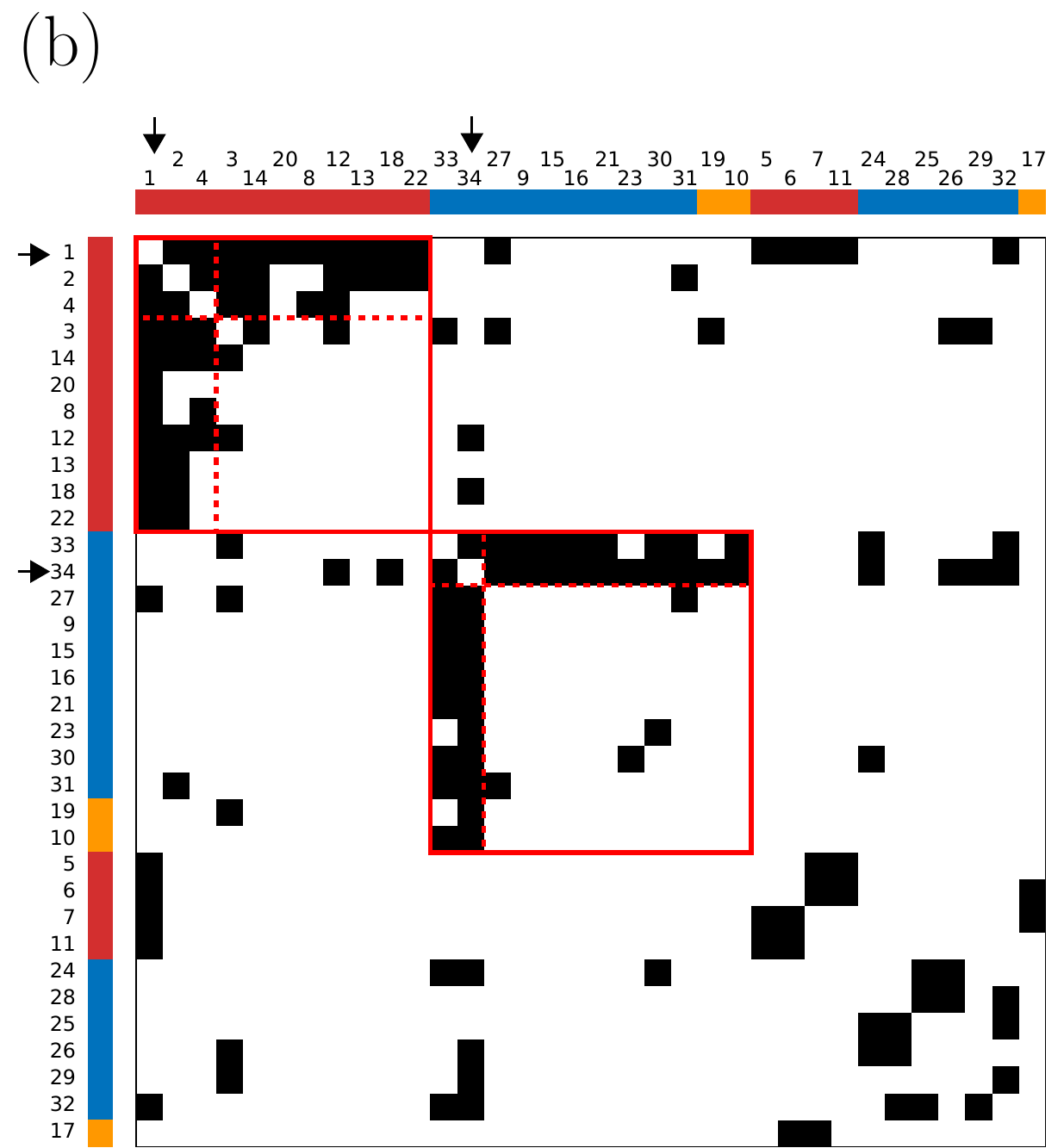}
        \end{minipage}
        \\
        \\
        \begin{minipage}{0.5\hsize}
    	\centering
            \includegraphics[width=\hsize]{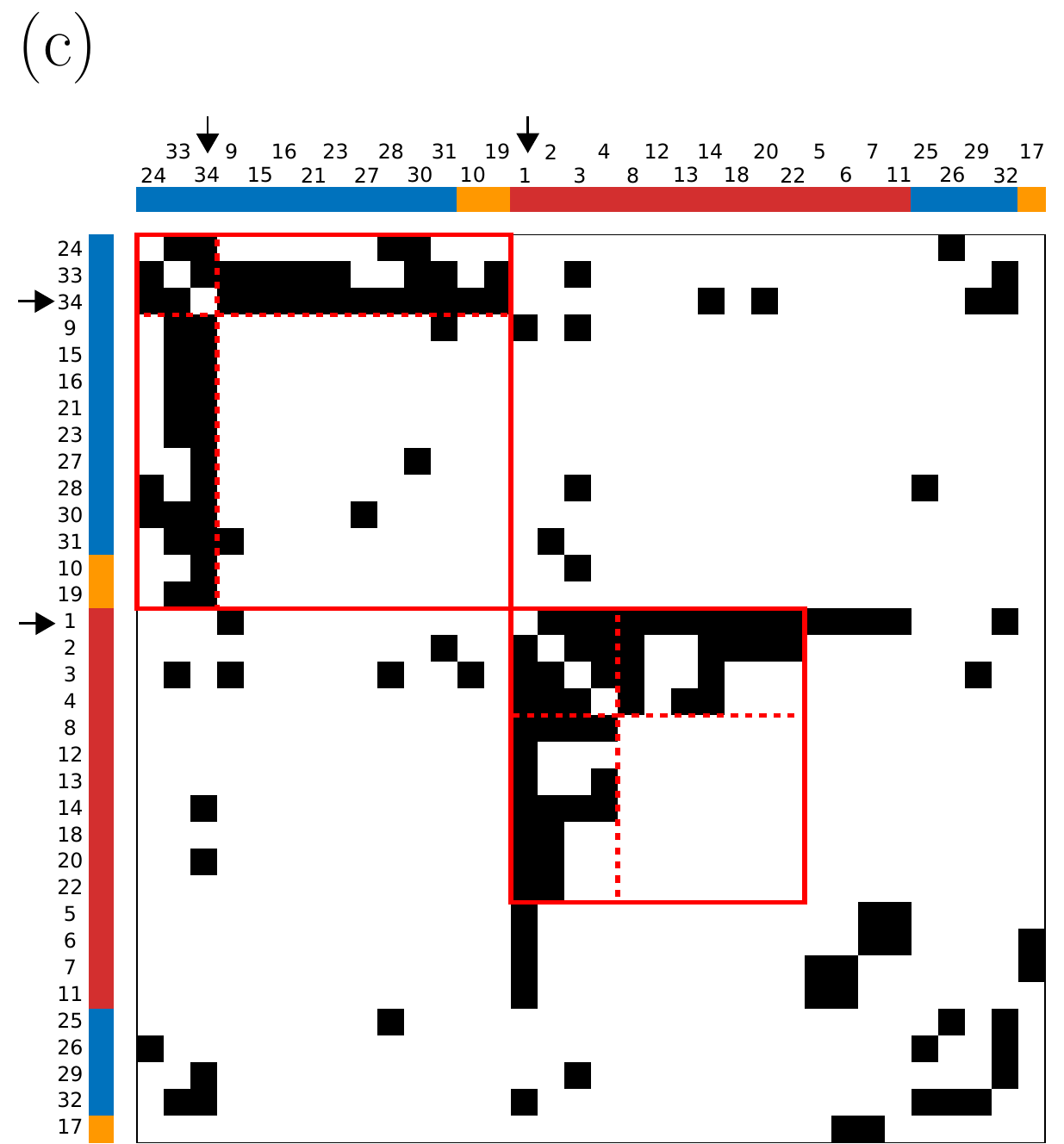}
        \end{minipage}
        & 
        \begin{minipage}{0.5\hsize}
    	\centering
            \includegraphics[width=0.9\hsize]{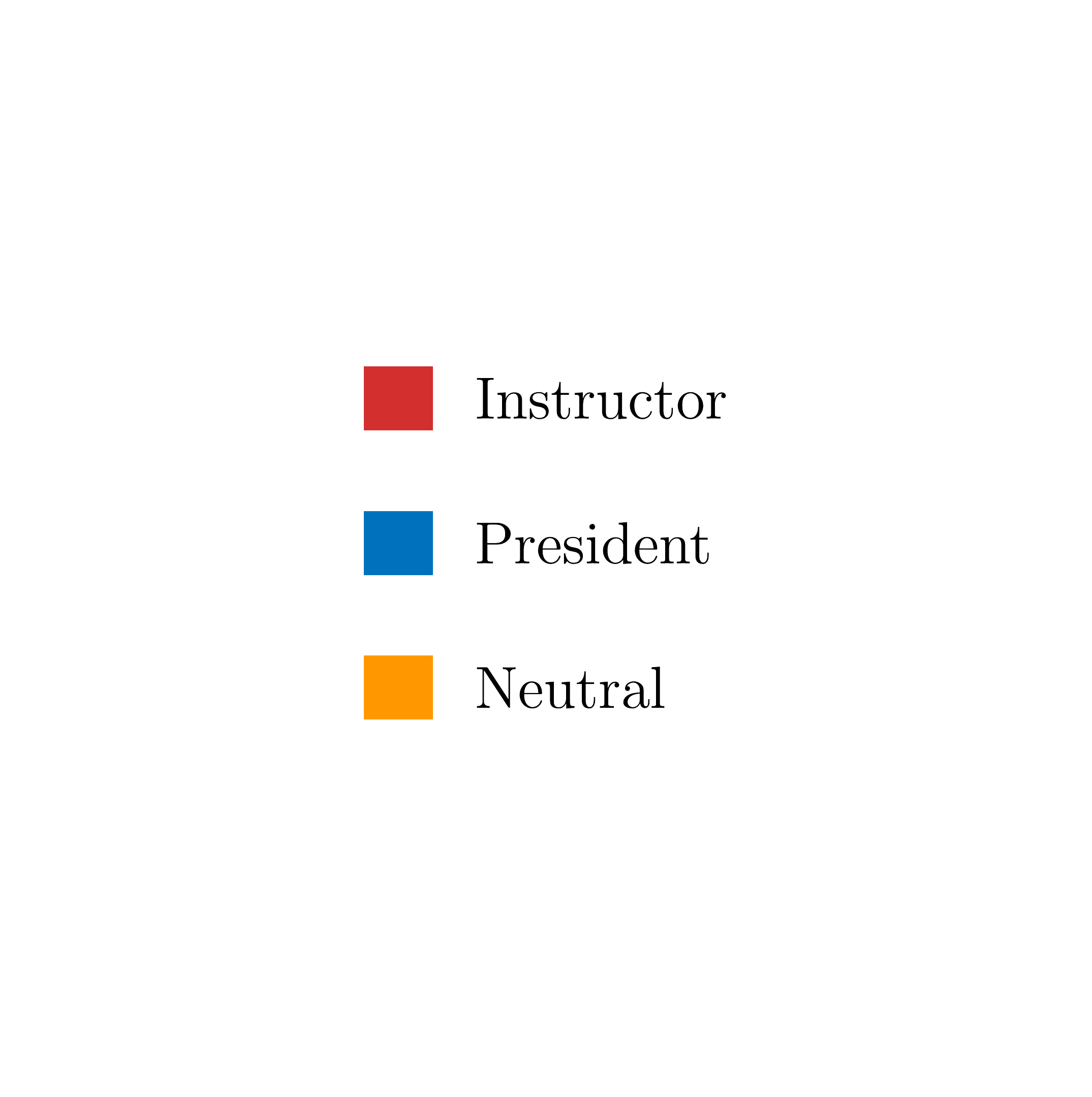}
        \end{minipage}
    \end{tabular}
    \caption{
        Core-periphery structure of the karate club network detected by ({a}) the BE--KL algorithm, ({b}) the two-step algorithm and ({c}) our algorithm.
        The filled and blank cells indicate $A_{ij}=1$ and $A_{ij}=0$, respectively. 
        The solid lines indicate the partition into core-periphery pairs.
        The dashed lines indicate the partition into the core and periphery within a core-periphery pair.
        The colour indicates the leaning of the members.
        The instructor (i.e., node 1) and president (i.e., node 34) are indicated by the arrows.
    }
    \label{fig:karate}
\end{figure}

The core-periphery structure detected by the three algorithms is shown in Fig.~\ref{fig:karate}. 
The BE--KL algorithm detects a single core-periphery pair such that both the instructor and president are core nodes (Fig.~\ref{fig:karate}(a)), neglecting the fissure of the club. 
The two-step algorithm detects two core-periphery pairs, each of which consists mostly of the members with the same leanings (Fig.~\ref{fig:karate}(b)).
In particular, the instructor and the president belong to the core of the different core-periphery pairs. 
Two neutral members, nodes 10 and 19, are assigned to the president's core-periphery pair, which does not agree with the self-reports by the members.
The residual nodes consist of the members on the instructor's side, those on the president's side and a neutral member. 
Our algorithm detects almost the same two core-periphery pairs as that detected by the two-step algorithm (Fig.~\ref{fig:karate}(c)).

Next, we compare the density of edges within core-periphery pairs. 
For each significant core-periphery pair, we compute the density of edges within the core, that of edges within the periphery and that of edges between the core and periphery.
Then we average each type of edge density over all significant core-periphery pairs (without weighing by the size of core-periphery pair when calculating the average).

We show the edge densities for the karate club network in Fig.~\ref{fig:density}(a).
For all algorithms, the average density of intra-core edges and that of core-periphery edges (i.e., edges connecting a core node and a peripheral node) are larger than the edge density for the entire network, $p=0.139$. 
The average density of intra-peripheral edges is smaller than $p = 0.139$. 
Therefore, we conclude that the structure detected by either of the three algorithms is consistent with the concept of core-periphery structure based on edge density.

\subsubsection{Political blog network}
The second example is a political blog network \cite{Adamic2005}, which has $N=1{,}222$ nodes and $M=16{,}714$ edges (edge density $p=0{.}0224$).
A node is a blog on the United States president election in 2004, and two blogs are adjacent if one blog cites the other blog on its front page. 
Each blog was labelled with one of the political leanings, liberal (586 blogs) or conservative (636 blogs), determined by automated categorisations by several weblog directories \cite{Adamic2005}. If a blog was uncategorised or classified to conflicting categories, then the authors of Ref.~\cite{Adamic2005} manually judged the political leaning.  

\begin{figure}
   \centering
   \begin{tabular}{cc}
   \begin{minipage}{0.5\hsize}
	\centering
        \includegraphics[width=\hsize]{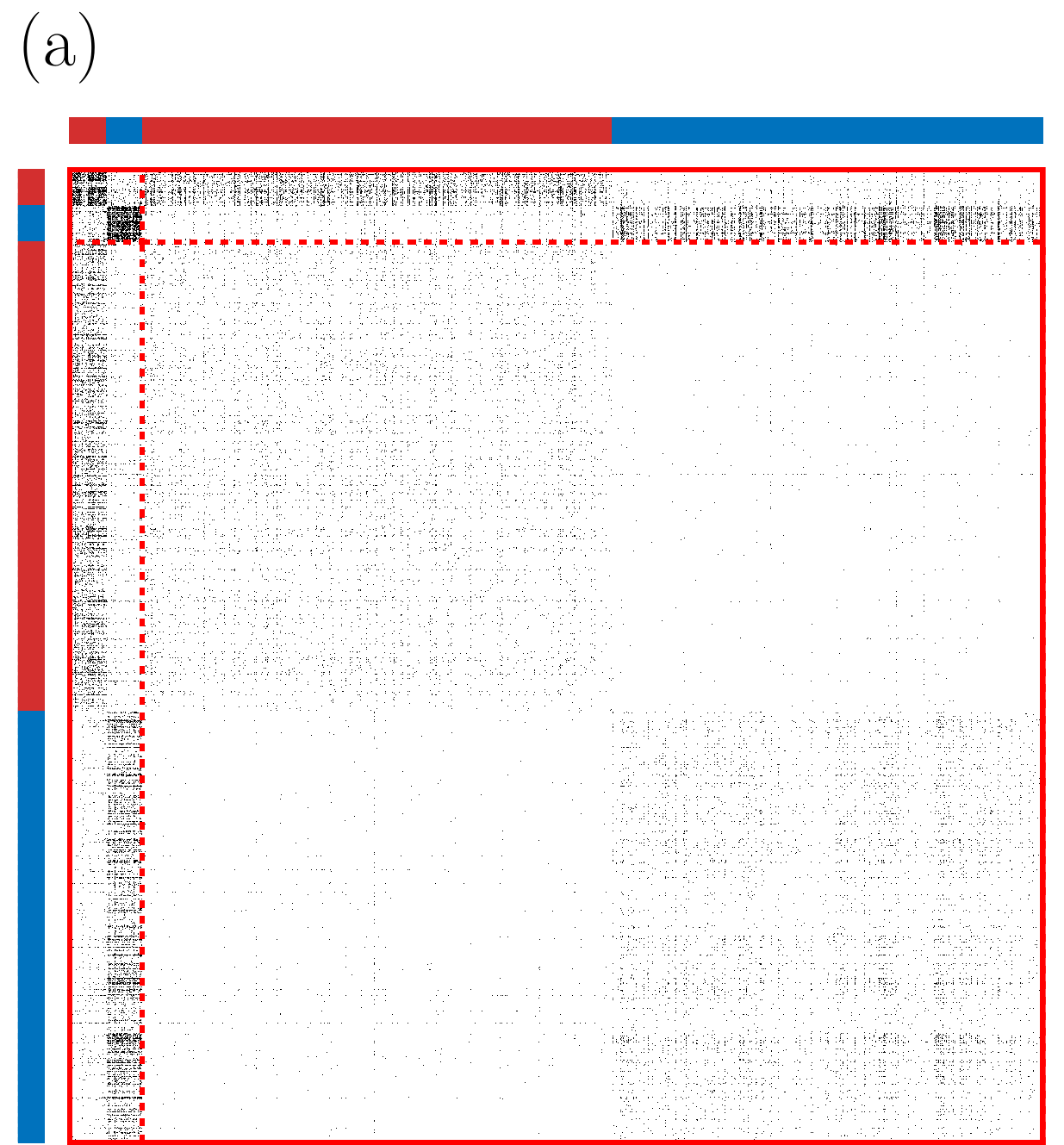}
    \end{minipage}
    & 
    \begin{minipage}{0.5\hsize}
	\centering
        \includegraphics[width=\hsize]{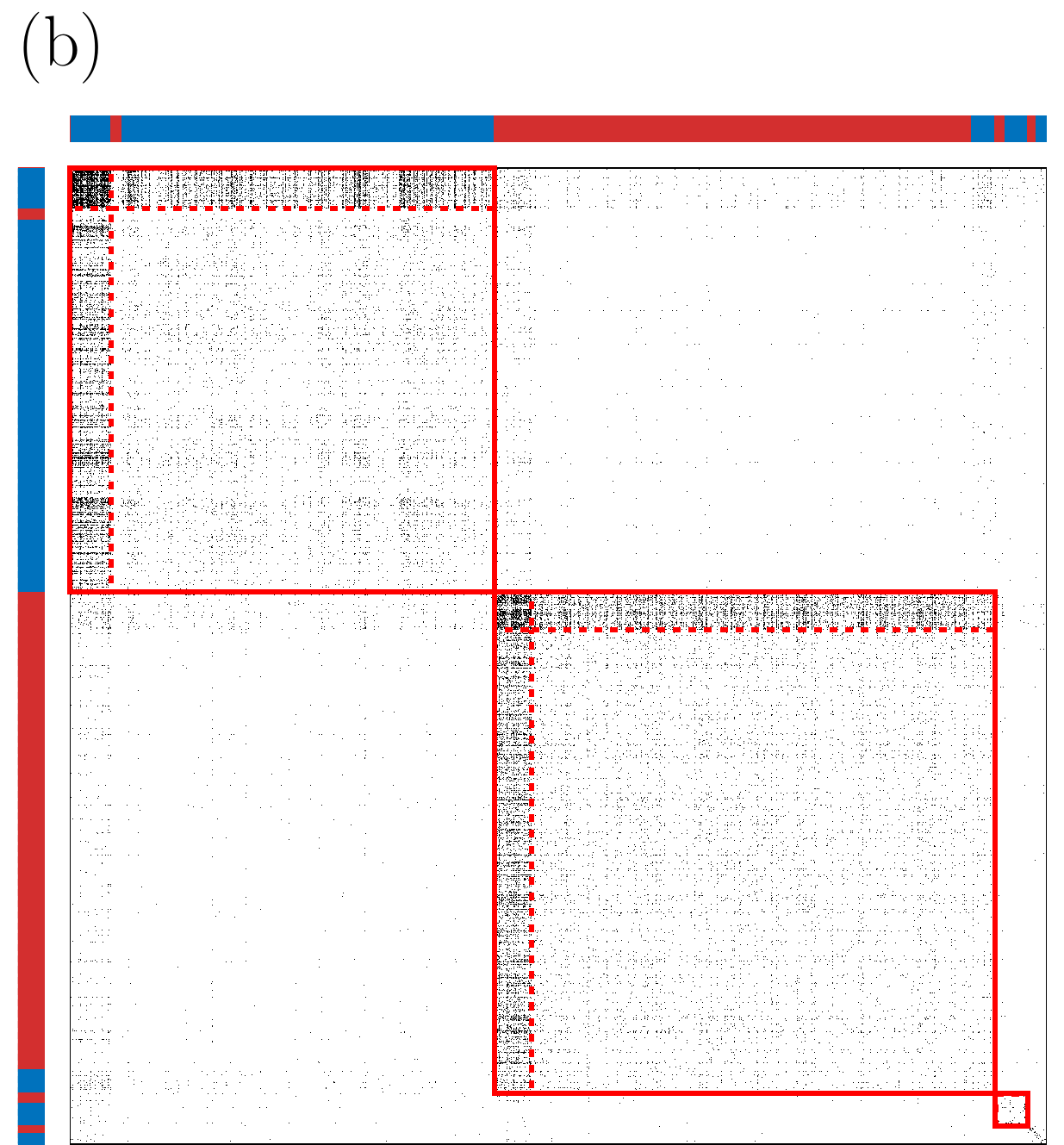}
    \end{minipage}
    \\
    \\
    \begin{minipage}{0.5\hsize}
	\centering
        \includegraphics[width=\hsize]{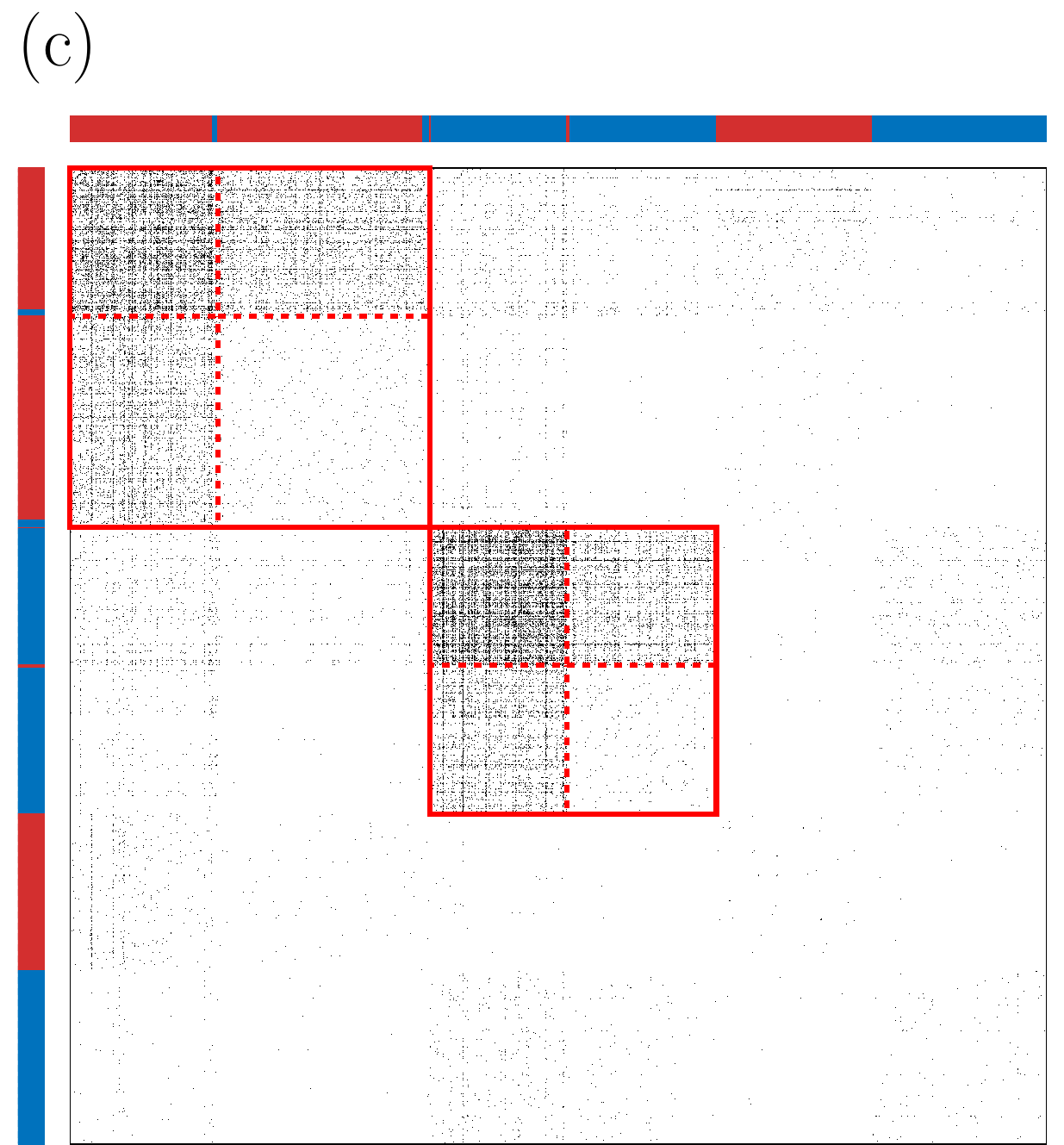}
    \end{minipage}
    & 
    \begin{minipage}{0.5\hsize}
	\centering
        \includegraphics[width=0.9\hsize]{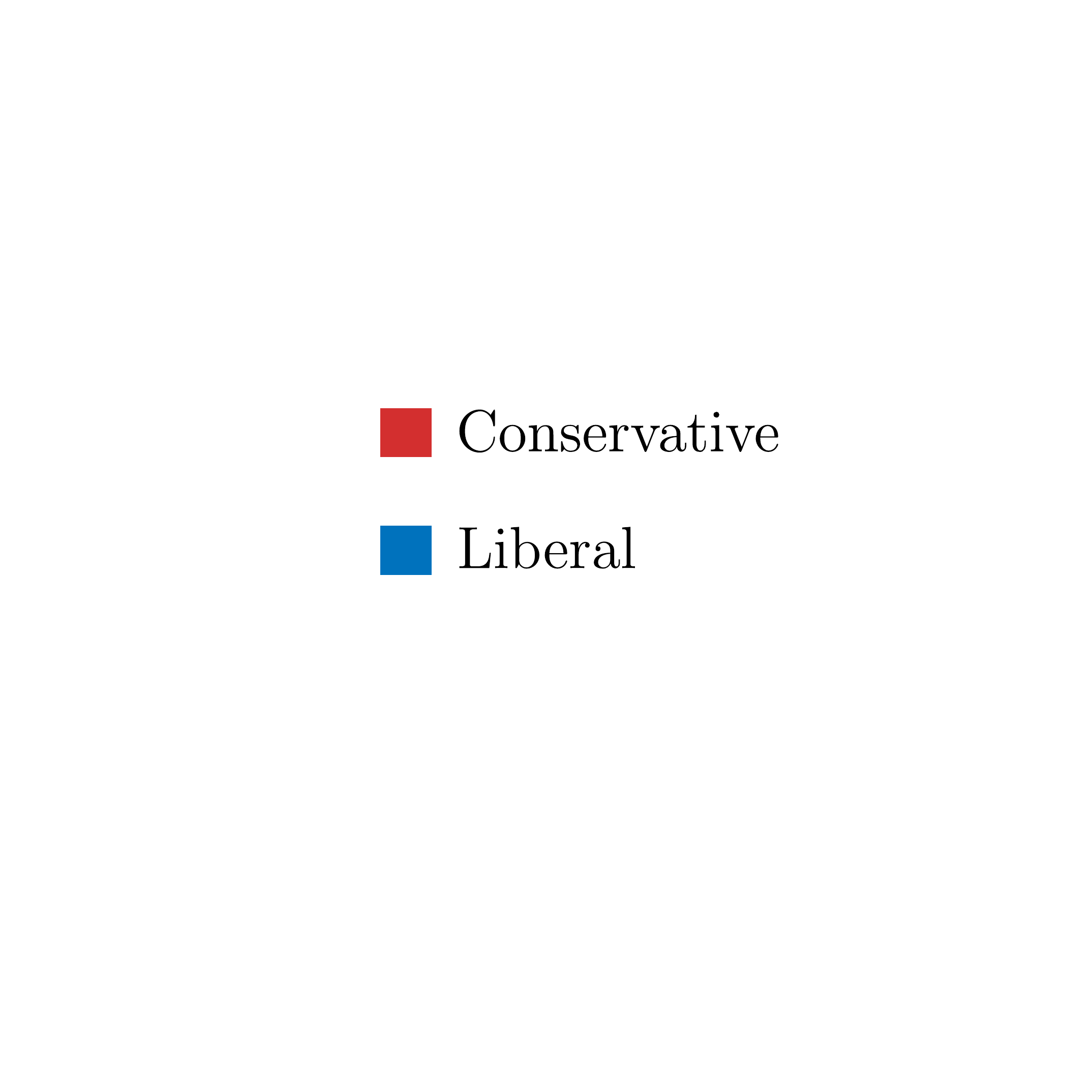}
    \end{minipage}
\end{tabular}
    \caption{
    Core-periphery structure of the blog network detected by ({a}) the BE--KL algorithm, ({b}) the two-step algorithm and ({c}) our algorithm.
    The red and blue indicate the conservative and liberal blogs, respectively. 
    }
    \label{fig:poliblog}
\end{figure}

The core-periphery structure detected by the three algorithms is shown in Fig.~\ref{fig:poliblog}.
The unique core detected by the BE--KL algorithm is a mixture of liberal and conservative blogs (Fig.~\ref{fig:poliblog}(a)).
The peripheral blogs are mostly adjacent to blogs with the same political leaning.
However, the structure detected by the BE--KL algorithm alone does not tell this unless we refer to the political leaning of the individual blogs.
A different algorithm for a single core-periphery structure yielded similar results for the same network \cite{Zhang2015}.

\begin{figure}
   \centering
   \begin{tabular}{cc}
    \begin{minipage}{0.5\hsize}
        \centering
        \includegraphics[width=\hsize]{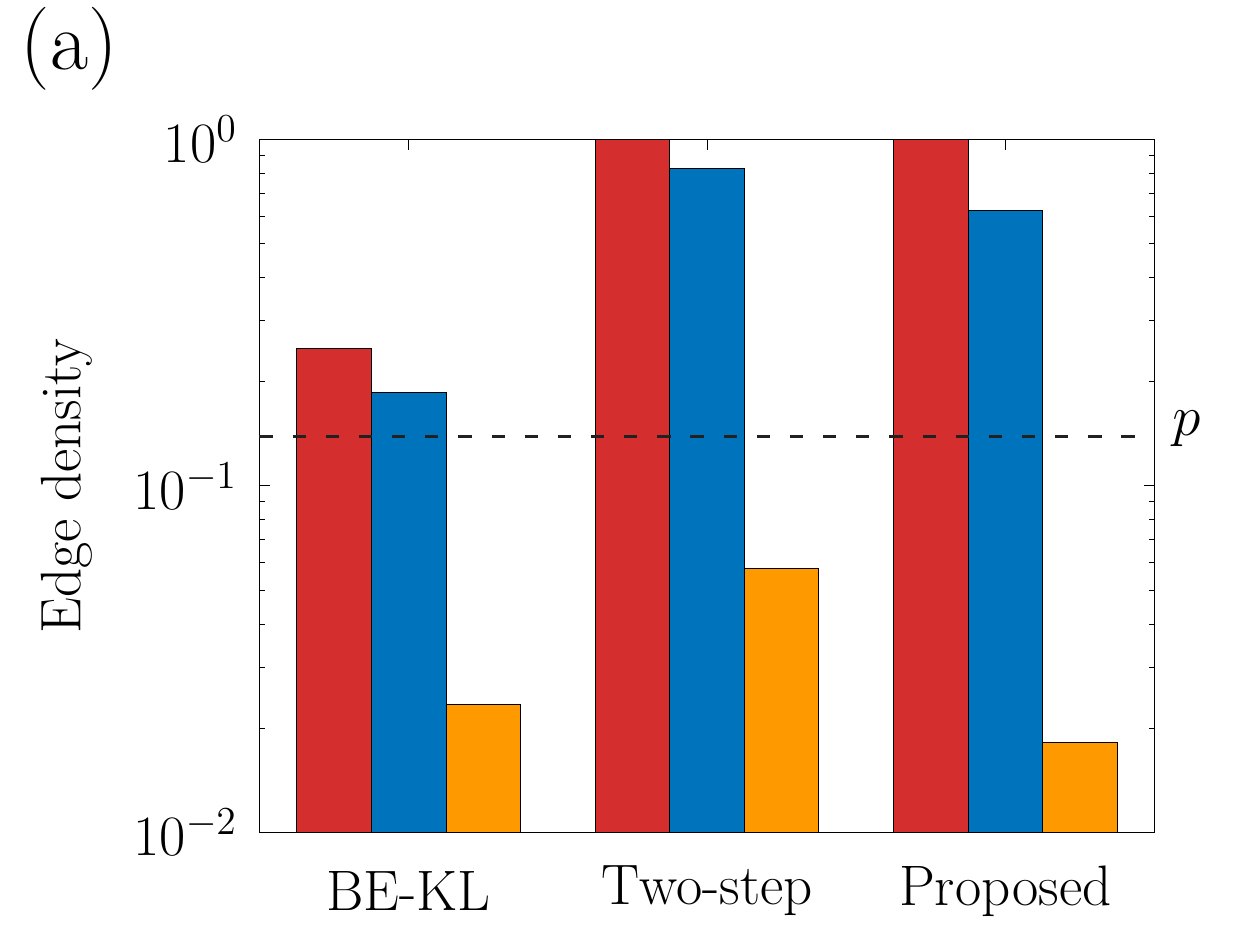} 
    \end{minipage}
    & 
    \begin{minipage}{0.5\hsize}
        \centering
        \includegraphics[width=\hsize]{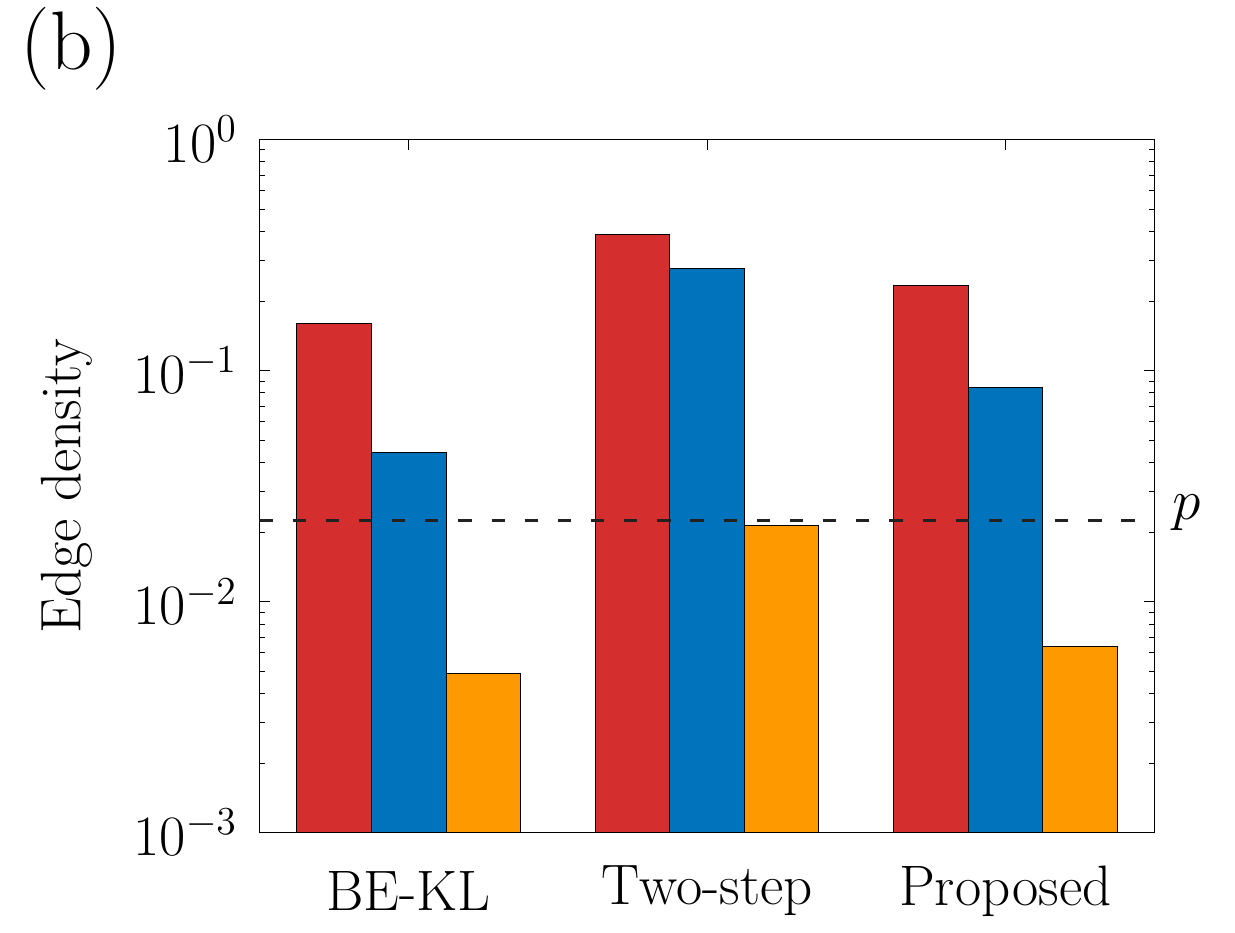} 
    \end{minipage}
    \\ 
    \\
    \begin{minipage}{0.5\hsize}
        \centering
        \includegraphics[width=\hsize]{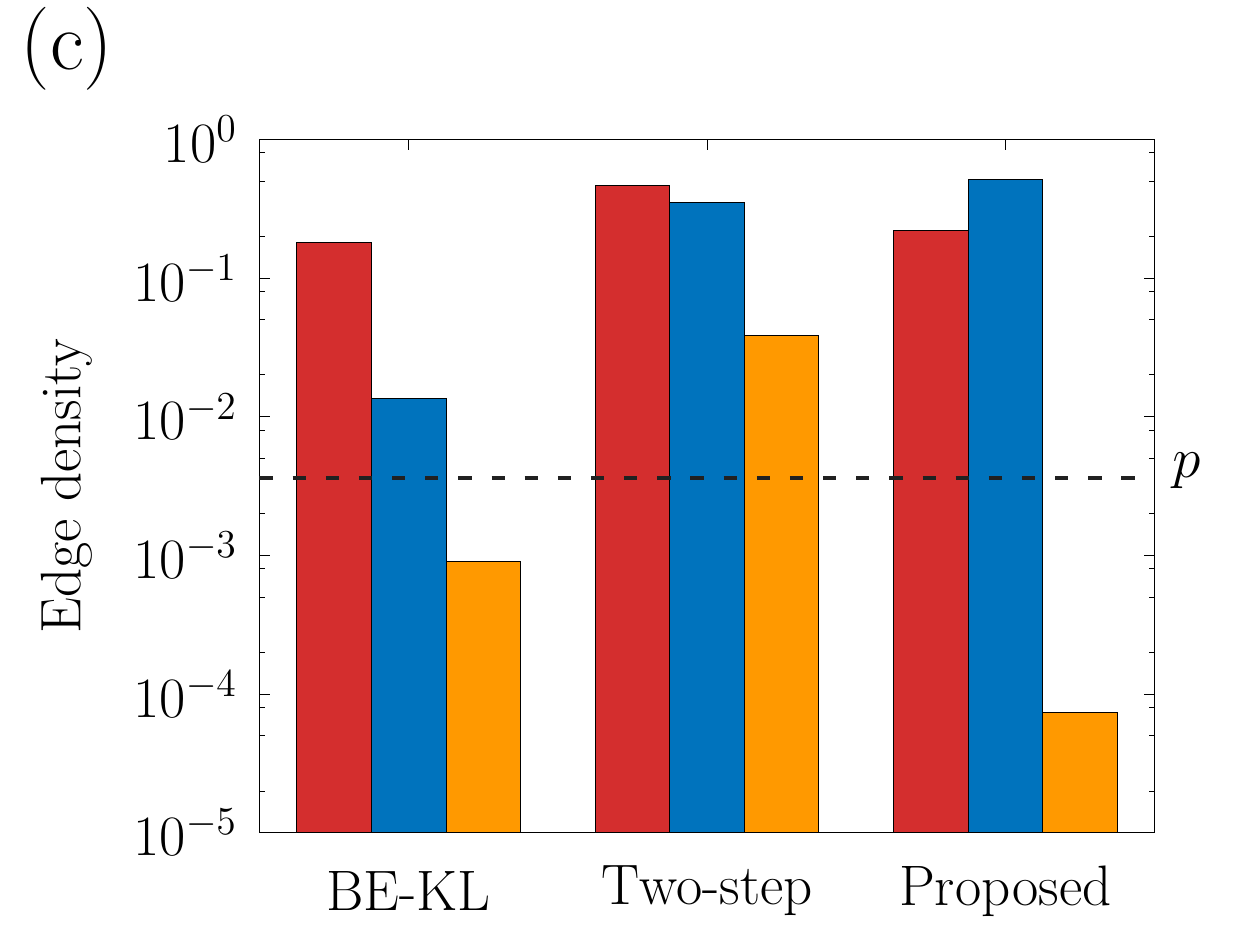} 
    \end{minipage}
    &
    \begin{minipage}{0.5\hsize}
        \centering
        \includegraphics[width=0.8\hsize]{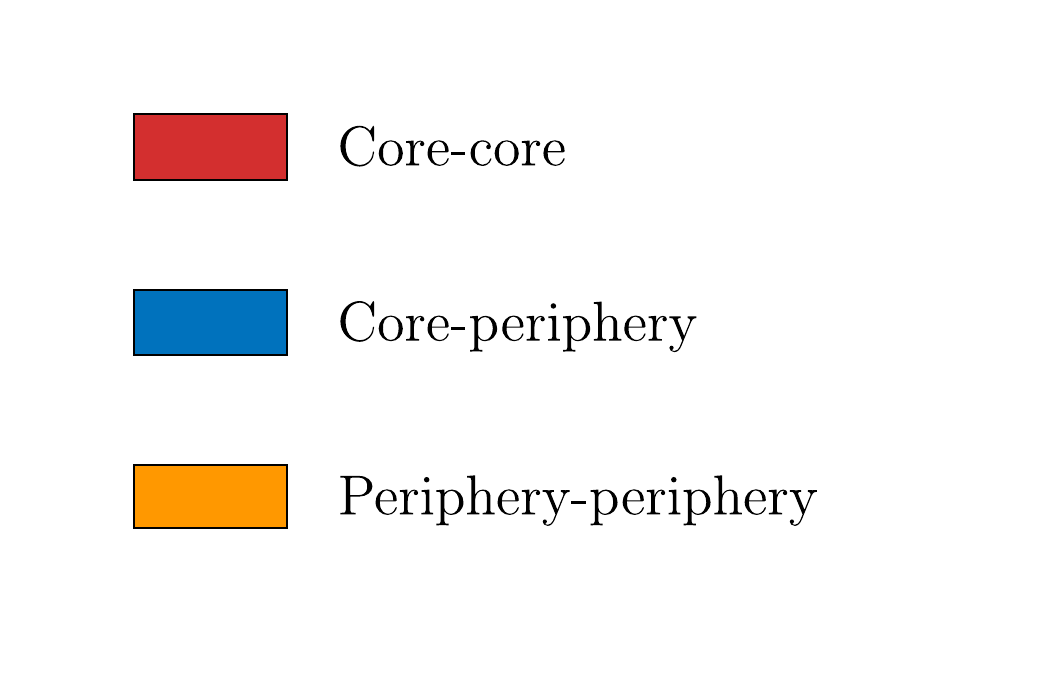} 
    \end{minipage}
   \end{tabular}
    \caption{
    Density of edges of different types within core-periphery pairs detected in (a) the karate club network, (b) blog network and (c) airport network. The dashed line indicates the edge density for the entire network, $p$.
    }
    \label{fig:density}
\end{figure}

The two-step algorithm detects three core-periphery pairs, each of which mostly comprises the blogs with the same political leanings (Fig.~\ref{fig:poliblog}(b)).
Two core-periphery pairs are much larger than the third one and have the opposite political leanings. 
The third small core-periphery pair is mainly composed of liberal blogs.
In each core-periphery pair, a majority of the peripheral nodes is densely interconnected, which is against the idealised core-periphery structure.
This is due to the community detection step that partitions a network into communities with dense intra-community edges.
In fact, the average density of intra-peripheral edges within a core-periphery pair is $0{.}0214$, which is approximately equal to the edge density for the entire network, $p=0{.}0224$ (Fig.~\ref{fig:density}(b)). 
This result indicates that the peripheral nodes are as densely adjacent to each other as expected for the \ER random graph, which contradicts the expectation from the core-periphery structure that peripheral nodes are rarely adjacent to each other.  
 
Our algorithm detects two core-periphery pairs, each of which mostly comprises the blogs with the same political leaning (Fig.~\ref{fig:poliblog}(c)).
The detected two core-periphery pairs are smaller than those detected by the two-step algorithm.
More nodes are classified as residual nodes than by the two-step algorithm.
The average density of intra-peripheral edges within a core-periphery pair  is $0{.}0064$ (Fig.~\ref{fig:density}(b)), which is smaller than the edge density for the entire network, $p=0{.}0224$, respecting the notion of the periphery.


\subsubsection{Airport network}
Our third example is a network of airports, which has $N=2{,}939$ nodes and $M=15{,}677$ edges (edge density $p=0{.}0036$) \cite{Openflight.org,ToreOpsahl}.
A node represents an airport. Two airports are adjacent if there is a direct commercial flight between them.

\begin{figure}
   \centering
   \begin{tabular}{cc}
   \begin{minipage}{0.5\hsize}
	\centering
        \includegraphics[width=\hsize]{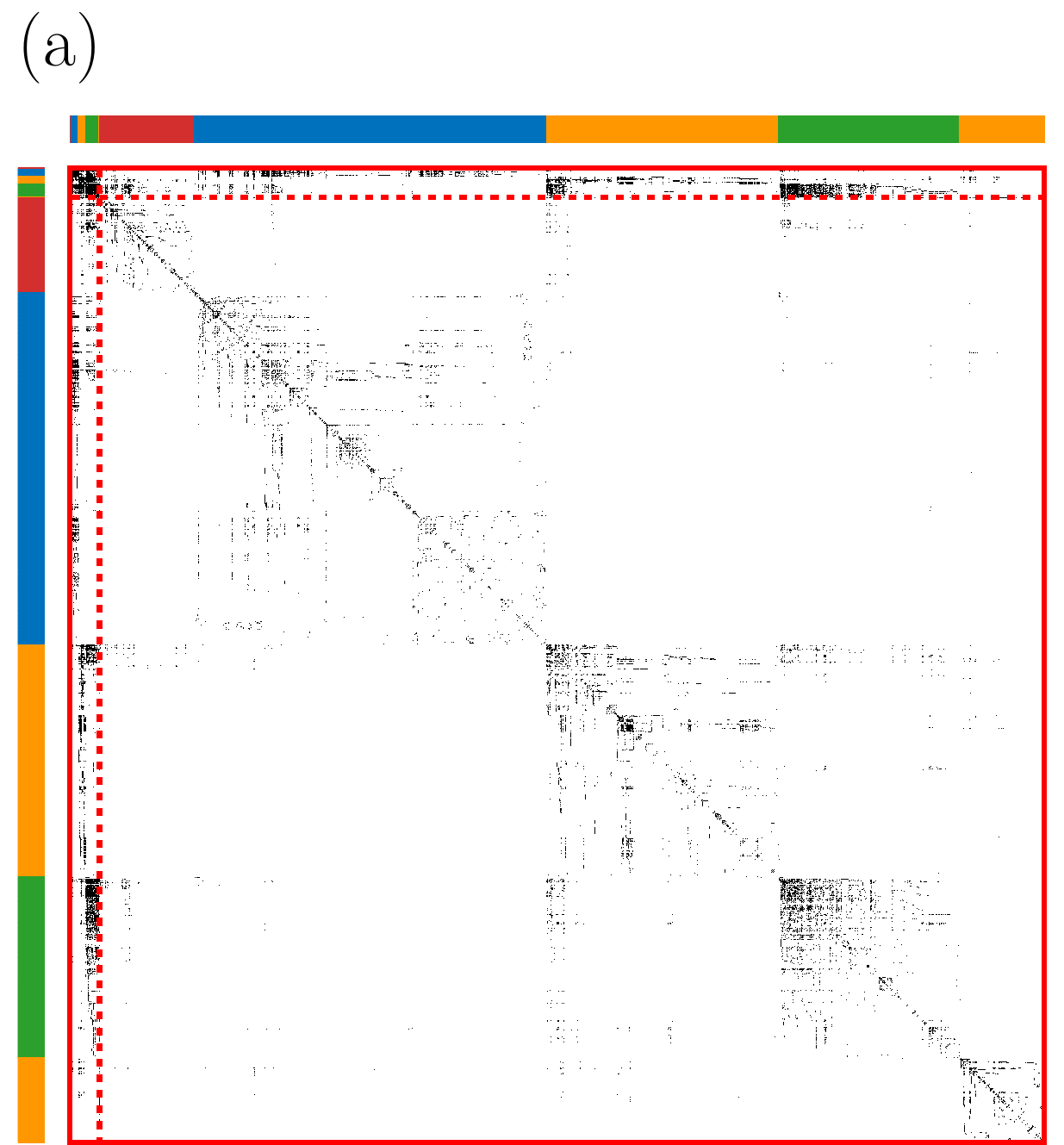}
    \end{minipage}
    & 
    \begin{minipage}{0.5\hsize}
	\centering
        \includegraphics[width=\hsize]{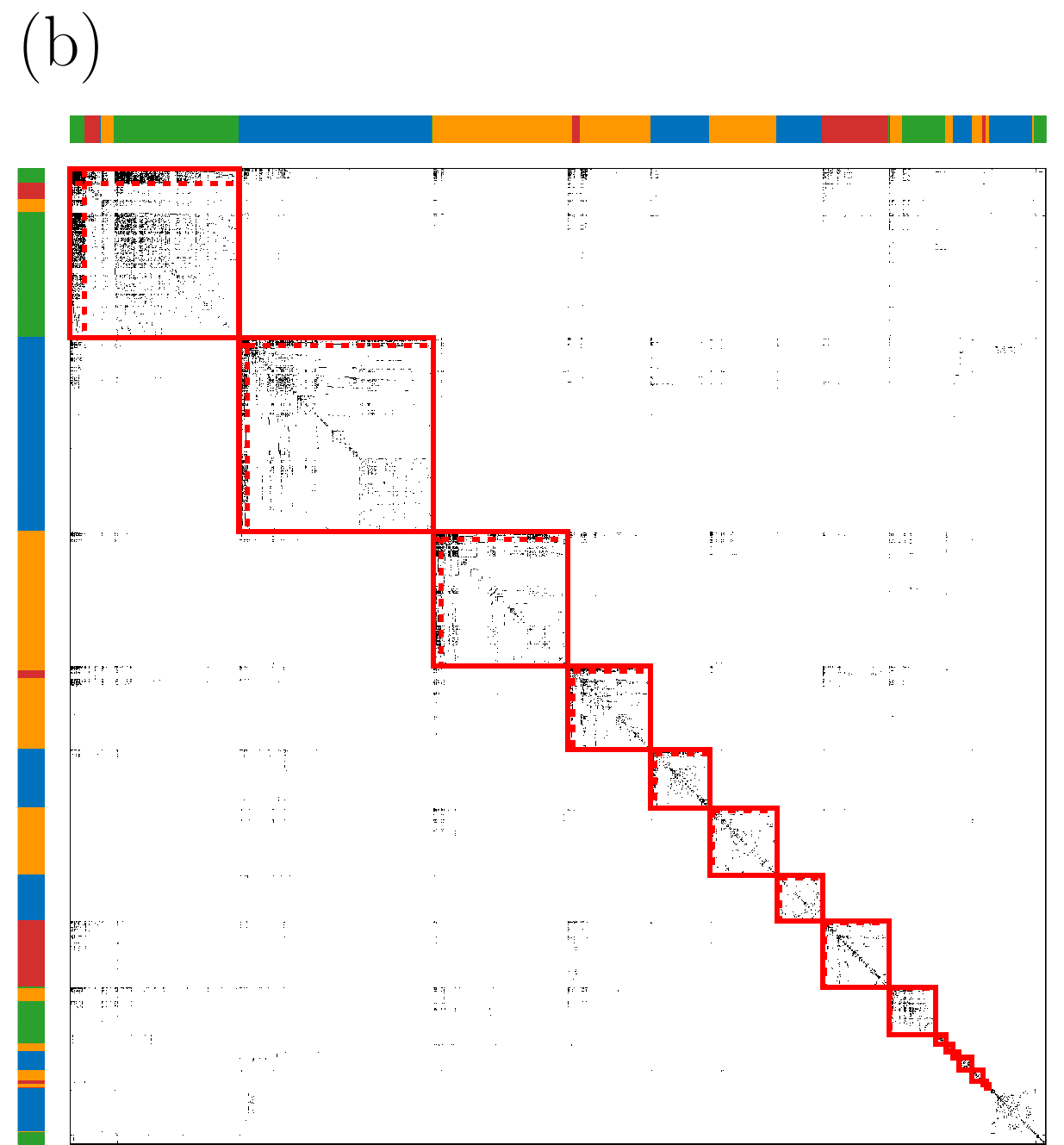}
    \end{minipage}
    \\
    \\
    \begin{minipage}{0.5\hsize}
	\centering
        \includegraphics[width=\hsize]{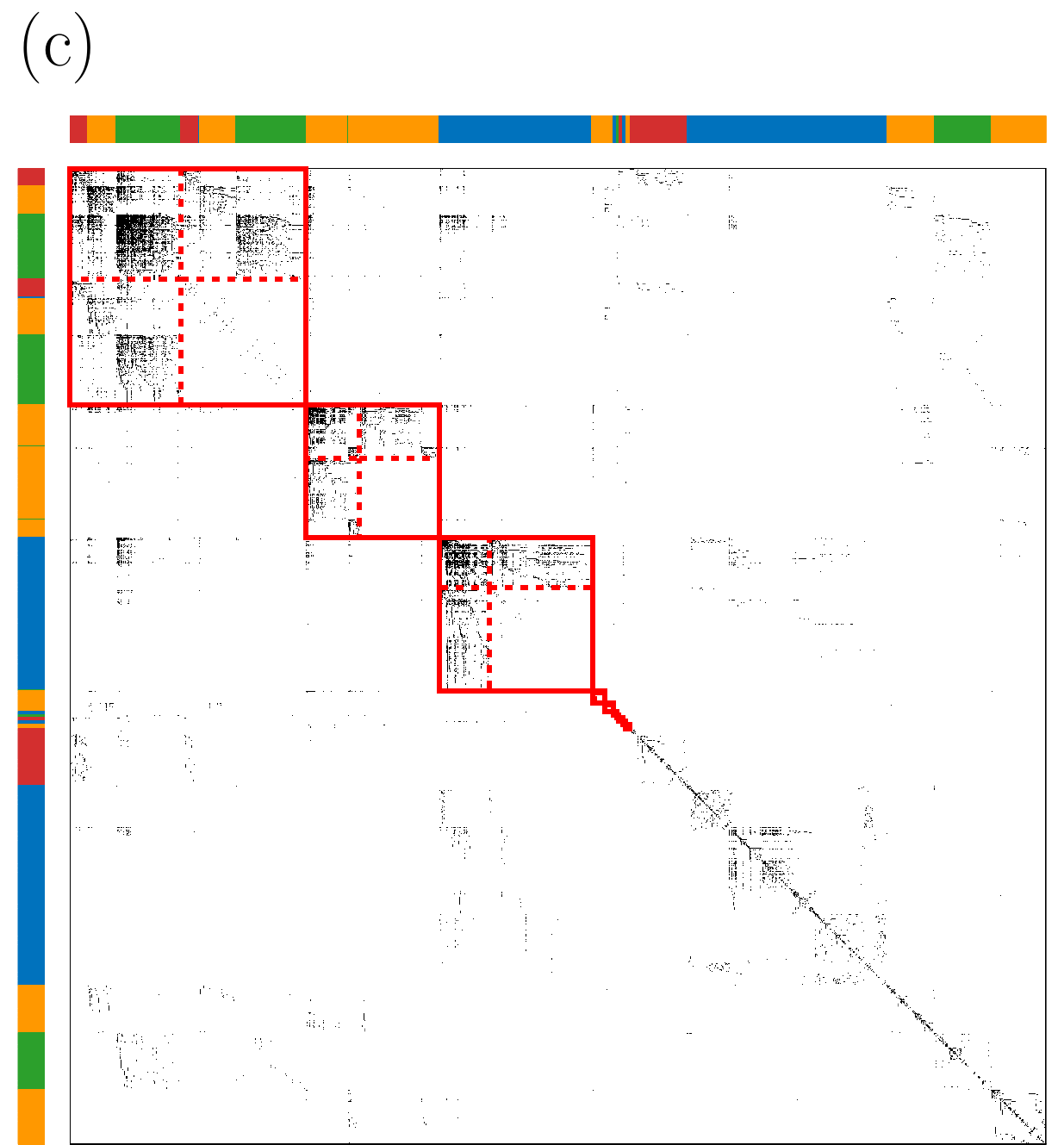}
    \end{minipage}
    & 
    \begin{minipage}{0.5\hsize}
	\centering
        \includegraphics[width=0.9\hsize]{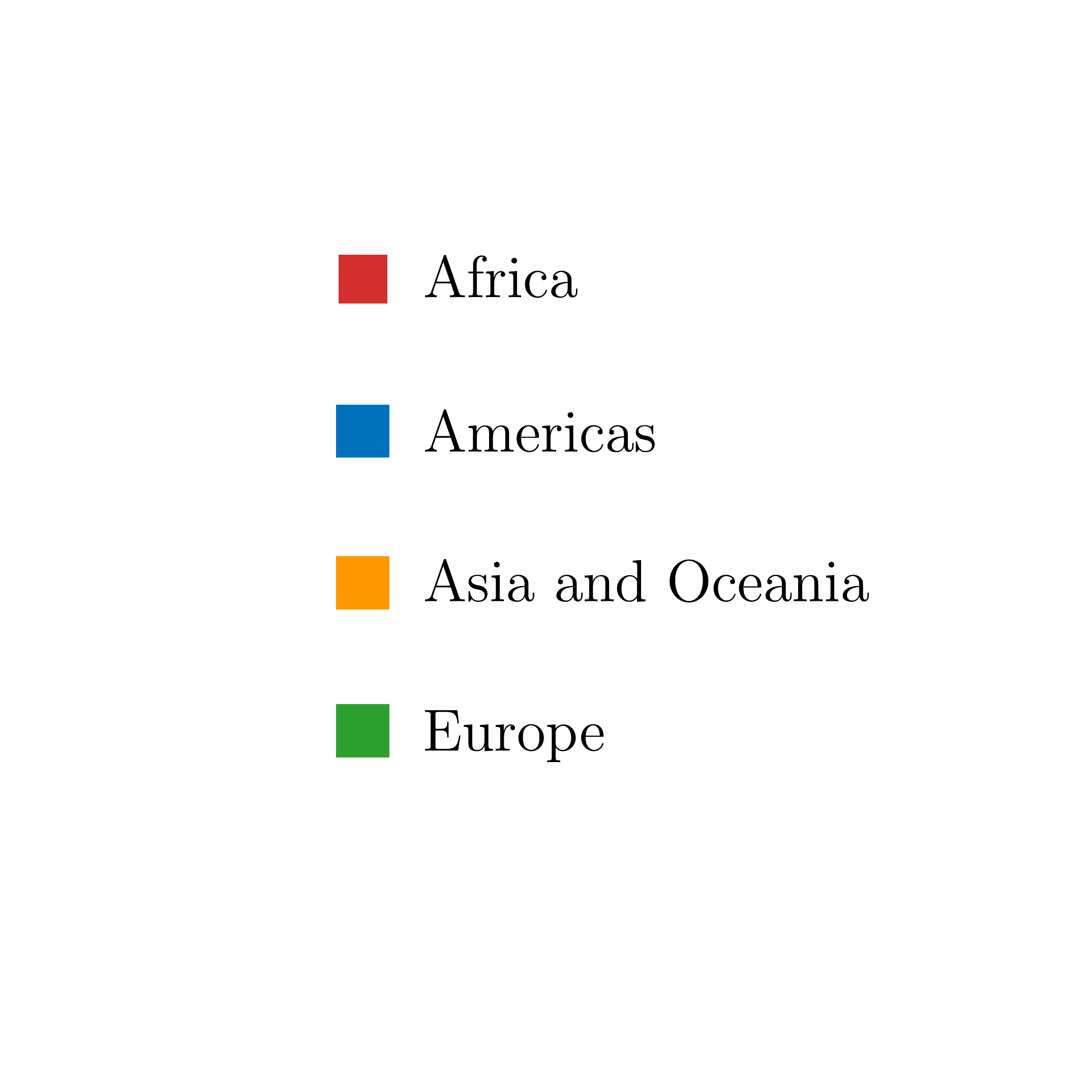}
    \end{minipage}
\end{tabular}
   \caption{
   	Core-periphery structure of the airport network detected by ({a}) the BE--KL algorithm, ({b}) the two-step algorithm and ({c}) our algorithm.
   	The colour indicates the geographical region. 
   }
   \label{fig:airport}
\end{figure}

Figure~\ref{fig:airport} shows the core-periphery structure detected by the three algorithms.
The BE--KL algorithm detects a dense core composed of 89 airports scattered in different geographical regions (Fig.~\ref{fig:airport}(a)).
The peripheral airports are rarely adjacent to the core airports in other regions. 
Furthermore, the peripheral airports tend to be adjacent to other peripheral airports in the same region, which is inconsistent with the notion of the periphery. 

The two-step algorithm detects 16 geographically concentrated core-periphery pairs (Fig.~\ref{fig:airport}(b)), in which 
some peripheral airports are densely interconnected within the core-periphery pairs. 
The average density of intra-peripheral edges within a core-periphery pair is $0.0383$, which is approximately 10 times larger than the edge density for the entire network, $p=0.0036$ (Fig.~\ref{fig:density}(c)).
Therefore, the structure detected by the two-step algorithm is not consistent with the concept of core-periphery structure with which peripheral nodes should be sparsely interconnected.

Our algorithm detects 10 geographically concentrated core-periphery pairs (Fig.~\ref{fig:airport}(c)).
The partition of the worldwide airport network into geographically distinct groups of airports found here is consistent with the previous results derived with community detection algorithms \cite{Guimera2005,Sales-Pardo2007}.
Compared to the two-step algorithm, the peripheral airports detected by our algorithm are not densely interconnected; the average density of intra-peripheral edges within a core-periphery pair is $0.000073$, which is smaller than the edge density for the entire network, $p=0.0036$  (Fig.~\ref{fig:density}(c)).

\begin{figure}
	\centering
	\includegraphics[width=\hsize]{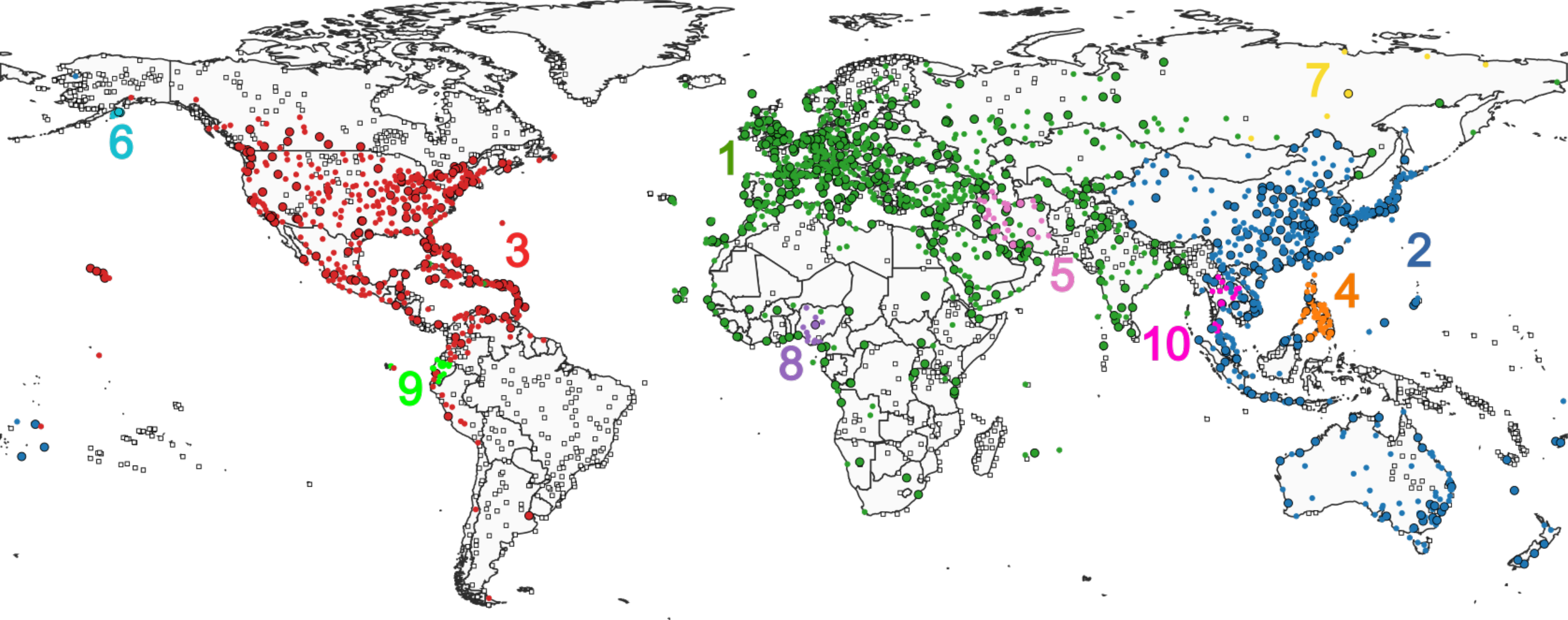}
	\caption{
		Location of the airports. 
		The large and small filled circles represent the core and peripheral airports, respectively.
		Each colour represents a core-periphery pair. The open squares represent residual airports. 
	}
\label{fig:map_world}
\end{figure}

We further analyse the core-periphery structure obtained by our algorithm. 
Figure~\ref{fig:map_world} maps the locations of the core and peripheral airports.
The three largest core-periphery pairs labelled 1, 2 and 3 are mainly based in Europe, East Asia and the United States, respectively.
The core-periphery pairs 1, 2 and 3 consist of the airports in 125, 35 and 47 countries, respectively.  
Each of the other core-periphery pairs labelled 4--10 consists of the airports in one country. 

\begin{figure}
    \centering
    \begin{tabular}{cc}
    	\begin{minipage}[t][][b]{0.5\hsize}
    		\centering
        	\includegraphics[width=\hsize]{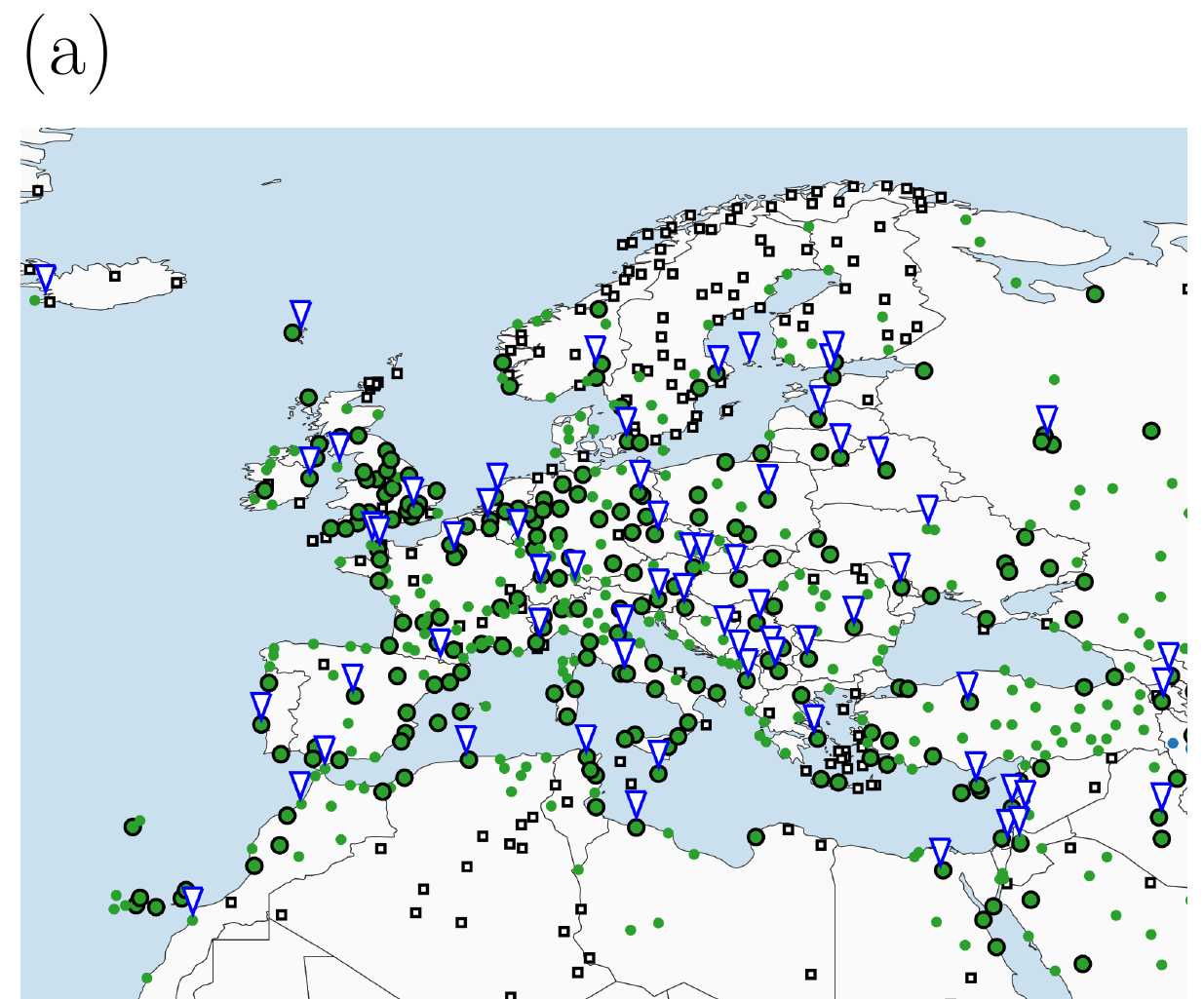} 
		\\
	        \includegraphics[width=\hsize]{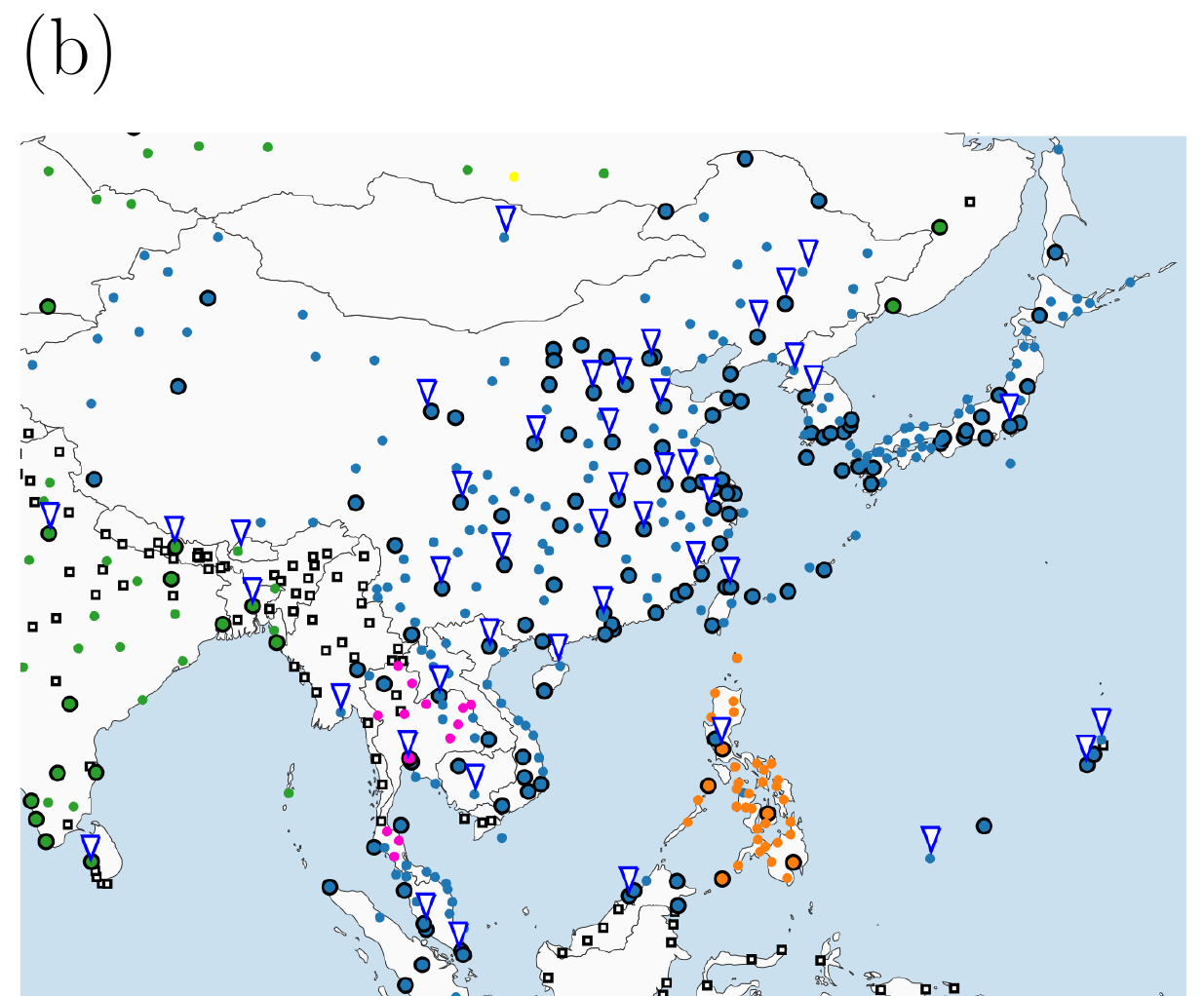} 
    	\end{minipage}
	&
 	 \begin{minipage}[t][][b]{0.5\hsize}
 	   	\centering
	        \includegraphics[width=\hsize]{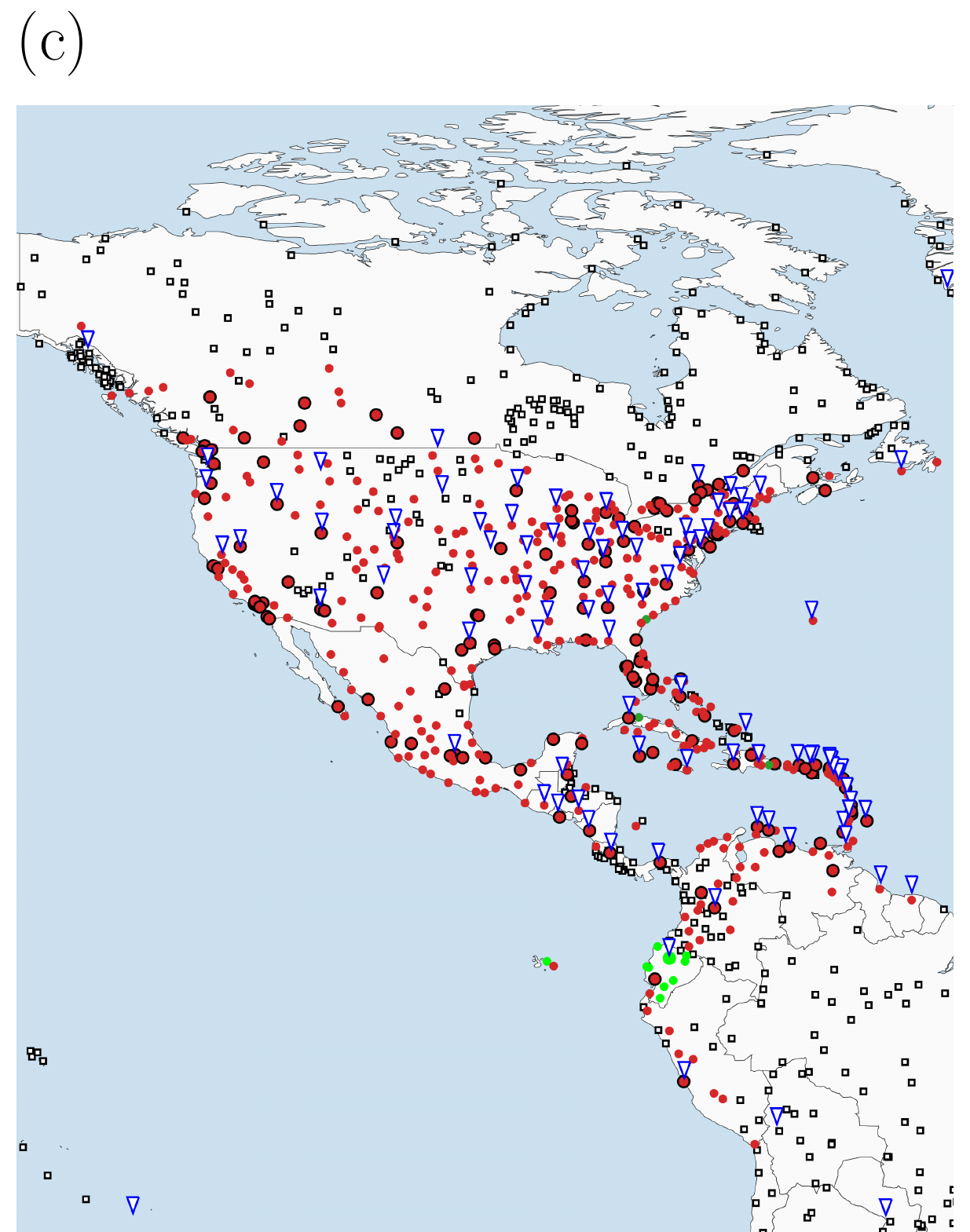} 
	    \end{minipage}
	\end{tabular}
    \caption{
    	Location of the airports in ({a}) Europe, ({b}) East Asia, ({c}) the contiguous United States and their surrounding areas. 
    	The large and small filled circles represent the core and the peripheral airports, respectively.
    	Each colour represents a core-periphery pair. The open squares represent residual airports. 
    	The inverted triangles indicate the location of metropolises, i.e., the capital cities of all countries, the provincial capitals of China and 
    	the state capitals of the United States.
    }
    \label{fig:map}
\end{figure}

The location of the airports and metropolises in Europe, East Asia, the United States and their surroundings are shown in Fig.~\ref{fig:map}.
Here the metropolis is defined as the capital city of all countries, 
the provincial capitals of China and the state capitals of the United States because China and the United States have many airports. 
Core-periphery pair 1 contains 333 core airports and 378 peripheral airports, of which 405 (57\%) airports are located in Europe (Fig.~\ref{fig:map}(a)).
However, this core-periphery pair excludes most airports in the Nordic countries (84 airports; 68\%).
There are 89 airports within 20 miles from a metropolis in Europe, among which there are 51 core airports (57\%), 28 peripheral airports (31\%) and 10 residual airports (11\%).
As a comparison, if we select the same number of the European airports with the largest degrees as that of the European core airports,
then 64 airports (72\%) are contained in the set of 89 airports within 20 miles from a metropolis, which is more than the number of  
the core airports (51 airports; see above) contained in the same set of airports.
This result indicates that hub metropolitan airports, which are common, are not necessarily core airports.
 
The second core-periphery pair contains 161 core airports and 240 peripheral airports, among which 217 (54\%) airports are located in East Asia (Fig.~\ref{fig:map}(b)).
In this core-periphery pair, 31 airports are located within 20 miles from a metropolis in East Asia, among which there are 23 core airports (74\%), eight peripheral airports and no residual airport (Fig.~\ref{fig:map}(b)). 

The third core-periphery pair contains 150 core airports and 312 peripheral airports, among which 210 (45\%) airports are located in the United States (Fig.~\ref{fig:map}(c)).
In this core-periphery pair, 71 airports are located within 20 miles from a metropolis in the United States, among which there are 29 core airports (41\%), 30 peripheral airports (42\%) and 12 residual airports (17\%) (Fig.~\ref{fig:map}(c)).
We have not found the partitioning of airports into core-periphery pairs corresponding to different major airline groups
(e.g., American Airlines,  Delta Airlines, Southwest Airlines and United Airlines in the United States). 

Table~\ref{ta:profile} lists the size of core-periphery pairs and the fractions of different types of edges.
The airports in a large core are not densely interconnected compared to those in small core-periphery pairs, probably due to the limited capacity of the airports (e.g., a small number of runways).
Core-periphery pairs 1, 2 and 3 contain hub airports in each region.
The other small core-periphery pairs consist of a small number of core airports, i.e., at most 20\% of the airports in each core-periphery pair. 
In these core-periphery pairs, most of the peripheral airports are adjacent to the core airports but not to other peripheral airports in the same core-periphery pair. 
This observation suggests that a small number of core airports relays most of the flights into these regions as gateway airports.
For example, the representative core airport (i.e., the core airport that has the largest number of neighbours in the core-periphery pair) in pair 4, MNL (Philippines), and that in pair 8, LOS (Nigeria), serve most of the domestic airports in the respective countries.
Such a structure is evident in small core-periphery pairs such as core-periphery pairs 4--10.

\begin{figure}
    \centering
   \begin{tabular}{cc}
    \begin{minipage}{0.5\hsize}
        \centering
        \includegraphics[height=250pt]{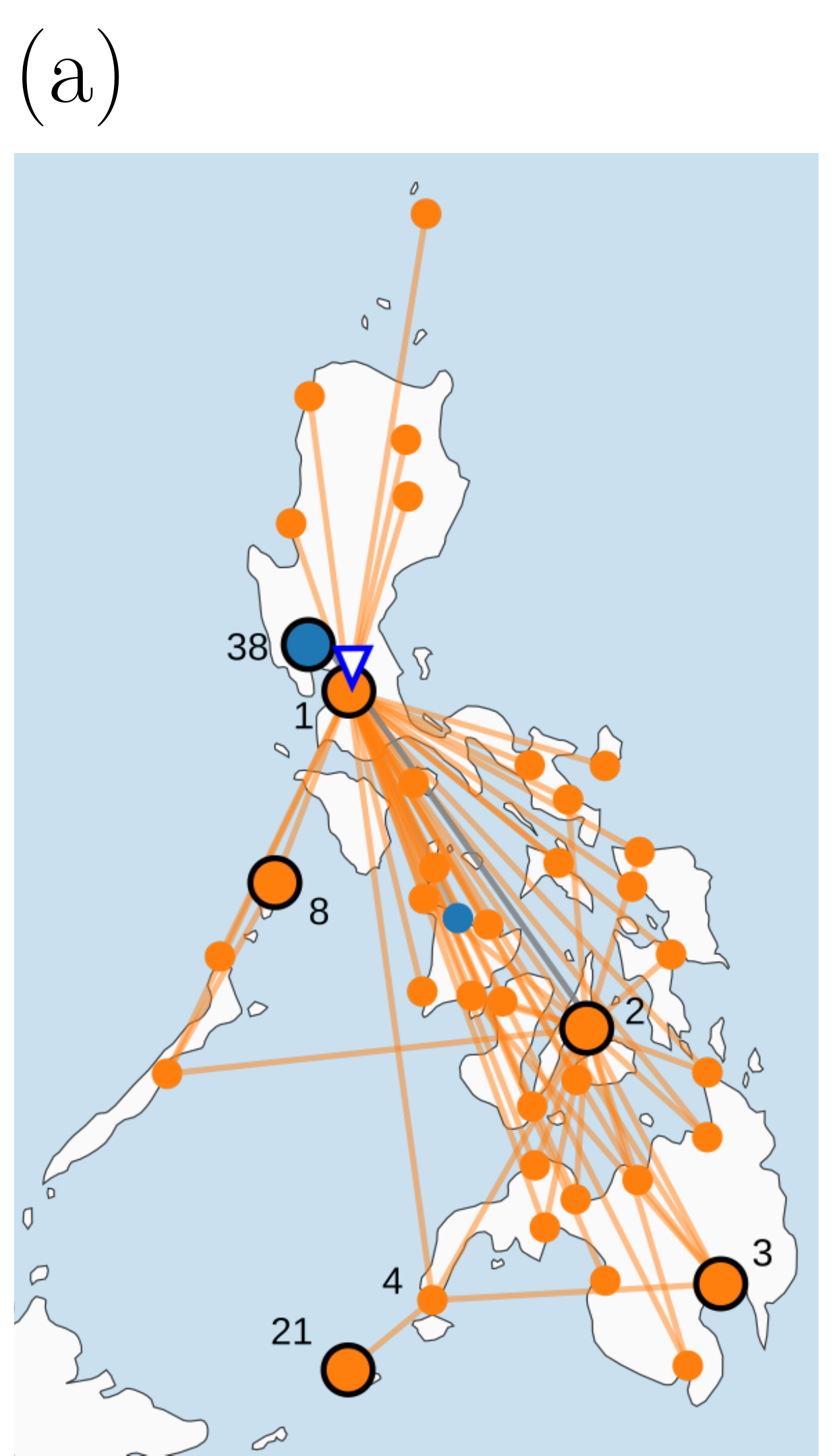} 
    \end{minipage}
    &
    \begin{minipage}{0.5\hsize}
        \centering
        \includegraphics[height=250pt]{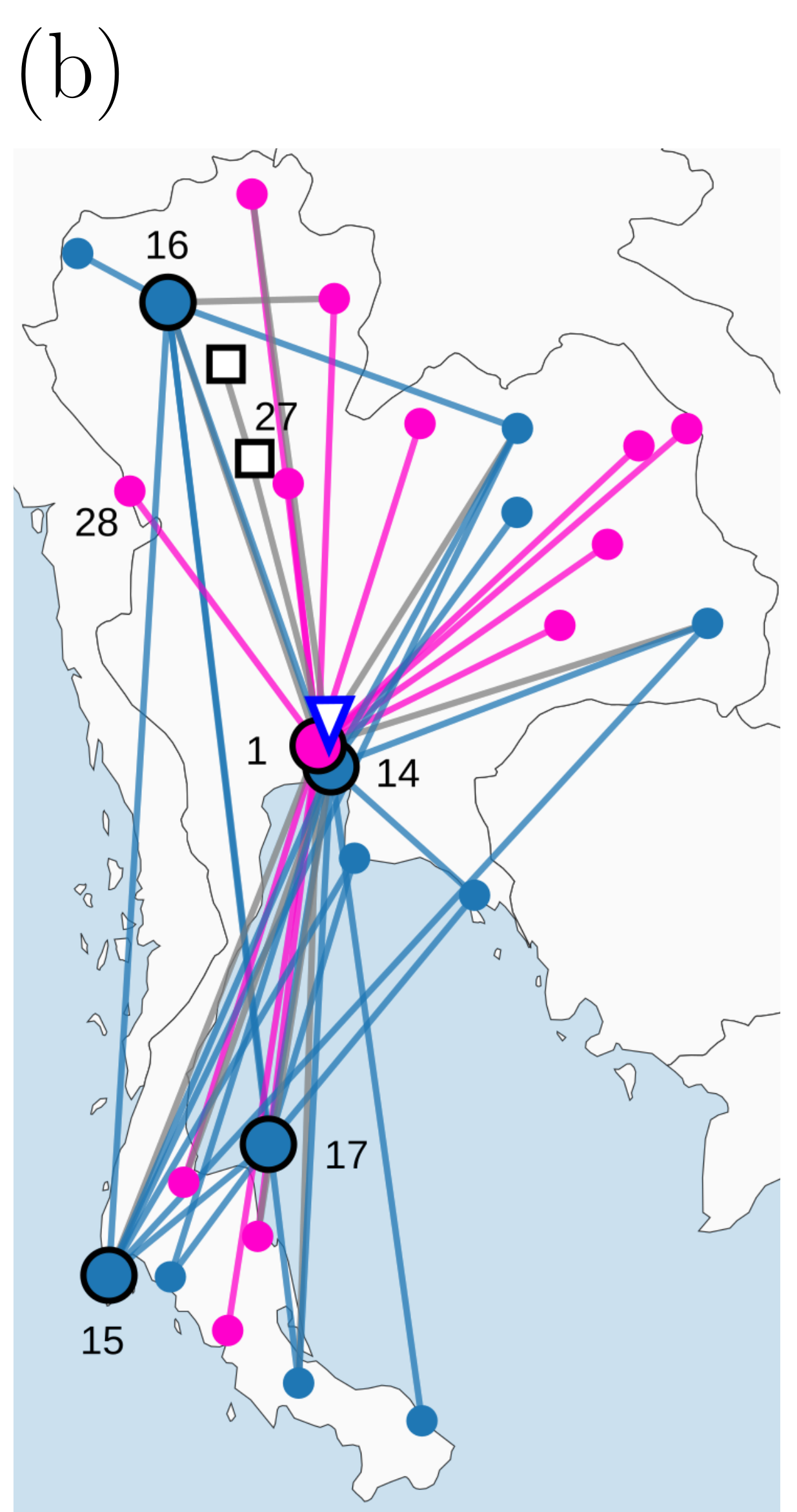} 
    \end{minipage}
\end{tabular}
\caption{
Airport network within  ({a}) the Philippines and ({b}) Thailand. 
The line colour indicates the core-periphery pair to which the two airports belong.
The edges connecting two airports in different core-periphery pairs are shown in grey.
The numbers attached to some airports indicate the IDs of the airports listed in Tables~\ref{ta:philippines} and \ref{ta:thailand}.
We only show the IDs of all core airports, some peripheral airports and all residual airports. 
}
\label{fig:philippines_thailand}
\end{figure}

The subnetwork within the Philippines is shown in Fig.~\ref{fig:philippines_thailand}(a); see Table~\ref{ta:philippines} for properties of all airports in the Philippines.
Most of the airports (34 airports; 92\%) in core-periphery pair 4 (shown in orange in Figs.~\ref{fig:map_world}, \ref{fig:map}(b) and \ref{fig:philippines_thailand}(a)) only serve domestic flights.
Core airport 1 (labelled in Fig.~\ref{fig:philippines_thailand}(a)) has most of the edges (41 edges; 84\%) between core-periphery pair 4 and the rest of the network.
Therefore, core airport 1 functions as a gateway airport in the Philippines.
Core airport 2 also functions as a gateway airport, but to a lesser extent than core airport 1 does.
Core-periphery pairs located in Alaska (core-periphery pair 6 in Table~\ref{ta:profile}), Russia (pair 7) and Ecuador (pair 9) also contain a few core airports serving as gateway airports in the respective regions (Appendix~\ref{sec:othercp}). 
Core airports 8 and 21 in the Philippines (Fig.~\ref{fig:philippines_thailand}(a)) have a small degree, which is counterintuitive.
Core nodes having degree one or two are also found in core-periphery pair 6 (Fig.~\ref{fig:small_cp}(c)). 
The airports 8 and 21 in the Philippines are adjacent to one peripheral airport 7 and 4, respectively.
If we assign airport 8 to the periphery, then two peripheral airports 7 and 8 would be adjacent.
Similarly, if we assign airport 21 to the periphery, then two peripheral airports 4 and 21 would be adjacent.
To avoid edges between peripheral nodes, our algorithm has identified airports 8 and 21 as core nodes.
However, airports 8 and 21 may be better regarded as peripheral airports given that they are not densely interconnected to other core airports.
Previous studies provided remedies for this issue \cite{Borgatti2000,Rombach2014} (see Section~\ref{sec:discussion} for further discussion).

The subnetwork within Thailand is shown in Fig.~\ref{fig:philippines_thailand}(b); see Table~\ref{ta:thailand} for properties of all airports in Thailand.
Two major airports 1 and 14 are located in the capital city, Bangkok, and belong to different core-periphery pairs.
All international airports in Thailand belong to core-periphery pair 2 (shown in blue in Figs.~\ref{fig:map_world}, \ref{fig:map}(b) and \ref{fig:philippines_thailand}(b)), including core airport 14.
Most of the domestic airports (13 airports; 59\%) belong to core-periphery pair 10 (shown in magenta), including core airport 1.
The subnetwork composed of core-periphery pair 10 is largely separated from the other airports in Thailand, which belong to core-periphery pair 2, and the rest of the world. 
The separation of the domestic and international airports and their respective subnetworks is also observed in the Philippines (Fig.~\ref{fig:philippines_thailand}(a)), Iran and Nigeria (Appendix~\ref{sec:othercp}).  
 
\subsection{Computation time}
\label{sec:cputime}
We implement the three algorithms in MATLAB and run simulations on a computer with Intel 2.6GHz Sandy Bridge processors and 4GB of memory.
The speed of an algorithm is measured by averaging the CPU time over 100 runs. 
We do not run the statistical test because it is a common process for the three algorithms.

The average CPU time of the three algorithms is compared in Table~\ref{ta:cputime}.
The BE--KL algorithm is the fastest on all synthetic networks and the karate club network. 
However, it is slower than our algorithm on the blog and airport networks.
The two-step algorithm is the slowest all but one network.
Our algorithm is approximately two times slower than the BE--KL algorithm on the synthetic and karate club networks.
However, on the blog and airport networks, it runs much faster than the BE--KL algorithm.
Our algorithm runs in ${\cal O}(rM)$ time, where $r$ is the number of rounds (Section~\ref{sec:algorithm}). 
Therefore, our algorithm is expected to be fast on sparse networks.

\section{Discussion}
\label{sec:discussion}
We proposed a scalable algorithm to detect multiple core-periphery pairs in networks by maximising a novel quality function $Q^{\text{cp}}$. 
The quality function $Q^{\text{cp}}$ compares the number of edges of different types in the given network with the expected number for an \ER random graph. %
The present algorithm reveals the groups of nodes sharing a common property (e.g., a group of members sharing the same political leaning) and different roles of nodes within the groups (e.g., leaders and followers in each group).
These properties are simultaneously revealed neither by partitioning of networks into a single core-periphery pair nor community detection.%

In the airport network, we have found several core nodes having degree one or two (e.g., airports 8 and 21 in Fig.~\ref{fig:philippines_thailand}(a)), 
which contradicts the intuition that core nodes would have a large degree.
Our algorithm assigned these nodes to a core to suppress the edges between peripheral nodes.
However, these nodes may be better regarded as peripheral nodes because they are adjacent to at most one core node.
One remedy is to weaken the suppression of the edges between peripheral nodes \cite{Borgatti2000,Rombach2014}.
Adapting this idea to the case of multiple core-periphery structure warrants future research.

Previous studies provided algorithms to detect multiple core-periphery pairs based on community detection algorithms.
Yang and Leskovec used an algorithm for detecting overlapping communities in networks \cite{Yang2012,Yang2013,Yang2014}.
They regarded the nodes belonging to many communities as core nodes and nodes belonging to few communities as peripheral nodes. 
The algorithm may detect densely interconnected peripheral nodes because the detected peripheral nodes in a single core-periphery pair belong to the same community.
In addition, a periphery may belong to multiple cores in these algorithms. 
These properties are shared by the algorithms proposed in Refs.~\cite{Sardana2016,Xiang2016}.
In contrast, our algorithm detects disjoint core-periphery pairs such that peripheral nodes are interconnected sparsely within each core-periphery pair and across different core-periphery pairs. 
Yan and Luo focussed on a different type of structure consisting of multiple cores and a single periphery \cite{Yan2016}.
In contrast, a core detected by our algorithm owns its exclusive periphery, including the case of an empty periphery.

\begin{figure}[tb]
    \centering
    \begin{minipage}{0.6\hsize}
    \centering
    \includegraphics[width=\hsize]{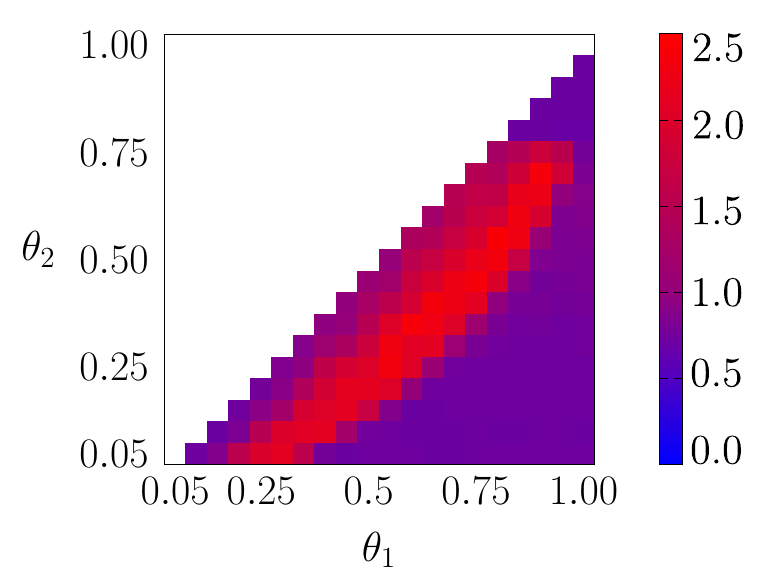}
    \end{minipage}
    \caption{
    The VI values between the true and inferred core-periphery structure obtained by the maximisation of $Q^{\text{cp}} _{\text{config}}$.
    }
    \label{fig:conf}
\end{figure}

We used the \ER random graph as the null model to define $Q^{\text{cp}}$. 
The \ER random graph model is also used for detecting communities in networks \cite{Reichardt2004,Reichardt2006,Karrer2011,Zhao2011}.
For example, the community detection algorithms based on the Potts model \cite{Reichardt2004,Reichardt2006} and stochastic block models without degree correction (i.e., without assuming a heterogeneous degree distribution) \cite{Karrer2011} use the \ER random graph model as a null model.
There are also null models that incorporate other properties of networks such as
degree heterogeneity \cite{Newman2004}, weighted edges \cite{Mastrandrea2014}, signed edges \cite{Traag2009}, correlations \cite{MacMahon2015}, bipartiteness \cite{Barber2007} and space embeddedness \cite{Expert2011}.
It is possible to incorporate these null models into the second term of the right-hand side of Eq.~(\ref{eq:proposed}).
For example, a popular null model is the configuration model, with which we randomly rewire edges while conserving the degree of each node \cite{Fortunato2010}.
With the configuration model, the quality function is given by $Q^{\text{cp}} _{\text{config}}=\sum_{i=1}^N \sum_{j=1}^{i}(A_{ij}-d_id_j/2M)(x_i+x_j-x_ix_j)\delta_{c_i,c_j}$.
As is the case in community detection \cite{Karrer2011}, the choice of the null model will affect the organisation of the detected core-periphery pairs.
To see this, we maximised $Q_{\text{config}} ^{\text{cp}}$ using a label switching heuristic (Section \ref{sec:algorithm}) for the synthetic networks with a single core-periphery pair (Fig.~\ref{fig:sb}(a)).
The VI values for $Q^{\text{cp}} _{\text{config}}$ were larger than 0.4 in the entire parameter region spanned by $\theta_1$ and $\theta_2$ (Fig.~\ref{fig:conf}), indicating that 
 the maximisation of $Q^{\text{cp}} _{\text{config}}$ did not enable us to detect the planted core-periphery pair.
This is due to the null-model term  $d_id_j/2M$ in $Q^{\text{cp}} _{\text{config}}$. 
In the synthetic networks with a planted single core-periphery pair, the planted core nodes have a large degree.
If we put the planted core nodes into the same core-periphery pair, then the null-model term $d_id_j/2M$ becomes large, giving a large decrement in $Q^{\text{cp}} _{\text{config}}$.
Therefore, the maximisation of $Q^{\text{cp}} _{\text{config}}$ assigned the planted core nodes to different core-periphery pairs (yielding $\delta_{c_i,c_j}=0$) or a periphery (yielding $x_i+x_j-x_ix_j =0$).

There are several lines of possible extensions of the present work.
First, we did not consider continuous versions of core-periphery structure, with which each node is assigned a strength (i.e., a coreness) value representing the belongingness of the node to a core \cite{Borgatti2000,DaSilva2008,Boyd2010,Rossa2013,Rossa2013,Rombach2014,Lee2014,Cucuringu2016}.
Continuous versions of core-periphery structure can reveal nested structure of cores (i.e., cores within a core), which discrete versions of the algorithms would not.
Borgatti and Everett generalised a discrete version of core-periphery structure defined by Eq.~(\ref{eq:bg}) to a continuous version
by replacing binary variables $x_i \in \{0,1\}$ ($1\leq i\leq N$) by continuous variables (e.g., $x_i \in [0,1]$) \cite{Borgatti2000}.
This approach may allow us to generalise our algorithm to continuous versions.

Second, we have ignored the weight and direction of edges.
It is straightforward to incorporate the weight of edges by redefining $A_{ij}$ on the right-hand side of Eq.~(\ref{eq:proposed}) as the weight of the edge.
In contrast, it is not easy to extend our algorithm to the case of directed networks. 
In the case of community detection, a natural extension is to allow an adjacency matrix ${\A}$ to be asymmetric \cite{Arenas2007,Leicht2008}, which is, in fact, problematic \cite{Kim2010}.
Therefore, we expect that the same pitfall exists in the detection of core-periphery pairs if we extend our algorithm to the case of directed networks by simply allowing $\A$ to be asymmetric.

Third, our quality function, $Q^{\text{cp}}$, has a similar mathematical form to modularity \cite{Newman2004,Fortunato2010}.
If we constrain all nodes to be core nodes, i.e.,  $x_i=1\ (1 \leq i \leq N)$, then $Q^{\text{cp}}$ is equivalent to the modularity based on the Potts model \cite{Reichardt2004,Reichardt2006}.
Another class of approach is based on the likelihood maximisation of stochastic block models.  
Stochastic block models are generative models of networks composed of blocks of nodes \cite{Karrer2011,Peixoto2017}.
In fact, stochastic block models for detecting a single core-periphery pair  \cite{Zhang2015} and multiple core-periphery pairs \cite{Tunc2015} have been proposed.
Modularity maximisation and likelihood maximisation are equivalent in a particular situation but are different in general \cite{Newman2016}.
Modularity has some limitations such as the incapacity of finding small communities \cite{Fortunato2007} and that of distinguishing non-random from random structure \cite{Guimera2004}.
Therefore, we may benefit from using the stochastic block models.

How multiple core-periphery pairs emerge is unclear. 
An economic mechanism explains the emergence of a single core-periphery pair in networks \cite{Verma2016}.
The authors considered the trade-offs between the profit obtained by connecting nodes and the cost for maintaining edges.
Core-periphery structure emerges if the cost is not extremely small or large relative to the profit \cite{Verma2016}.
Given their results, multiple core-periphery pairs may emerge when the cost of intergroup edges is significantly larger than the cost of intragroup edges.  
For example, in airport networks, interregional flights would be more costly than intraregional flights due to the different fuel expense and tax.
Exploration of dynamical or economic mechanisms behind the formation of multiple core-periphery pairs in a network warrants future work.


\bibliographystyle{apsrev4-1}
%
\appendix
\section{\Tunc--Verma algorithm}
\label{sec:birkan}
We evaluated the performance of the \Tunc--Verma (TV) algorithm \cite{Tunc2015} using the synthetic networks used in Section~\ref{sec:experiment}.
The Python code provided by the authors  \cite{Tunc2015} is not fast enough.
Therefore, we reimplement their algorithm based on the original code by changing the data structure and replacing some functions by faster inbuilt functions in MATLAB.
We did not change the algorithm itself including the parameters.
We refer the original and reimplemented algorithms as the TV and TV$^{+}$, respectively. 
With the TV and TV$^{+}$ algorithms, each node belongs to multiple core-periphery pairs.
Therefore, we assign each node to the core-periphery pair to which the degree of belonging \cite{Tunc2015} is the largest.
The TV and TV$^{+}$ algorithms can infer the number of core-periphery pairs. 
However, to save their computational time, we provide the correct number of core-periphery pairs in the synthetic networks to these algorithms. 
We set $\theta_1=0.9$ and $\theta_2=0.05$, use the same $\Theta$ and $\pi$ parameter values as those used in the main text and vary $N \in \left\{10, 20, \ldots, 400\right\}$. 

\begin{figure}
    \centering
   \begin{tabular}{cc}
    \begin{minipage}{0.5\hsize}
        \centering
        \includegraphics[width=\hsize]{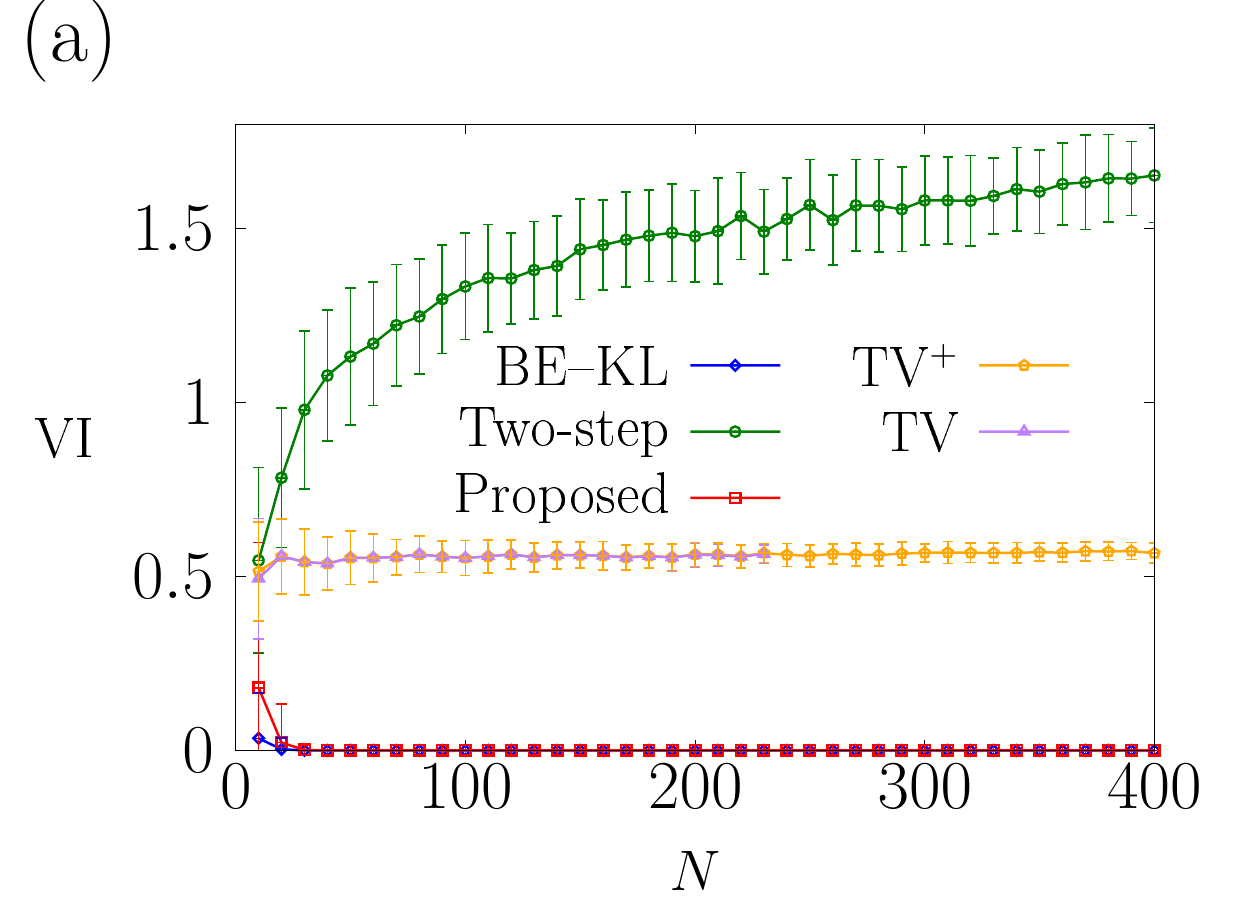} 
    \end{minipage}
    & 
    \begin{minipage}{0.5\hsize}
        \centering
        \includegraphics[width=\hsize]{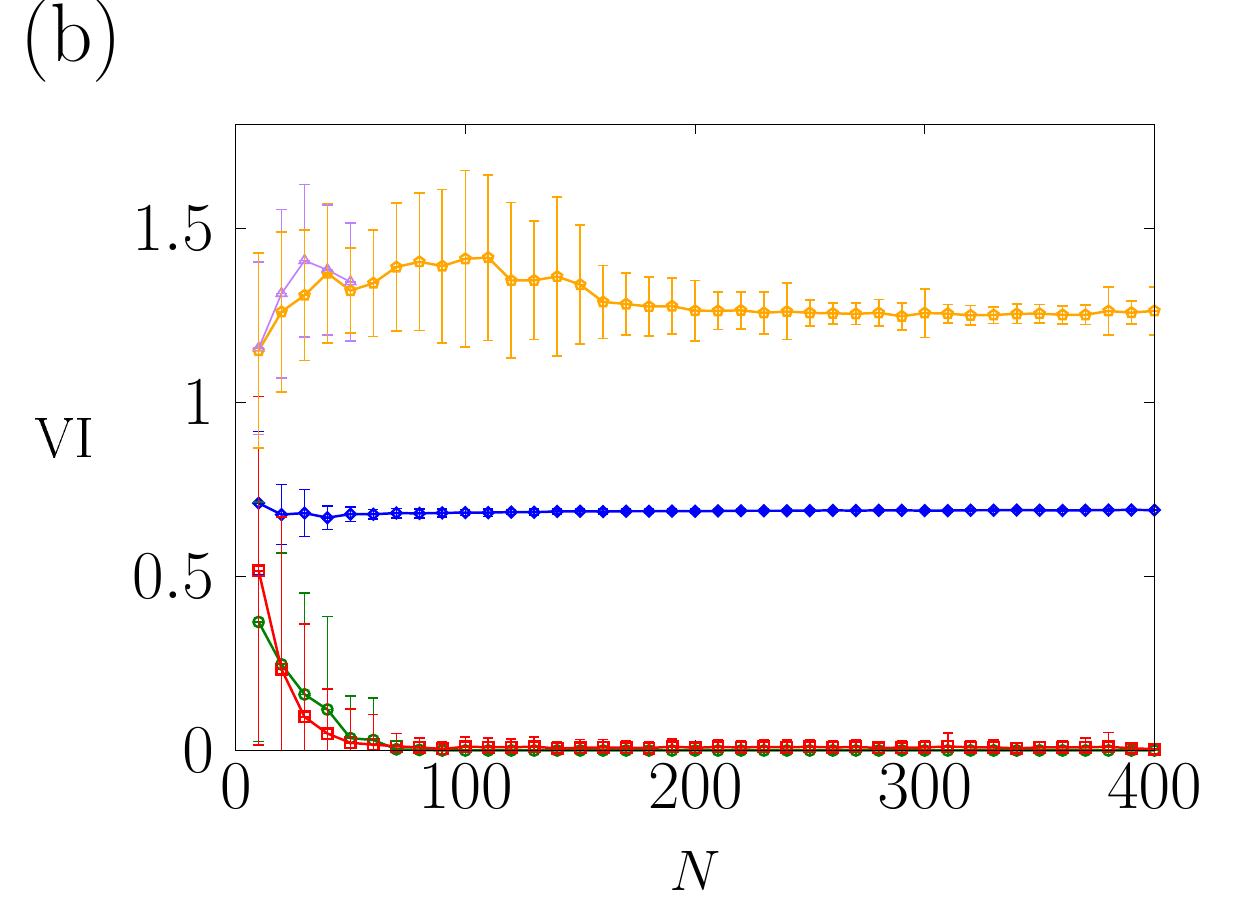} 
    \end{minipage}
   \end{tabular}
    \caption{
        The VI values between the true and inferred core-periphery structure for the five algorithms. 
        Panels ({a}) and ({b}) correspond to the networks whose structure is shown in Figs.~\ref{fig:sb}(a) and \ref{fig:sb}(b), respectively.
        The error bars indicate the $\pm1$ standard deviation.
    }
    \label{fig:birkan}
\end{figure}

Figure~\ref{fig:birkan}(a) shows the VI for the networks composed of a single core-periphery pair, whose structure is schematically shown in Fig.~\ref{fig:sb}(a).
The VI for the BE--KL and our algorithms is approximately equal to zero except for small $N$.
The VI for the two-step algorithm increases as $N$ increases.
The VI for the TV algorithm is large for $N\leq 230$.
The TV algorithm did not terminate for $N\ge240$. 
As expected, the VI values for the TV$^{+}$ algorithm are similar to those for the TV algorithm and similarly large for $N\ge 240$. 

Figure~\ref{fig:birkan}(b) shows the results for the networks composed of two core-periphery pairs, whose structure is schematically shown in Fig.~\ref{fig:sb}(b).
The VI for the BE--KL algorithm is large for all values of $N$ because the BE--KL algorithm is not designed for multiple core-periphery pairs.
The VI values for the two-step and our algorithms are comparable and close to zero for $N\geq 50$.
The VI for the TV and TV$^{+}$ algorithms is larger than that for the other algorithms.
The TV algorithm did not terminate for $N \ge 60$. 

In both types of the synthetic networks, the TV and TV$^{+}$ algorithms do not infer the true planted core-periphery structure for the entire range of $N$.
\Tunc\ and Verma applied their algorithm to dense and weighted networks.
In contrast, here we have analysed sparse and unweighted networks, which might be a main reason for the poor performance of these algorithms.

Next, we measure the speed of the TV$^{+}$ algorithm (which is faster than the TV algorithm) for the synthetic networks used in the main text (Section~\ref{sec:cputime}).
For the synthetic networks, we set $N=400$, $\theta_1=0.9$ and $\theta_2 = 0.05$. 
For the synthetic network with a single core-periphery pair (Fig.~\ref{fig:sb}(a)), the TV$^{+}$ algorithm requires 29.6 seconds on average.
This is 83 times slower than our algorithm (Table~\ref{ta:cputime}).
For the synthetic network with two core-periphery pairs (Fig.~\ref{fig:sb}(b)),
the TV$^{+}$ algorithm requires 81.1 seconds on average, which is 338 times slower than our algorithm (Table~\ref{ta:cputime}).
On the karate club network, for which we set the number of the core-periphery pair to two, 
the TV$^+$ algorithm requires 298.3 seconds on average, which is 11,932 times slower than our algorithm.  
For the blog and the airport networks, the TV$^{+}$ algorithm did not terminate in 12 hours. 

\section{Divisive algorithm}
\label{sec:divisive}
The divisive algorithm \cite{Girvan2002,Newman2004,Fortunato2010,Flores2013} partitions the nodes into communities using the Louvain algorithm \cite{Blondel2008} and then partitions the nodes in each community into core and peripheral nodes using the BE--KL algorithm. 
We apply the statistical test (Section \ref{sec:stest}) to the core-periphery pairs detected by the divisive algorithm.   

\begin{figure}
    \centering
	\begin{tabular}{ccc}
	    \begin{minipage}{0.45\hsize}
		\centering
	    \includegraphics[width=\hsize]{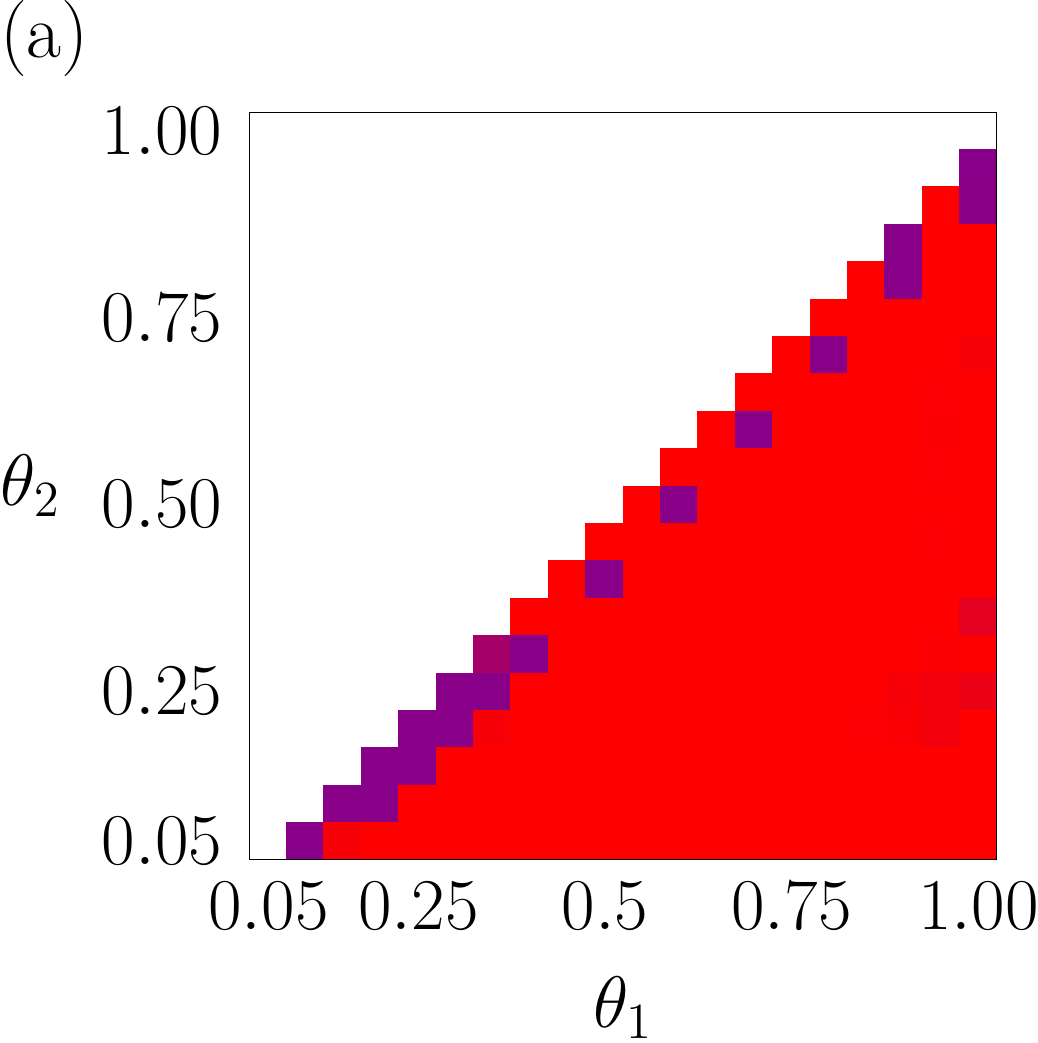}
	    \end{minipage}
	    &
	    \begin{minipage}{0.45\hsize}
		\centering
	    \includegraphics[width=\hsize]{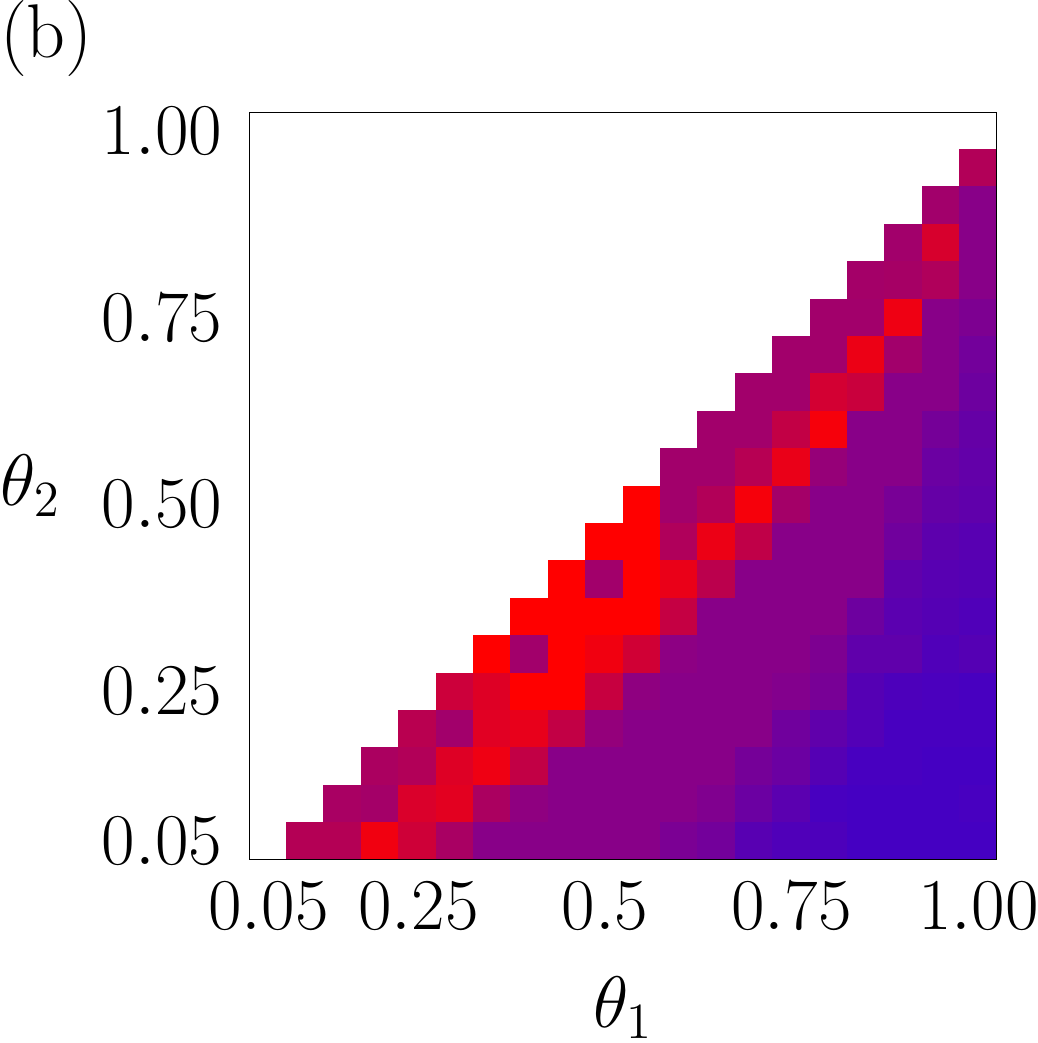}
	    \end{minipage}
	    &
	    \multirow{2}{*}{
	        \begin{minipage}{0.12\hsize}
	        \centering
	        \hspace{-.25\hsize}
	        \vspace{-.55\hsize}
	            \includegraphics[height=4\hsize]{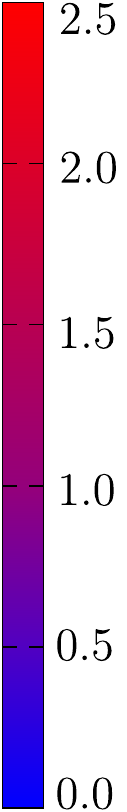}
	        \end{minipage}
	    }
	    \\
	    \\
	    \begin{minipage}{0.45\hsize}
		\centering
	    \includegraphics[width=\hsize]{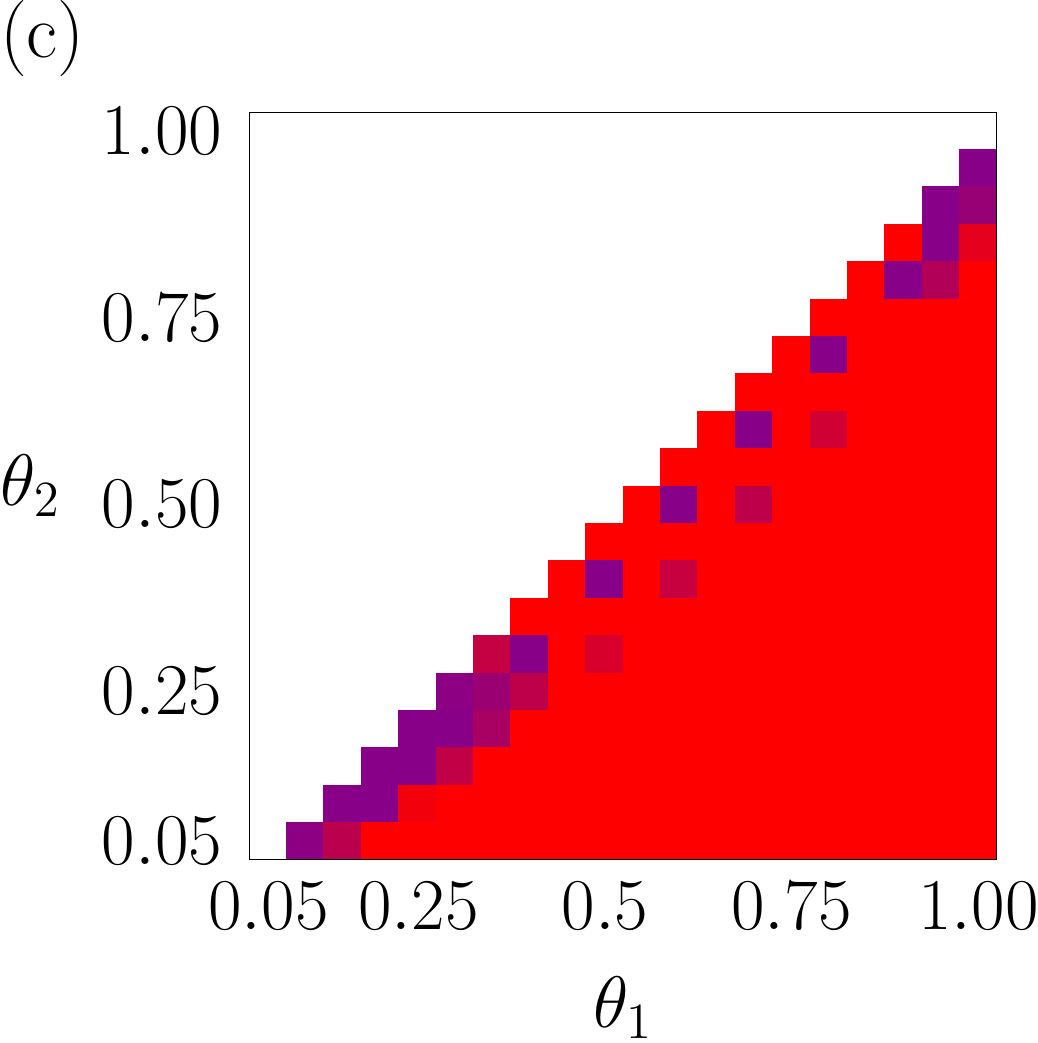}
	    \end{minipage}
	    &
	    \begin{minipage}{0.45\hsize}
		\centering
	    \includegraphics[width=\hsize]{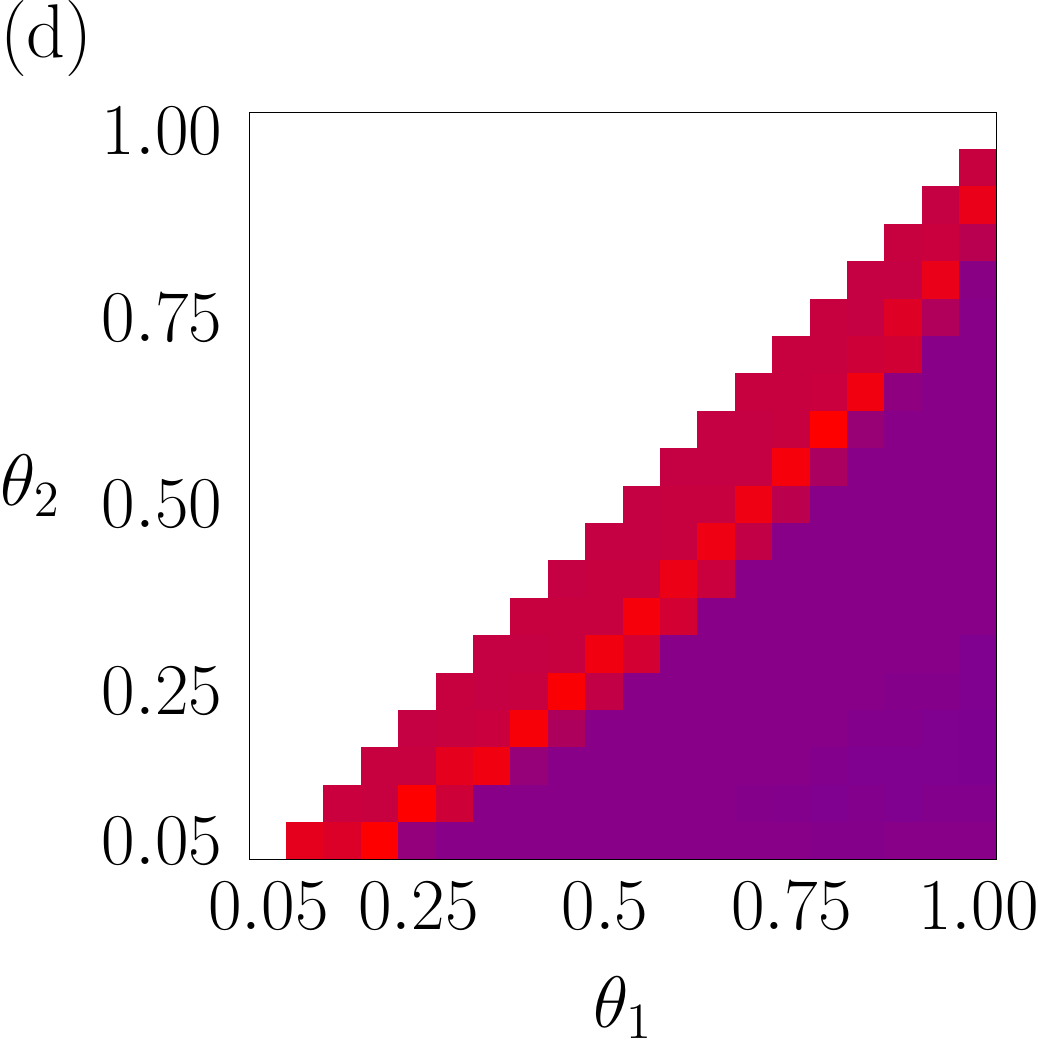}
	    \end{minipage}
	\end{tabular}
\caption{
    The VI values between the true and inferred core-periphery structure for the divisive algorithm. 
    Panels ({a}), ({b}), ({c}) and ({d}) show the VI values for the synthetic networks shown in Figs.~\ref{fig:sb}(a), \ref{fig:sb}(b), \ref{fig:sb}(c) and \ref{fig:sb}(d), respectively.
}
\label{fig:divisive_synthe}
\end{figure}

The VI values for the synthetic networks are shown in Fig.~\ref{fig:divisive_synthe}.
For the synthetic network with a single core-periphery pair (Fig.~\ref{fig:sb}(a)), the VI values are large in the entire $\theta_1$-$\theta_2$ parameter space (Fig.~\ref{fig:divisive_synthe}(a)) because the divisive algorithm divides the planted single core-periphery pair into multiple core-periphery pairs.
For the synthetic network with two core-periphery pairs (Fig.~\ref{fig:sb}(b)), the VI values are larger than those for the proposed algorithm for most $\theta_1$ and $\theta_2$ values (Figs.~\ref{fig:exp}(f) and \ref{fig:divisive_synthe}(b)).
For the synthetic network with a single core-periphery pair and residual nodes (Fig.~\ref{fig:sb}(c)), the VI values are larger than those for the proposed algorithm for most $\theta_1$ and $\theta_2$ values (Figs.~\ref{fig:exp}(i) and \ref{fig:divisive_synthe}(c)).
For the synthetic network with two core-periphery pairs and residual nodes (Fig.~\ref{fig:sb}(d)), 
the VI values are larger than those for the proposed algorithm in the entire parameter region of $\theta_1$ and $\theta_2$ (Figs.~\ref{fig:exp}($\ell$) and \ref{fig:divisive_synthe}(d)).

\begin{figure}
   \centering
   \begin{tabular}{cc}
    \multicolumn{2}{c}{
    \begin{minipage}{0.53\hsize}
        \centering
        \includegraphics[width=\hsize]{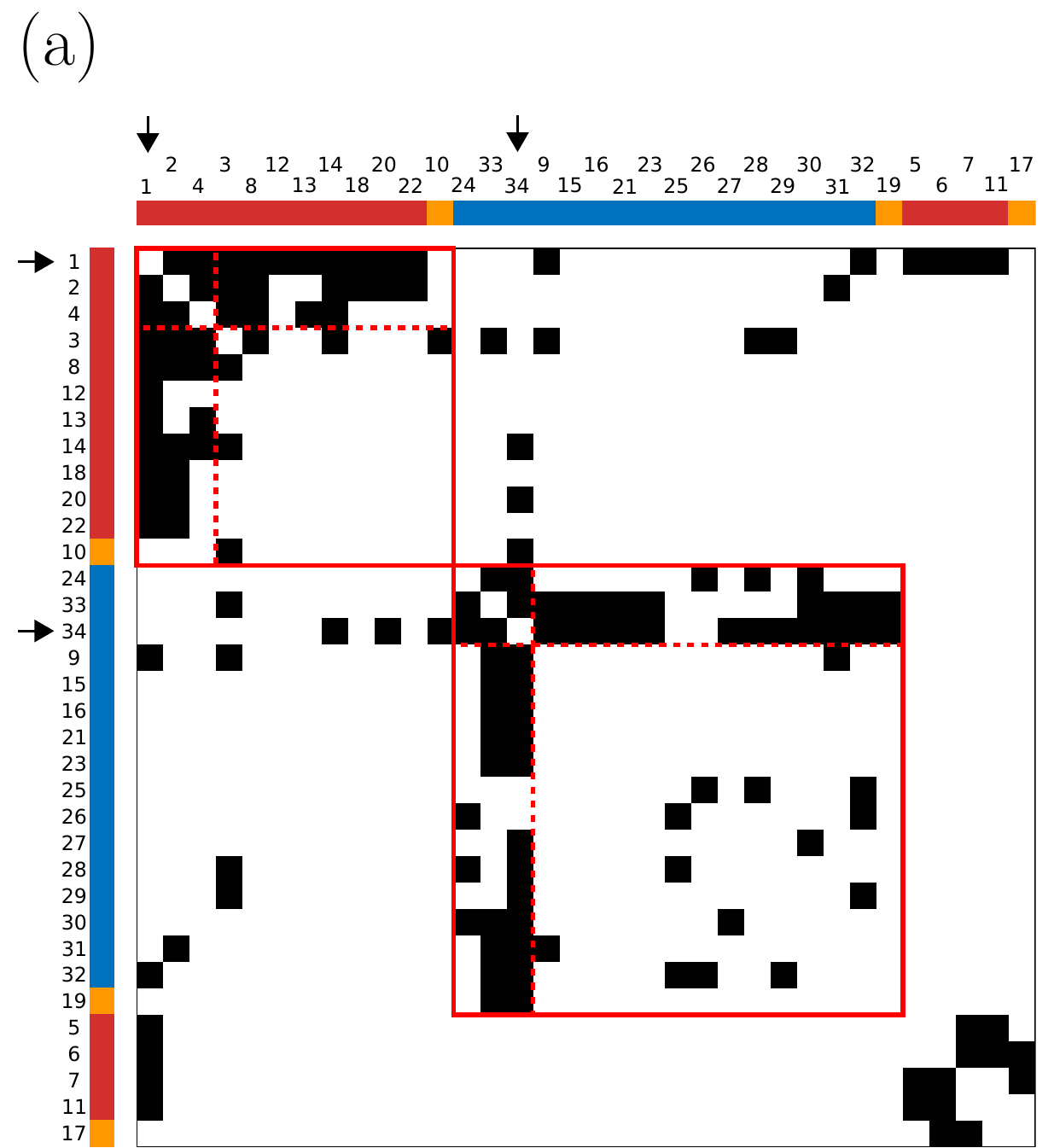} 
    \end{minipage}
    }
    \\ 
    \\ 
    \begin{minipage}{0.5\hsize}
        \centering
        \includegraphics[width=0.9\hsize]{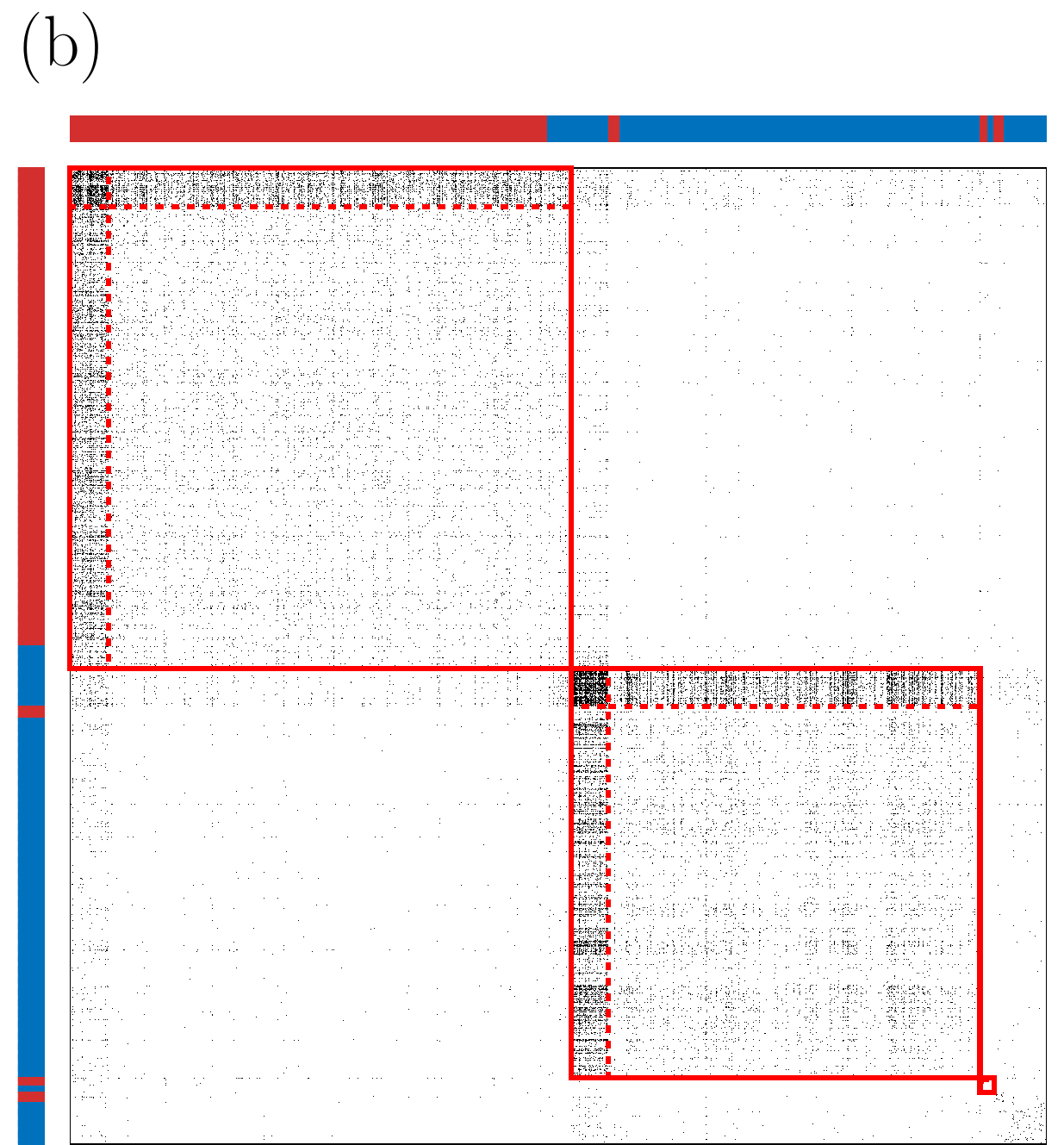} 
    \end{minipage}
    &
    \hspace{-1.3em}
    \begin{minipage}{0.5\hsize}
        \centering
        \includegraphics[width=0.9\hsize]{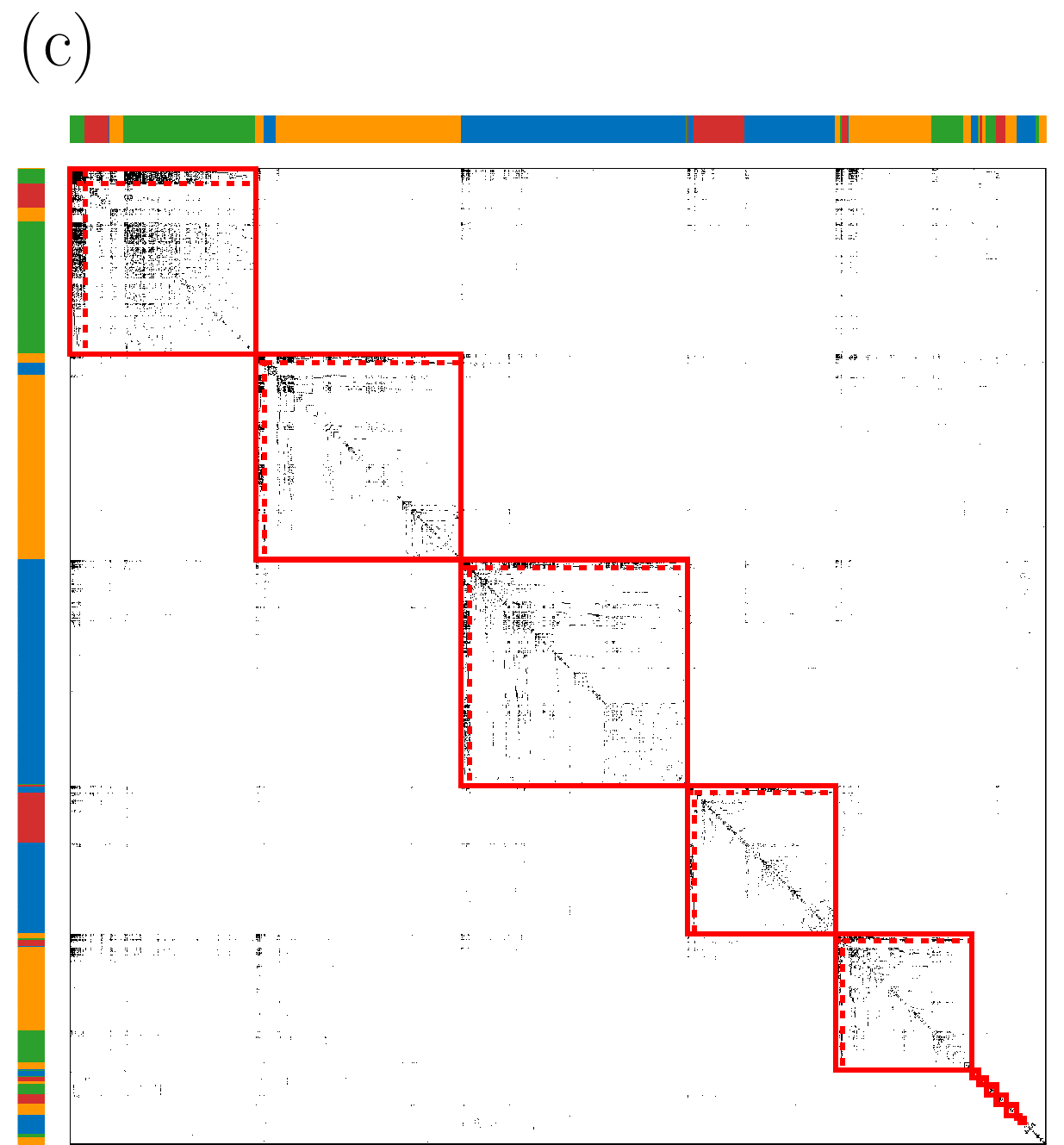} 
    \end{minipage}
   \end{tabular}
    
    \caption{
    Core-periphery structure identified by the divisive algorithm in  
    (a) the karate club network, (b) blog network and (c) airport network.
    }
    \label{fig:divisive}
\end{figure}

The core-periphery structure in the karate club detected by the divisive algorithm is shown in Fig.~\ref{fig:divisive}(a).
The divisive algorithm detects two core-periphery pairs, each of which mostly consists of the members supporting the same leader (Fig.~\ref{fig:divisive}(a)).
The average density of intra-core edges and that of core-periphery edges within a core-periphery are $1.000$ and $0.619$, respectively, which are larger than the edge density for the entire network, $p=0.139$.
The average density of intra-peripheral edges within a core-periphery pair is $0{.}080$, which is smaller than the edge density for the entire network, $p=0{.}139$. 
Therefore, the core-periphery structure in the karate network detected by the divisive algorithm is consistent with the concept of core-periphery structure based on edge density. 

In the blog network, the divisive algorithm detects three core-periphery pairs, each of which mostly comprises the blogs with the same political leaning (Fig.~\ref{fig:divisive}(b)).
Two core-periphery pairs are much larger than the third one and have the opposite political leanings. 
The divisive algorithm identifies more residual nodes than the two-step algorithm.  
The average density of intra-core edges and that of core-periphery edges within a core-periphery pair are $0.3964$ and $0.2906$, respectively, which are larger than the edge density for the entire network, $p=0.0224$.
The average density of intra-peripheral edges within a core-periphery pair is $0{.}0138$, which is smaller than the edge density for the entire network, $p=0{.}0224$.
Therefore, the core-periphery structure in the blog network detected by the divisive algorithm is consistent with the concept of core-periphery structure based on edge density. 

In the airport network, the divisive algorithm identifies 12 core-periphery pairs, each of which mostly consists of the airports located in the same geographical region (Fig.~\ref{fig:divisive}(c)).
The average density of intra-core edges and that of core-periphery edges within a core-periphery pair are $0.6335$ and $0.2978$, respectively, which are larger than the edge density for the entire network, $p=0.0036$.
The average density of intra-peripheral edges within a core-periphery pair is $0{.}0419$, which is larger than the edge density for the entire network, $p=0{.}0036$.
Therefore, the core-periphery structure in the airport network detected by the divisive algorithm is inconsistent with the concept of core-periphery structure based on edge density. 

\section{Properties of airports in core-periphery pairs 4--10}
\label{sec:othercp}
\begin{figure}[tb]
    \centering
   \begin{tabular}{cc}
    \begin{minipage}[t][][b]{0.5\hsize}
        \centering
        \includegraphics[width=\hsize]{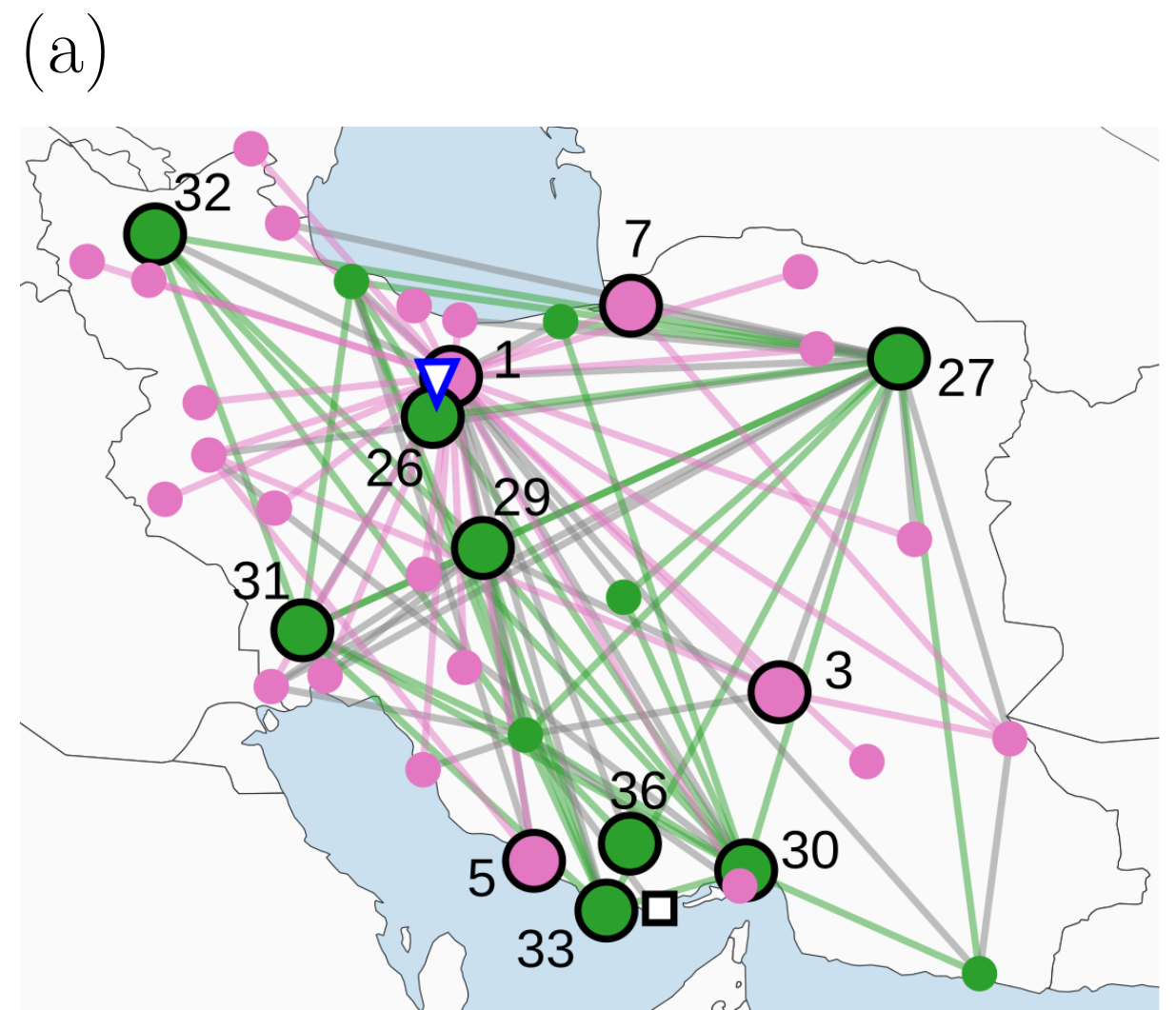} 
    \end{minipage}
    &
    \begin{minipage}[t][][b]{0.5\hsize}
        \centering
        \includegraphics[width=\hsize]{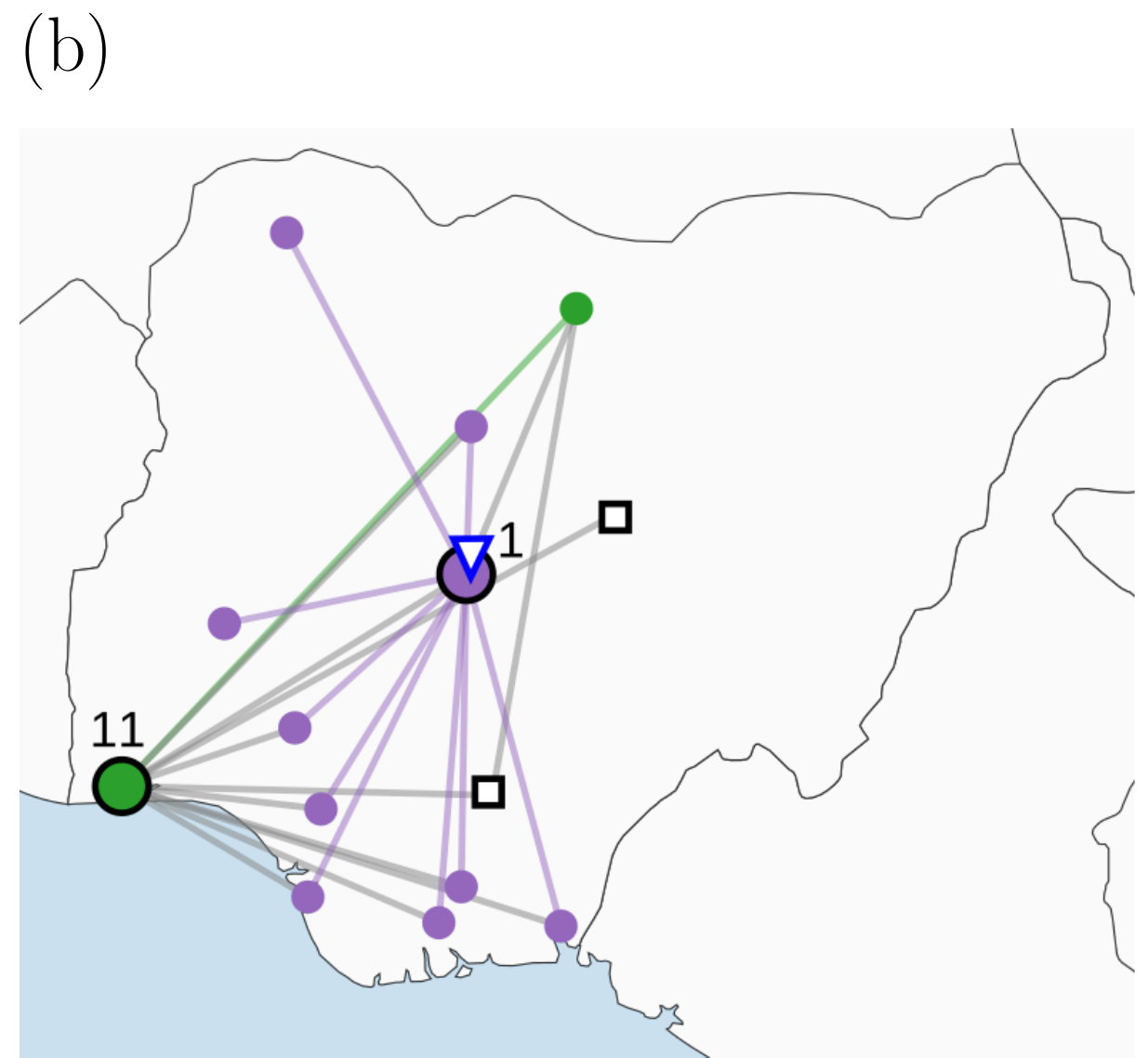} 
    \end{minipage}
    \\
    \begin{minipage}[t][][b]{0.5\hsize}
        \centering
        \includegraphics[width=\hsize]{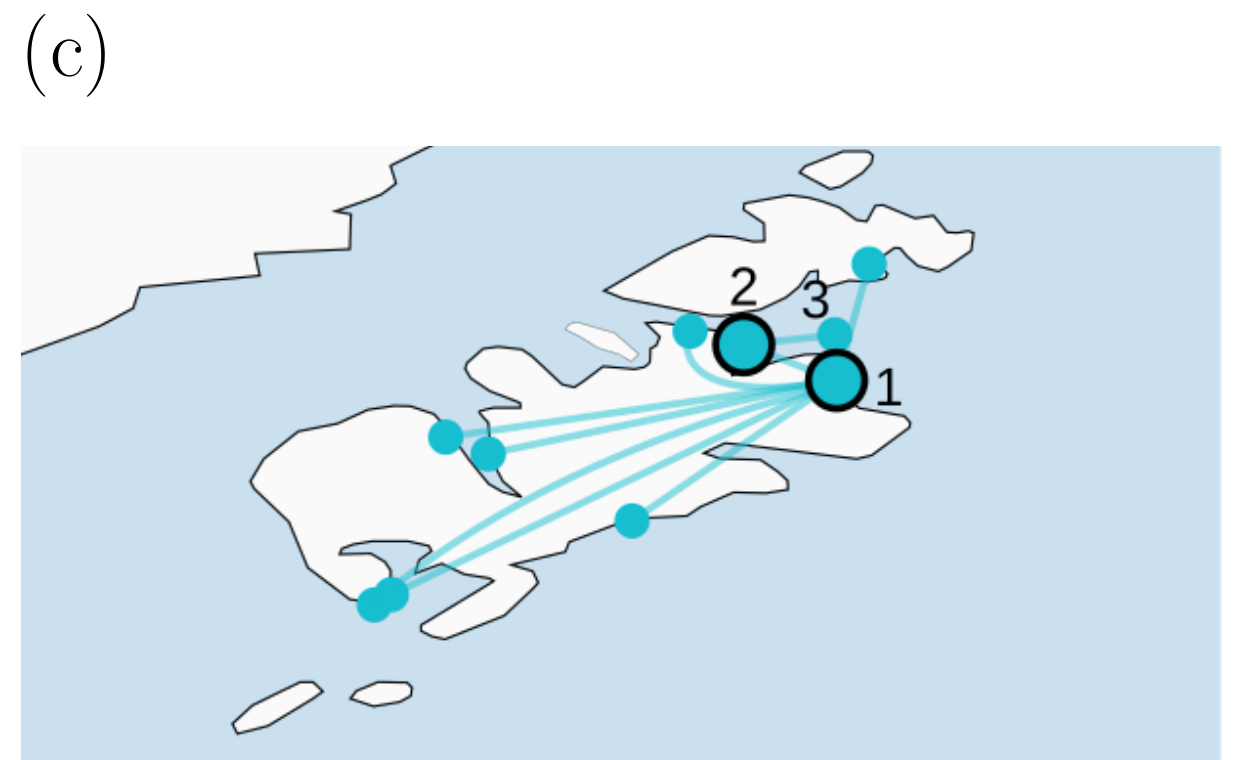} 
    \end{minipage}
    &
    \begin{minipage}[t][][b]{0.5\hsize}
        \centering
        \includegraphics[width=\hsize]{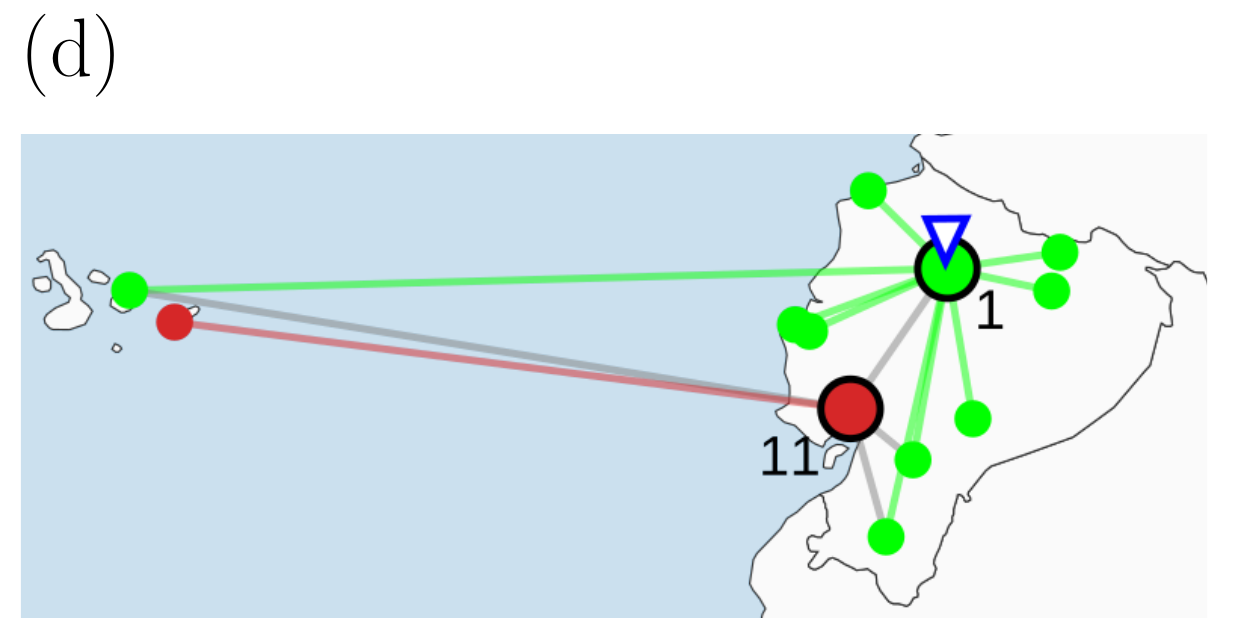} 
    \end{minipage}
    \\
     \\
    \multicolumn{2}{c}{
    \begin{minipage}{\hsize}
        \centering
        \includegraphics[width=\hsize]{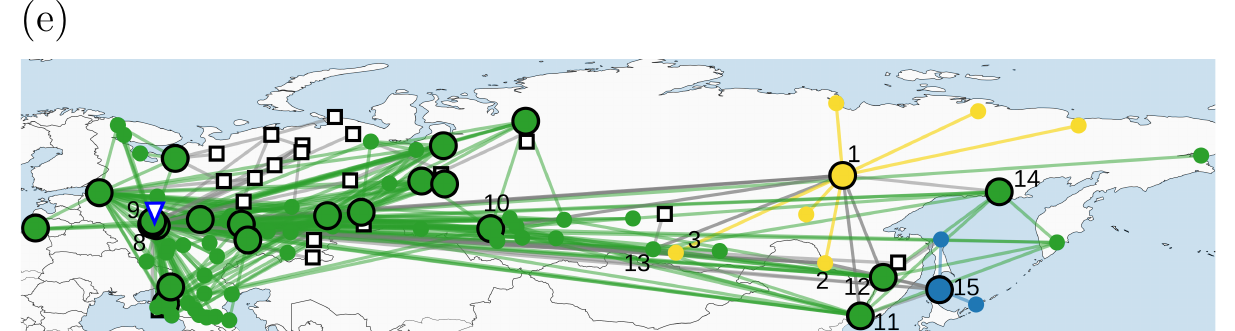} 
    \end{minipage}
    }
\end{tabular}

\caption{
Airport network within ({a}) Iran, ({b}) Nigeria, ({c}) core-periphery pair 6 based in Alaska, ({d}) Ecuador and ({e}) Russia. 
The line colour indicates the core-periphery pair to which the two airports belong.
The edges connecting two airports in different core-periphery pairs are shown in grey.
The numbers attached to some airports indicate the IDs of the airports listed in Tables~\ref{ta:iran}--\ref{ta:russia}.
We only show the IDs of the core airports, some peripheral airports and some residual airports. 
}
\label{fig:small_cp}
\end{figure}

Our algorithm separates the international and domestic airports in the Philippines, Iran and Nigeria into different core-periphery pairs.
In Iran, the airports mainly serving the international and domestic flights belong to core-periphery pairs 1 and 5, respectively (Table~\ref{ta:iran}).
In Nigeria, the airports mainly serving the international and domestic flights belong to core-periphery pairs 1 and 8, respectively (Table~\ref{ta:nigeria}). 
There is no clear geographical division of the international and domestic airports in Iran and Nigeria (Figs.~\ref{fig:small_cp}(a) and (b)).

Some core airports in Alaska, Ecuador and Russia serve as gateway airports in the respective regions. 
In core-periphery pair 6 based in Alaska, 
core airport 1 is adjacent to all the other airports in this core-periphery pair (Fig.~\ref{fig:small_cp}(c)). 
In addition, only core airport 1 has an edge to the rest of the network (Table~\ref{ta:alaska}) and therefore is the unique gateway airport for this core-periphery pair.
In Ecuador, most of the airports (10 airports; 83\%) are adjacent to airport 1, which is the unique core airport in core-periphery pair 9 (Fig.~\ref{fig:small_cp}(d)).
This core airport has most of the edges (10 edges; 77\%) between core-periphery pair 9 and the rest of the network (Table~\ref{ta:ecuador}).
Therefore, core airport 1 serves as a gateway airport in Ecuador. 
Airport 11 also functions as a gateway airport in Ecuador. 
The Russian airports belong to core-periphery pair 1, 2 or 7 (Fig.~\ref{fig:small_cp}(e)).
Most of the airports in core-periphery 7 are located in Russian Far East. 
In core-periphery 7, all peripheral airports are adjacent to core airport 1.
The core airport 1 has most of the edges (eight edges; 67\%) between core-periphery pair 7 and the rest of the network (Table~\ref{ta:russia}).
Therefore, core airport 1 serves as a gateway airport for this core-periphery pair.
There is no clear separation between the domestic and international airports into different core-periphery pairs in Russia.

\newpage
\clearpage

\clearpage

\clearpage

\clearpage

\clearpage
\clearpage 
\clearpage 
\clearpage
\clearpage
\clearpage
\clearpage
\clearpage
\clearpage
\clearpage
\clearpage
\begin{table}
\caption{
Properties of the core-periphery pairs in the airport network.
The core-periphery pairs are ordered according to the number of the core nodes.
The core nodes that have the largest number of neighbours within the same core-periphery pair are shown as representative core nodes.
C--C, C--P, and  P--P denote core--core, core--periphery and periphery--periphery edges, respectively.
}
\label{ta:profile}
\vspace{0.5em} 
\scalebox{1}{
\begin{tabular}{cccccccclll} \hline
\multirow{2}{*}{Pair} &
\multicolumn{2}{c}{Number of nodes} & & 
\multicolumn{3}{c}{Fraction of edges within a pair} & & \multicolumn{3}{c}{Representative core nodes} \\ \cline{2-3} \cline{5-7} \cline{9-11} 
    & Core &  Periphery & & \multirow{1}{4em}{\centering C--C} & \multirow{1}{4em}{\centering C--P} & \multirow{1}{4em}{\centering P--P} & & IFTA & City & Country \\ \hline \hline 
 &  &  &   &  &  &  &&  FRA & Frankfurt & Germany\\
1 & 333 & 378 &   & 0.0825 & 0.0194 & 0.0005 &&  CDG & Paris & France\\
 &  &  &   &  &  &  &&  AMS & Amsterdam & Netherlands\\ \hline
 &  &  &   &  &  &  &&  PEK & Beijing & China\\
2 & 161 & 240 &   & 0.0964 & 0.0234 & 0.0000 &&  CAN & Guangzhou & China\\
 &  &  &   &  &  &  &&  PVG & Shanghai & China\\ \hline
 &  &  &   &  &  &  &&  ATL & Atlanta & USA\\
3 & 150 & 312 &   & 0.1372 & 0.0265 & 0.0003 &&  LAS & Las Vegas & USA\\
 &  &  &   &  &  &  &&  MCO & Orlando & USA\\ \hline
4 & 5 & 32 &   & 0.4000 & 0.3312 & 0.0000 &&  MNL & Manila & Philippines\\ \hline
5 & 4 & 21 &   & 0.5000 & 0.2976 & 0.0000 &&  THR & Teheran & Iran\\ \hline
6 & 2 & 8 &   & 1.0000 & 0.5625 & 0.0000 &&  ADQ & Kodiak & USA\\ \hline
7 & 1 & 6 &   & 0.0000 & 1.0000 & 0.0000 &&  YKS & Yakutsk & Russia\\ \hline
8 & 1 & 9 &   & 0.0000 & 1.0000 & 0.0000 &&  LOS & Lagos & Nigeria\\ \hline
9 & 1 & 9 &   & 0.0000 & 1.0000 & 0.0000 &&  UIO & Quito & Ecuador\\ \hline
10 & 1 & 12 &   & 0.0000 & 1.0000 & 0.0000 &&  DMK & Bangkok & Thailand\\ \hline
\end{tabular}
}
\end{table} \clearpage
\begin{table}
\centering
\caption{
Properties of the airports in the Phillipines. The airports are sorted in the descending order of the total number of edges.
The internal edge is defined as that between two airports within the same core-periphery pair.
The external edge is defined as that between an airport in the focal core-periphery pair and an airport in a different core-periphery pair or a residual airport. 
}
\label{ta:philippines}
\scalebox{0.6}{
\begin{tabular}{clclcccc}
\multirow{3}{*}{ID} & \multirow{3}{*}{IFTA} & \multirow{3}{2.5em}{\centering Pair} & \multirow{3}{2.5em}{Type} &\multicolumn{4}{c}{Number of edges} \\ \cline{5-8} 
 & & & &  \multirow{2}{4em}{\centering Domestic} &  \multirow{2}{4em}{\centering Inter-national} & \multirow{2}{4em}{\centering Internal} & \multirow{2}{4em}{\centering External} \\ \ 
 & & & &   &   &  &  \\ \hline \hline 
1 & MNL & 4 & core & 36 & 40 & 35 & 41\\
2 & CEB & 4 & core & 19 & 6 & 18 & 7\\
3 & DVO & 4 & core & 5 & 1 & 5 & 1\\
4 & ZAM & 4 & periphery & 4 & 0 & 4 & 0\\
5 & CGY & 4 & periphery & 3 & 0 & 3 & 0\\
6 & ILO & 4 & periphery & 3 & 0 & 3 & 0\\
7 & PPS & 4 & periphery & 3 & 0 & 3 & 0\\
8 & USU & 4 & core & 2 & 0 & 2 & 0\\
9 & PAG & 4 & periphery & 2 & 0 & 2 & 0\\
10 & TAC & 4 & periphery & 2 & 0 & 2 & 0\\
11 & BCD & 4 & periphery & 2 & 0 & 2 & 0\\
12 & DGT & 4 & periphery & 2 & 0 & 2 & 0\\
13 & MPH & 4 & periphery & 2 & 0 & 2 & 0\\
14 & BXU & 4 & periphery & 2 & 0 & 2 & 0\\
15 & DPL & 4 & periphery & 2 & 0 & 2 & 0\\
16 & LGP & 4 & periphery & 2 & 0 & 2 & 0\\
17 & OZC & 4 & periphery & 2 & 0 & 2 & 0\\
18 & GES & 4 & periphery & 2 & 0 & 2 & 0\\
19 & SUG & 4 & periphery & 2 & 0 & 2 & 0\\
20 & CRM & 4 & periphery & 2 & 0 & 2 & 0\\
21 & JOL & 4 & core & 1 & 0 & 1 & 0\\
22 & CBO & 4 & periphery & 1 & 0 & 1 & 0\\
23 & SJI & 4 & periphery & 1 & 0 & 1 & 0\\
24 & TAG & 4 & periphery & 1 & 0 & 1 & 0\\
25 & LAO & 4 & periphery & 1 & 0 & 1 & 0\\
26 & ENI & 4 & periphery & 1 & 0 & 1 & 0\\
27 & WNP & 4 & periphery & 1 & 0 & 1 & 0\\
28 & BSO & 4 & periphery & 1 & 0 & 1 & 0\\
29 & SFE & 4 & periphery & 1 & 0 & 1 & 0\\
30 & TUG & 4 & periphery & 1 & 0 & 1 & 0\\
31 & VRC & 4 & periphery & 1 & 0 & 1 & 0\\
32 & CYP & 4 & periphery & 1 & 0 & 1 & 0\\
33 & MBT & 4 & periphery & 1 & 0 & 1 & 0\\
34 & RXS & 4 & periphery & 1 & 0 & 1 & 0\\
35 & CYZ & 4 & periphery & 1 & 0 & 1 & 0\\
36 & TBH & 4 & periphery & 1 & 0 & 1 & 0\\
37 & MRQ & 4 & periphery & 1 & 0 & 1 & 0\\
38 & CRK & 2 & core & 1 & 8 & 8 & 1\\
39 & KLO & 2 & periphery & 1 & 3 & 3 & 1\\ \hline
\end{tabular}
}
\end{table}\clearpage
\begin{table}
\centering
\caption{
Properties of the airports in Thailand. 
}
\label{ta:thailand}
\scalebox{0.95}{
\begin{tabular}{clclcccc}
\multirow{3}{*}{ID} & \multirow{3}{*}{IFTA} & \multirow{3}{2.5em}{\centering Pair} & \multirow{3}{2.5em}{Type} &\multicolumn{4}{c}{Number of edges} \\ \cline{5-8} 
 & & & &  \multirow{2}{4em}{\centering Domestic} &  \multirow{2}{4em}{\centering Inter-national} & \multirow{2}{4em}{\centering Internal} & \multirow{2}{4em}{\centering External} \\ \ 
 & & & &   &   &  &  \\ \hline \hline 
1 & DMK & 10 & core & 17 & 0 & 12 & 5\\
2 & CEI & 10 & periphery & 2 & 0 & 1 & 1\\
3 & NST & 10 & periphery & 2 & 0 & 1 & 1\\
4 & URT & 10 & periphery & 2 & 0 & 1 & 1\\
5 & NNT & 10 & periphery & 2 & 0 & 1 & 1\\
6 & PHS & 10 & periphery & 1 & 0 & 1 & 0\\
7 & TST & 10 & periphery & 1 & 0 & 1 & 0\\
8 & SNO & 10 & periphery & 1 & 0 & 1 & 0\\
9 & LOE & 10 & periphery & 1 & 0 & 1 & 0\\
10 & KOP & 10 & periphery & 1 & 0 & 1 & 0\\
11 & ROI & 10 & periphery & 1 & 0 & 1 & 0\\
12 & BFV & 10 & periphery & 1 & 0 & 1 & 0\\
13 & MAQ & 10 & periphery & 1 & 0 & 1 & 0\\
14 & BKK & 2 & core & 14 & 122 & 69 & 67\\
15 & HKT & 2 & core & 7 & 21 & 23 & 5\\
16 & CNX & 2 & core & 8 & 6 & 12 & 2\\
17 & USM & 2 & core & 6 & 3 & 9 & 0\\
18 & HDY & 2 & periphery & 3 & 2 & 4 & 1\\
19 & UTH & 2 & periphery & 4 & 0 & 3 & 1\\
20 & KBV & 2 & periphery & 2 & 2 & 4 & 0\\
21 & UBP & 2 & periphery & 3 & 0 & 2 & 1\\
22 & UTP & 2 & periphery & 2 & 0 & 2 & 0\\
23 & TDX & 2 & periphery & 2 & 0 & 2 & 0\\
24 & NAW & 2 & periphery & 1 & 0 & 1 & 0\\
25 & KKC & 2 & periphery & 1 & 0 & 1 & 0\\
26 & HGN & 2 & periphery & 1 & 0 & 1 & 0\\
27 & THS & - & residual & 2 & 0 & 0 & 2\\
28 & LPT & - & residual & 1 & 0 & 0 & 1\\ \hline
\end{tabular}
}
\end{table} \clearpage
\begin{table}
\centering
\caption{
The average CPU time of the three algorithms on different networks.
We generate synthetic networks 1--4 using the stochastic block models schematically shown in Figs.~\ref{fig:sb}(a), \ref{fig:sb}(b), \ref{fig:sb}(c) and \ref{fig:sb}(d), respectively.
For each of them, we set $\theta_1=0.9$ and $\theta_2=0.05$ and measure the CPU time for one generated network.
}
\label{ta:cputime}
\begin{tabular}{lccccc}
\multirow{2}{2em}{Network} & \multirow{2}{2em}{\centering $N$} & \multirow{2}{2em}{\centering $M$} & \multicolumn{3}{c}{Average CPU time [second]} \\ \cline{4-6} 
& & & BE--KL & Two-step & Proposed \\ \hline \hline 
Synthetic~1 & 400 & 31331 &  0.186 & 1.076 & 0.356\\
Synthetic~2 & 400 & 16355 & 0.162 & 0.343 & 0.240\\
Synthetic~3 & 400 & 19308 & 0.187 & 1.117 & 0.366\\
Synthetic~4 & 400 & 10939 & 0.193 & 0.392 & 0.251\\
Karate & 34 & 78  & 0.006 & 0.023 & 0.025\\
Blog & 1222 & 0.022 & 5.076 & 9.080 & 1.321\\
Airport & 2939 & 15677 & 51.57 & 67.17 & 5.709\\ \hline
\end{tabular}
\end{table}
\begin{table}[tb]
\centering
\caption{
    Properties of airports in Iran.
}
\label{ta:iran}
\scalebox{0.7}{
\begin{tabular}{clclcccc}
\multirow{3}{*}{ID} & \multirow{3}{*}{IFTA} & \multirow{3}{2.5em}{\centering Pair} & \multirow{3}{2.5em}{Type} &\multicolumn{4}{c}{Number of edges} \\ \cline{5-8} 
 & & & &  \multirow{2}{4em}{\centering Domestic} &  \multirow{2}{4em}{\centering Inter-national} & \multirow{2}{4em}{\centering Internal} & \multirow{2}{4em}{\centering External} \\ \ 
 & & & &   &   &  &  \\ \hline \hline 
1 & THR & 5 & core & 36 & 4 & 24 & 16\\
2 & ZAH & 5 & periphery & 5 & 2 & 3 & 4\\
3 & KER & 5 & core & 6 & 0 & 3 & 3\\
4 & KSH & 5 & periphery & 5 & 0 & 3 & 2\\
5 & PGU & 5 & core & 4 & 0 & 2 & 2\\
6 & ABD & 5 & periphery & 4 & 0 & 1 & 3\\
7 & GBT & 5 & core & 3 & 0 & 2 & 1\\
8 & MRX & 5 & periphery & 3 & 0 & 1 & 2\\
9 & BUZ & 5 & periphery & 2 & 1 & 1 & 2\\
10 & OMH & 5 & periphery & 1 & 1 & 1 & 1\\
11 & XBJ & 5 & periphery & 2 & 0 & 1 & 1\\
12 & GSM & 5 & periphery & 2 & 0 & 1 & 1\\
13 & NSH & 5 & periphery & 2 & 0 & 1 & 1\\
14 & ADU & 5 & periphery & 2 & 0 & 1 & 1\\
15 & CQD & 5 & periphery & 1 & 1 & 1 & 1\\
16 & SDG & 5 & periphery & 1 & 0 & 1 & 0\\
17 & RZR & 5 & periphery & 1 & 0 & 1 & 0\\
18 & KHD & 5 & periphery & 1 & 0 & 1 & 0\\
19 & BXR & 5 & periphery & 1 & 0 & 1 & 0\\
20 & BJB & 5 & periphery & 1 & 0 & 1 & 0\\
21 & AFZ & 5 & periphery & 1 & 0 & 1 & 0\\
22 & ACP & 5 & periphery & 1 & 0 & 1 & 0\\
23 & IIL & 5 & periphery & 1 & 0 & 1 & 0\\
24 & PFQ & 5 & periphery & 1 & 0 & 1 & 0\\
25 & YES & 5 & periphery & 1 & 0 & 1 & 0\\
26 & IKA & 1 & core & 1 & 43 & 40 & 4\\
27 & MHD & 1 & core & 21 & 17 & 26 & 12\\
28 & SYZ & 1 & periphery & 14 & 9 & 17 & 6\\
29 & IFN & 1 & core & 11 & 5 & 11 & 5\\
30 & BND & 1 & core & 12 & 2 & 12 & 2\\
31 & AWZ & 1 & core & 8 & 5 & 12 & 1\\
32 & TBZ & 1 & core & 6 & 4 & 9 & 1\\
33 & KIH & 1 & core & 7 & 1 & 7 & 1\\
34 & RAS & 1 & periphery & 7 & 1 & 6 & 2\\
35 & ZBR & 1 & periphery & 4 & 2 & 4 & 2\\
36 & LRR & 1 & core & 2 & 3 & 4 & 1\\
37 & AZD & 1 & periphery & 3 & 1 & 3 & 1\\
38 & SRY & 1 & periphery & 3 & 1 & 3 & 1\\
39 & BDH & - & residual & 1 & 0 & 0 & 1\\ \hline
\end{tabular}
}
\end{table}\clearpage
\begin{table}[tb]
\centering
\caption{
    Properties of the airports in Nigeria.
}
\label{ta:nigeria}
\scalebox{1}{
\begin{tabular}{clclcccc}
\multirow{3}{*}{ID} & \multirow{3}{*}{IFTA} & \multirow{3}{2.5em}{\centering Pair} & \multirow{3}{2.5em}{Type} &\multicolumn{4}{c}{Number of edges} \\ \cline{5-8} 
 & & & &  \multirow{2}{4em}{\centering Domestic} &  \multirow{2}{4em}{\centering Inter-national} & \multirow{2}{4em}{\centering Internal} & \multirow{2}{4em}{\centering External} \\ \ 
 & & & &   &   &  &  \\ \hline \hline 
1 & ABV & 8 & core & 11 & 8 & 9 & 10\\
2 & AKR & 8 & periphery & 2 & 0 & 1 & 1\\
3 & BNI & 8 & periphery & 2 & 0 & 1 & 1\\
4 & CBQ & 8 & periphery & 2 & 0 & 1 & 1\\
5 & PHC & 8 & periphery & 2 & 0 & 1 & 1\\
6 & QOW & 8 & periphery & 2 & 0 & 1 & 1\\
7 & QRW & 8 & periphery & 2 & 0 & 1 & 1\\
8 & KAD & 8 & periphery & 2 & 0 & 1 & 1\\
9 & ILR & 8 & periphery & 1 & 0 & 1 & 0\\
10 & SKO & 8 & periphery & 1 & 0 & 1 & 0\\
11 & LOS & 1 & core & 11 & 23 & 23 & 11\\
12 & KAN & 1 & periphery & 3 & 5 & 6 & 2\\
13 & ENU & - & residual & 2 & 0 & 0 & 2\\ 
14 & JOS & - & residual & 1 & 0 & 0 & 1\\\hline
\end{tabular}
}
\end{table}\clearpage
\begin{table}[tb]
\centering
\caption{
    Properties of the airports in core-periphery pair 6 based in Alaska. 
}
\label{ta:alaska}
\scalebox{1}{
\begin{tabular}{clclcccc}
\multirow{3}{*}{ID} & \multirow{3}{*}{IFTA} & \multirow{3}{2.5em}{\centering Pair} & \multirow{3}{2.5em}{Type} &\multicolumn{4}{c}{Number of edges} \\ \cline{5-8} 
 & & & &  \multirow{2}{4em}{\centering Domestic} &  \multirow{2}{4em}{\centering Inter-national} & \multirow{2}{4em}{\centering Internal} & \multirow{2}{4em}{\centering External} \\ \ 
 & & & &   &   &  &  \\ \hline \hline 
1 & ADQ & 6 & core & 10 & 0 & 9 & 1\\
2 & ORI & 6 & core & 2 & 0 & 2 & 0\\
3 & KOZ & 6 & periphery & 2 & 0 & 2 & 0\\
4 & AKK & 6 & periphery & 1 & 0 & 1 & 0\\
5 & KLN & 6 & periphery & 1 & 0 & 1 & 0\\
6 & OLH & 6 & periphery & 1 & 0 & 1 & 0\\
7 & ALZ & 6 & periphery & 1 & 0 & 1 & 0\\
8 & AOS & 6 & periphery & 1 & 0 & 1 & 0\\
9 & KKB & 6 & periphery & 1 & 0 & 1 & 0\\
10 & KPY & 6 & periphery & 1 & 0 & 1 & 0\\ \hline
\end{tabular}
}
\end{table}\clearpage
\begin{table}
\centering
\caption{
    Properties of the airports in Ecuador. 
}
\label{ta:ecuador}
\scalebox{1}{
\begin{tabular}{clclcccc}
\multirow{3}{*}{ID} & \multirow{3}{*}{IFTA} & \multirow{3}{2.5em}{\centering Pair} & \multirow{3}{2.5em}{Type} &\multicolumn{4}{c}{Number of edges} \\ \cline{5-8} 
 & & & &  \multirow{2}{4em}{\centering Domestic} &  \multirow{2}{4em}{\centering Inter-national} & \multirow{2}{4em}{\centering Internal} & \multirow{2}{4em}{\centering External} \\ \ 
 & & & &   &   &  &  \\ \hline \hline 
1 & UIO & 9 & core & 10 & 9 & 9 & 10\\
2 & CUE & 9 & periphery & 2 & 0 & 1 & 1\\
3 & GPS & 9 & periphery & 2 & 0 & 1 & 1\\
4 & LOH & 9 & periphery & 2 & 0 & 1 & 1\\
5 & OCC & 9 & periphery & 1 & 0 & 1 & 0\\
6 & XMS & 9 & periphery & 1 & 0 & 1 & 0\\
7 & MEC & 9 & periphery & 1 & 0 & 1 & 0\\
8 & PVO & 9 & periphery & 1 & 0 & 1 & 0\\
9 & ESM & 9 & periphery & 1 & 0 & 1 & 0\\
10 & LGQ & 9 & periphery & 1 & 0 & 1 & 0\\
11 & GYE & 3 & core & 5 & 11 & 11 & 5\\
12 & SCY & 3 & periphery & 1 & 0 & 1 & 0\\ \hline
\end{tabular}
}
\end{table}\clearpage
\begin{table}
\centering
\caption{
Properties of the airports in Russia. 
Only the airports in core-periphery pair 7 and those in other core-periphery pairs that are adjacent to core-periphery pair 7 are shown. 
}
\label{ta:russia}
\scalebox{1}{
\begin{tabular}{clclcccc}
\multirow{3}{*}{ID} & \multirow{3}{*}{IFTA} & \multirow{3}{2.5em}{\centering Pair} & \multirow{3}{2.5em}{Type} &\multicolumn{4}{c}{Number of edges} \\ \cline{5-8} 
 & & & &  \multirow{2}{4em}{\centering Domestic} &  \multirow{2}{4em}{\centering Inter-national} & \multirow{2}{4em}{\centering Internal} & \multirow{2}{4em}{\centering External} \\ \ 
 & & & &   &   &  &  \\ \hline \hline 
1 & YKS & 7 & core & 13 & 1 & 6 & 8\\
2 & BQS & 7 & periphery & 4 & 0 & 1 & 3\\
3 & UUD & 7 & periphery & 2 & 0 & 1 & 1\\
4 & CKH & 7 & periphery & 1 & 0 & 1 & 0\\
5 & CYX & 7 & periphery & 1 & 0 & 1 & 0\\
6 & IKS & 7 & periphery & 1 & 0 & 1 & 0\\
7 & NER & 7 & periphery & 1 & 0 & 1 & 0\\
8 & DME & 1 & core & 62 & 97 & 134 & 25\\
9 & VKO & 1 & core & 34 & 15 & 42 & 7\\
10 & OVB & 1 & core & 14 & 18 & 27 & 5\\
11 & VVO & 1 & core & 13 & 7 & 11 & 9\\
12 & KHV & 1 & core & 14 & 5 & 11 & 8\\
13 & IKT & 1 & periphery & 9 & 5 & 10 & 4\\
14 & GDX & 1 & core & 8 & 0 & 7 & 1\\
15 & UUS & 2 & core & 7 & 6 & 8 & 5\\ \hline
\end{tabular}
}
\end{table}
\end{document}